\documentclass[letter]{article}
\title{Methods and Models for Interpretable Linear Classification}
\author{Berk Ustun and Cynthia Rudin}

\usepackage{amsfonts,bbm,bm,courier}
\usepackage{hyperref}
\usepackage{etex}
\usepackage{relsize}
\usepackage{rotating}

\usepackage[numbers,square]{natbib}

\usepackage{amsmath,amssymb}
\usepackage{mathtools}
\usepackage{eqnarray}

\usepackage{array}
\usepackage{tabularx}
\usepackage{multirow}
\usepackage{hhline}
\usepackage{booktabs}
\newcommand{\cell}[2]{\setlength{\tabcolsep}{0pt}\begin{tabular}{#1}#2 \end{tabular}}
\newcommand{\bfcell}[2]{\setlength{\tabcolsep}{0pt}\textbf{\begin{tabular}{#1}#2\end{tabular}}}
\newcommand{\ddcell}[1]{\setlength{\tabcolsep}{0pt}\begin{tabular}{l>{\quad}l}#1\end{tabular}}

\usepackage{comment}
\usepackage{verbatim}
\usepackage{fancyvrb}
\usepackage{colortbl}
\usepackage{placeins}

\usepackage{enumitem} 

\usepackage{algorithm}
\usepackage{algorithmic}

\usepackage{graphicx}
\usepackage{epstopdf} 
\usepackage{subfigure}
\usepackage[font=small,labelfont=bf,width=.75\textwidth]{caption}
\usepackage{tikz,pgfplots}
\usetikzlibrary{shapes,arrows}
\usetikzlibrary{matrix}
\pgfplotsset{compat=newest}
\pgfplotsset{plot coordinates/math parser=false}
\newlength\fheight
\newlength\fwidth
\newcommand{\tabdataname}[1]{\begin{tabular}{c}\texttt{#1}\end{tabular}}

\newtheorem{thm}{Theorem}

\newtheorem{corollary}{Corollary}
\newenvironment{proof}[1][Proof]{\begin{trivlist}
\item[\hskip \labelsep {\bfseries #1}]}{\end{trivlist}}

\newcommand{\qed}{\nobreak \ifvmode \relax \else
      \ifdim\lastskip<1.5em \hskip-\lastskip
      \hskip1.5em plus0em minus0.5em \fi \nobreak
      \vrule height0.75em width0.5em depth0.25em\fi}

\newcommand{\textds}[1]{\texttt{#1}}

\DeclareMathOperator*{\argmin}{argmin}
\newcommand{\lzero}{L_0}
\newcommand{\lone}{L_1}
\newcommand{\ltwo}{L_2}

\newcommand{\vnorm}[1]{\left\|#1\right\|}
\newcommand{\indic}[1]{\mathbbm{1}\left[#1\right]}
\newcommand{\ZeroOneLoss}[1]{\frac{1}{N}\sum_{i=1}^N \indic{y_i #1^T \xb_i \leq 0}}
\newcommand{\LogLoss}[1]{\frac{1}{N}\sum_{i=1}^N \log (1 + \exp(- y_i #1^T\xb_i))}

\newcommand{\for}{\textnormal{ for }}
\newcommand{\sign}[1]{\textnormal{sign}\left(#1\right)}
\newcommand{\lambdab}{\bm{\lambda}}
\newcommand{\xb}{\bm{x}}
\newcommand{\rhob}{\bm{\rho}}

\newcommand{\iplus}{\mathcal{I}^{+}} 
\newcommand{\iminus}{\mathcal{I}^{-}}
\newcommand{\wplus}{W^{+}} 
\newcommand{\wminus}{W^{-}}
\newcommand{\nplus}{N^{+}}
\newcommand{\nminus}{N^{-}}

\newcommand{\data}{\mathcal{D}}
\newcommand{\Lset}{\mathcal{L}}
\newcommand{\X}{\mathcal{X}}
\newcommand{\Y}{\mathcal{Y}}
\newcommand{\F}{\mathcal{F}}
\newcommand{\I}{\mathcal{I}}
\newcommand{\J}{\mathcal{J}}

\newcommand{\R}{\mathbb{R}}

\newcommand{\Z}{\mathbb{Z}}
\newcommand{\B}{\{0,1\}}

\newcommand{\OriginalP}{\mathcal{P}}

\newcommand{\ProxyP}[1]{\tilde{\mathcal{P}}_{#1}}
\newcommand{\CutPoints}[1]{\mathcal{H}^{#1}}

\newcommand{\IntPen}{\Phi}
\newcommand{\loss}{\psi}
\newcommand{\Loss}[1]{\textnormal{Loss}\left(#1\right)}
\newcommand{\dLoss}[1]{\left( \nabla \textnormal{Loss}\right)_{#1}}
\newcommand{\ApproxLoss}[1]{\widetilde{\textnormal{Loss}}\left(#1\right)}

\newcommand{\st}{\textnormal{s.t.}}
\newcommand{\mprange}[3]{{#1}={#2}\textnormal{,...,}{#3}}
\newcommand{\mpdes}[1]{\textit{\tiny #1}}

\newcommand{\pkg}[1]{{\fontseries{b}\selectfont #1}} 

\begin{document}
\maketitle
\abstract{
We present an integer programming framework to build accurate and interpretable discrete linear classification models. Unlike existing approaches, our framework is designed to provide practitioners with the control and flexibility they need to tailor accurate and interpretable models for a domain of choice. To this end, our framework can produce models that are fully optimized for accuracy, by minimizing the 0--1 classification loss, and that address multiple aspects of interpretability, by incorporating a range of discrete constraints and penalty functions. We use our framework to produce models that are difficult to create with existing methods, such as scoring systems and M-of-N rule tables. In addition, we propose specially designed optimization methods to improve the scalability of our framework through decomposition and data reduction. We show that discrete linear classifiers can attain the training accuracy of any other linear classifier, and provide an Occam's Razor type argument as to why the use of small discrete coefficients can provide better generalization. We demonstrate the performance and flexibility of our framework through numerical experiments and a case study in which we construct a highly tailored clinical tool for sleep apnea diagnosis.
}

\maketitle

%

%
\section{Introduction}\label{Sec::Introduction}
\begin{quote}
\textit{``Each time one of our favorite [machine learning\ldots] approaches has been applied in industry, \break the [interpretability\ldots ] of the results, though ill-defined, has been a decisive factor of choice."}
\par \hspace{5cm} --- Yves Kodratoff, \textit{The Comprehensibility Manifesto}
\end{quote}
Possibly \textit{the} greatest obstacle in the deployment of predictive models is the fact that humans simply do not trust them. Consider a case where the prediction of a black-box machine learning model disagrees with a doctor's intuition on a high-stakes medical decision: if only we could explain how the model combined various input variables to generate its prediction, then we could use this information to validate its prediction, and convince the doctor to make the right decision. 

Recent research in statistics and machine learning has primarily focused on designing accurate and scalable black-box models to address complex automation problems such as spam prediction and computer vision \citep{Freitas:2014ic}. In turn, the goal of creating interpretable models -- once recognized as a holy grail in the fields of expert systems and artificial intelligence -- has been neglected over the last two decades. Even so, interpretable models are far more likely to be accepted across numerous domains because they are easy to explain, easy to troubleshoot, and capable of producing insights from data. These domains include credit scoring \citep{martens2007comprehensible}, crime prediction \citep{steinhart2006juvenile,andrade2009handbook,pitfall}, national defense \citep{abs2002marine}, marketing \citep{hauser2010disjunctions,verbeke2011building}, medical diagnosis  \citep{tian2011adaptive,van2012mathematical}, and scientific discovery \citep{sun2006accurate,freitas2010importance,haury2011influence}. 

Interpretable models provide ``a \textit{qualitative understanding} of the relationship between joint values of the input variables and the resulting predicted response value," \citep{ESL}. The process of creating models that convey such qualitative understanding, however, is inherently complicated due to the fact that interpretability is a \textit{subjective} and \textit{multifaceted} notion \citep{kodratoff1994comprehensibility,pazzani2000knowledge,Freitas:2014ic}. Models that are highly interpretable to one audience may be completely uninterpretable to others due to differences in their affinity for certain types of knowledge representation, their exposure to the data, and/or their domain expertise \citep{kodratoff1994comprehensibility,martens2008building,Freitas:2014ic}. 
In practice, the interpretability of a predictive model is therefore often addressed through a \textit{tailoring} process, in which practitioners adjust \textit{multiple} qualities such as:
\begin{enumerate}[leftmargin=0.45cm,topsep=2pt,parsep=2pt]

\item \textit{Sparsity}: According to \citet{miller1984selection}, humans can only handle a few cognitive entities at once ($7\pm 2$). In statistics, sparsity refers to the number of terms in a model and constitutes the standard measure of model complexity \citep{sommer1996theory,ruping2006learning}. Sparsity has drawbacks as a measure of interpretability because models that are too sparse are thought to oversimplify complicated problems \citep{Freitas:2014ic}. 

\item \textit{Expository Power}: Humans are seriously limited in estimating the association between three or more variables \citep{jennings1982informal}. Linear models help us gauge the influence of one input variable with respect to the others by comparing their coefficients. Many medical scoring systems \citep[e.g.][]{antman2000timi} and criminology risk assessment tools \citep[e.g.][]{steinhart2006juvenile,psych} enhance the expository power of linear models by using integer coefficients. This approach has recently been adopted by \citet{chevaleyre2013rounding} and \citet{carrizosaDILSVM13}.

\item \textit{Monotonicity}: \citet{ruping2006learning} warns that humans tend to find a fact understandable if they are already aware of it. He illustrates this idea using the statement ``rhinoceroses can fly," - a very understandable assertion that no one would believe. Unfortunately, the signs of coefficients in many linear models are at odds with views of domain experts due to the correlation between variables. In turn, recent approaches to interpretable predictive modeling have sought to produce models with monotonicity constraints so that the relationship between input variables and the predicted response value matches the views of domain experts \citep{ben1995monotonicity,pazzani2001acceptance,verbeke2011building, martens2011performance}. 

\end{enumerate}

State-of-the-art methods for linear classification were not designed for building interpretable predictive models. These methods were primarily designed to be scalable -- making approximations in how they measure accuracy (i.e. by using surrogate loss functions, such as the logistic loss), how they measure interpretability (by using proxy measures, such as the $\lone$-norm), or their optimization process (by using heuristics). Methods that use approximations produce models that are not fully optimized for accuracy or interpretability. Moreover, they provide practitioners with poor control in the training process, as practitioners have to perform extensive tuning in order to obtain a model that satisfies even simple constraints on accuracy and interpretability.

In this paper, we introduce a framework for building accurate and interpretable predictive models. Our framework uses integer programming (IP) to produce linear classification models with discrete coefficients. Unlike existing methods, our approach is primarily designed to help practitioners tailor accurate predictive models for their domain of choice. To this end, our framework avoids approximations and provides an unprecedented level of flexibility and control in the training process, allowing practitioners to: (i) optimize the 0--1 classification loss, which produces models that are highly accurate, completely robust to outliers, and that achieve the best learning-theoretic guarantee on accuracy; (ii) control the balance between accuracy and interpretability via meaningful regularization parameters that can be set purposefully, without the extensive tuning required of existing methods; (iii) incorporate preferences and constraints on a wide range of model qualities including sensitivity, specificity, sparsity, monotonicity, coefficient values, and feature composition. 

We illustrate how our framework can create a wide range of linear and rule-based models that are difficult to produce using existing methods, such as scoring systems and rule tables. In addition, we pair our models with specially designed optimization methods to assist with scalability, such as \textit{data reduction}, which eliminates some of the training data prior to the heavier integer programming computation, and \textit{loss decomposition}, which provides a means to train our models with any convex loss function (using an IP solver) and with polynomial running time in the number of examples.  We present theoretical results to show that our discrete linear classifiers can attain the training accuracy of any other linear classifier, and provide an Occam's Razor type argument as to why the use of small discrete coefficients can provide better generalization. We demonstrate the flexibility of our approach on a real-world problem by building a tailored clinical tool for sleep apnea diagnosis. Lastly, we present numerical experiments to show that our framework can produce accurate and interpretable models for many real-world datasets in minutes.
%
%
\subsection{Related Work}

Interpretability is a widely-used yet ``ill-defined" concept in the literature \citep{pazzani2000knowledge,kodratoff1994comprehensibility}. In this paper, we view interpretability as a notion that not only governs how easy it is to understand a predictive model in a particular domain, but also governs how likely it is for a predictive model to be used in that domain. Our view is aligned with many related works, which may refer to it using related terms and concepts such as \textit{comprehensibility} \citep{kodratoff1994comprehensibility,Freitas:2014ic}, \textit{acceptability} \citep{martens2008building}, and \textit{justifiability} \citep{martens2008building,martens2011performance}.

A comprehensive review on the interpretability of popular classification models can be found in \cite{Freitas:2014ic}. In assessing the interpretability of classification models, we distinguish between \textit{transparent} models, which provide a textual or visual representation of the relationship between input variables and the predicted outcome, and \textit{black-box} models, which do not. Popular transparent classification models include linear models (addressed in this work), decision trees \citep{quinlan1986induction,utgoff1989incremental,quinlan1993c4}, decision lists \citep{rivest1987learning,LethamRuMcMa13}, and decision tables \citep{kohavi1995power}. The interpretability of transparent models is usually improved by tuning sparsity \citep{tibshirani1996regression,zou2005regularization,efron2004least,hesterberg2008least,quinlan1999simplifying,breslow1997simplifying,kohavi1997wrappers,guyon2003introduction}, by ensuring monotonic relationships between certain input variables and the predicted outcome \citep{ben1995monotonicity,pazzani2001acceptance,verbeke2011building, martens2011performance}, and by restricting coefficients to a small set of integer values \citep{carrizosaDILSVM13,chevaleyre2013rounding}. Popular black-box models include artificial neural networks \citep{turing2004intelligent}, support vector machines \citep{vapnik1998statistical}, and ensemble models such as random forests \citep{breimanRF} and AdaBoost \citep{freund1997decision}. The interpretability of black-box models is mainly improved by auxiliary methods that extract rules and prototype examples to illustrate the relationship between input variables and the predicted outcome \citep{meinshausen2010node,van2007seeing,fung2005rule,martens2007comprehensible,bien2011prototype}. These rules and prototypes are useful for troubleshooting and generating insights, but do not allow practitioners to tailor models that fit the accuracy and interpretability constraints of a given domain.
 
In practice, training interpretable models involves a tailoring process that requires control over multiple qualities of models, and could even require control over qualities that have not been addressed in the literature \citep{pazzani2000knowledge}. Many existing methods do not include controls over multiple interpretability-related qualities, forcing practitioners to train a model that is either sparse, expository, or monotonic (see e.g. Section \ref{Sec::Demonstrations}). Further, existing methods make approximations in the way they measure accuracy (i.e. by using a convex surrogate loss function, \citealt{bartlett2003large}), the way they measure interpretability (by using proxy measures, such as an $\lone$-norm), or the way they train their models (i.e. by using heuristic procedures, such as rounding). Approximations result in a poor trade-off between accuracy and interpretability \citep{bratko1997machine,freitas2004critical,pitfall}. Linear classifiers that minimize surrogate loss functions, for instance, are not robust to outliers \citep{li2007optimizing,nguyen2013algorithms}. Similarly, linear classifiers that regularize the $\lone$-norm are only guaranteed to match the correct sparse solution (i.e. the one that minimizes the number of non-zero coefficients) under very restrictive conditions that are rarely satisfied in practice \citep{zhao2007model,liu2009estimation}. Performance issues aside, methods that use approximations often provide poor control over interpretability as they require extensive tuning of free parameters. \citet{pazzani2000knowledge}, for example, mentions ``we must adjust the available parameters with indirect control over these criteria until we satisfy the domain expert." 

There is a ``conflicting... [and unfounded].. set of claims in the literature as to which [type of model]... is easiest to understand," \citep{pazzani2000knowledge}. Methods that are specifically designed to be interpretable often market a specific and limited brand of interpretability, that may not produce models that are interpretable across all domains. This is consistent with field studies on interpretability, which often conclude that different predictive models are ``most" interpretable in different domains for different reasons \citep{subramanian1992comparison,kohavi1998targeting,allahyari2011user,huysmans2011empirical}. Even recent methods that are specifically designed to produce interpretable models make approximations. \citet{carrizosaDILSVM13}, for instance, use a MIP-based approach to produce discrete linear classification models that highlight the agreement of features and the outcome using a Likert scale, but train these models with the hinge loss (i.e. an approximate measure of accuracy). Similarly, \citet{chevaleyre2013rounding}, propose discrete linear classification as a way to create M-of-N rule tables \citep{Towell:1993tx}, but train these models using randomized rounding (i.e. an approximate means of optimization).

Our paper is about an integer programming (IP) framework to train linear classification models with \textit{discrete} linear coefficients with any discrete or convex loss function. Our use of integer programming is meant to avoid the use of approximations, and provides practitioners with flexibility and control in the training process. Mixed-integer programming (MIP) has been previously applied to classification problems, but not in this way \citep[see e.g.][]{lee2009classification,fan2009deterministic,carrizosa2013supervised}. Many MIP approaches deal with the \textit{misclassification minimization} problem, which trains a linear classification model with \textit{real} coefficients by minimizing the 0--1 loss \citep{Rubin1990,mangasarian1994misclassification,asparoukhov1997mathematical,rubin2009mixed}. Early attempts at misclassification minimization were only feasible for tiny datasets with at most $N=200$ examples \citep{joachimsthaler1990mathematical,erenguc1990survey}. Accordingly, a large body of work has focused on improving the scalability of misclassification minimization by modifying formulations \citep{brooks2011support}, applying heuristics \citep{Rubin1990,yanev1999combinatorial,Asparouhov2004}, and designing specialized algorithms \citep{soltysik1994warmack,Rubin1997,nguyen2013algorithms}. Recent progress in commercial MIP software has made it possible to solve exponentially larger problems \citep{bixby2004mixed,bixby2007progress}, and ushered in new MIP classification models that involve feature selection \citep{glen1999integer,GoldbergEc2012,guan2009mixed,nguyen2012general}, or the creation of a reserved-judgement region \citep{brooks2007mixed,brooks2010analysis}. 

Our framework can produce discrete linear models that attain the training accuracy of any other linear classifier (see Section \ref{Sec::DiscretizationBounds}). In addition, it can reproduce many interpretable linear models in the literature, such as those of \citet{tian2011adaptive,chevaleyre2013rounding,carrizosa2010binarized}, and \citet{carrizosaDILSVM13}, often providing substantial improvements in terms of accuracy, flexibility, control, and scalability. Our framework addresses many unresolved challenges that have been brought up in the literature such as:
\begin{itemize}[leftmargin=0.45cm,topsep=0pt,parsep=0pt]
\item the ability to control the trade-off between accuracy and interpretability (the need for which is mentioned by \citealt{bradley1999mathematical}, and addressed in Section \ref{Sec::Framework});
\item the ability to incorporate hard constraints on model size (the need for which is mentioned by \citealt{schwabacher2001discovering}, and addressed in Section \ref{Sec::FeatureBasedConstraints});
\item the ability to train models that scale to large databases (the need for which is mentioned by \citealt{bradley1999mathematical}, and addressed in Section \ref{Sec::Decomposition});
\item the need algorithms to remove redundant or irrelevant data (which is mentioned by \citealt{bradley1999mathematical}, and addressed in Section \ref{Sec::Reduction});
\end{itemize}
\clearpage
\section{Framework}\label{Sec::Framework}
We start with a dataset of $N$ training examples $\data_N = \{(\xb_i,y_i)\}_{i=1}^N$ where each $\xb_i \in \X \subseteq \R^{P+1}$ denotes a vector of features $[1, x_{i,1},\ldots,x_{i,P}]^T$ and each $y_i \in \Y = \{-1,1\}$ denotes a class label. We consider linear classification models of the form $y=\sign{\lambdab^T\xb}$, where $\lambdab \subseteq \R^{P+1}$ denotes a vector of coefficients $[\lambda_0, \lambda_1,\ldots,\lambda_P]^T$ and $\lambda_0$ denotes an intercept term. We determine the coefficients of our models by solving an optimization problem of the form:
\begin{align}
\begin{split}\label{Eq::InterpretabilityFrameworkLinear}
\min_{\lambdab} & \qquad \Loss{\lambdab;\data_N} + C \cdot \IntPen(\lambdab) \\ 
\st & \qquad \lambdab \in \Lset.
\end{split}
\end{align}
Here:
the \textit{loss function} $\Loss{\lambdab;\data_N}: \R^{P+1} \times (\X\times\Y)^N \to \R$ penalizes misclassifications;
the \textit{interpretability penalty function} $\IntPen(\lambdab): \R^{P+1} \to \R$ induces \textit{soft} interpretability-related qualities that are desirable but may be sacrificed for greater accuracy;
the \textit{interpretability set} $\Lset$ encodes \textit{hard} interpretability-related qualities that are absolutely required;
and the \textit{regularization parameter} $C$ controls the balance between accuracy and soft interpretability-related qualities.

We make the following assumptions, without loss of generality: 
(i) the interpretability set is specified component-wise so that $\Lset = \left\{\lambdab: \lambda_j \in \Lset_j \subseteq \R \for j = 0,\ldots,P \right\}$;
(ii) the interpretability set contains the null vector so that $\bf{0} \in \Lset$;
(iii) the interpretability penalty is additively separable so that $\IntPen(\lambdab) = \sum_{j=0}^P\IntPen(\lambda_j)$
(iv) the intercept is never penalized so that $\IntPen(\lambda_0) = 0$;
(v) the loss function for the data is an average over losses for the training examples so that $$\Loss{\lambdab;\data_N} = \frac{1}{N}\sum_{i=1}^N \Loss{\lambdab;(\xb_i,y_i)}.$$ 
\subsection{On Accuracy, Flexibility and Scalability}
We formulate the optimization problem in \eqref{Eq::InterpretabilityFrameworkLinear} as an integer program (IP) with discrete variables. Discrete variables provide us with flexibility and control by letting us directly formulate objectives and constraints in terms of quantities that we care about, without the use of approximations. Using discrete variables, for example, we can train models with the 0--1 loss function, $$\Loss{\lambdab;\data_N}=\ZeroOneLoss{\lambdab},$$ which directly measures the error rate. Similarly, we can regularize models with the $\lzero$-penalty, $\IntPen({\lambdab})=\vnorm{\lambdab}_0,$ which directly measures the number of non-zero coefficients. 

Classifiers that minimize the 0--1 loss are highly accurate, robust to outliers, and provide the best learning-theoretic guarantee on predictive accuracy \citep[see also][]{nguyen2013algorithms}. Because of this, classifiers that minimize a 0--1 loss and a user-defined interpretability penalty attain the best-possible trade-off between accuracy and interpretability: when we train models with the 0--1 loss function and the $\ell_0$-penalty, for example, we only sacrifice classification accuracy to attain higher sparsity, and vice versa. There are no additional sources of bias due to computational shortcuts, such as a convex surrogate loss function, or $\lone$-regularization term on the coefficients. As we will show, these shortcuts can hinder accuracy, interpretability and control.

Our framework produces linear models whose coefficients lie within a discrete interpretability set $\Lset$. Using discrete interpretability, we can train models that are difficult to create with the state-of-the-art, such as scoring systems and rule tables (see Section \ref{Sec::Models}), and also encode complicated accuracy and interpretability-related constraints without tuning free parameters (see Section \ref{Sec::Demonstrations}). In theory, we can craft $\Lset$ so that a linear classifier with discrete coefficients $\lambdab \in \Lset$ is at least as accurate as any linear classifier with real coefficients $\rho \in \R^{P+1}$ (see Section \ref{Sec::DiscretizationBounds}). In practice, we find that linear classifiers with coefficients $\lambdab \in \Lset$ are highly accurate even when the coefficients are restricted to a small discrete $\Lset$ set, and we demonstrate this via numerical experiments in Section \ref{Sec::NumericalExperiments}. While there is a sacrifice made in accuracy for restricting coefficients to a discrete set, our experiments show that this sacrifice is not that bad. In fact, a worse sacrifice is often made by using approximation measures to induce accuracy or sparsity. 
%

\begin{figure}[htbp]
\centering
\includegraphics[width=0.31\textwidth,trim=2mm 2.5mm 0mm 2.5mm,clip=true]{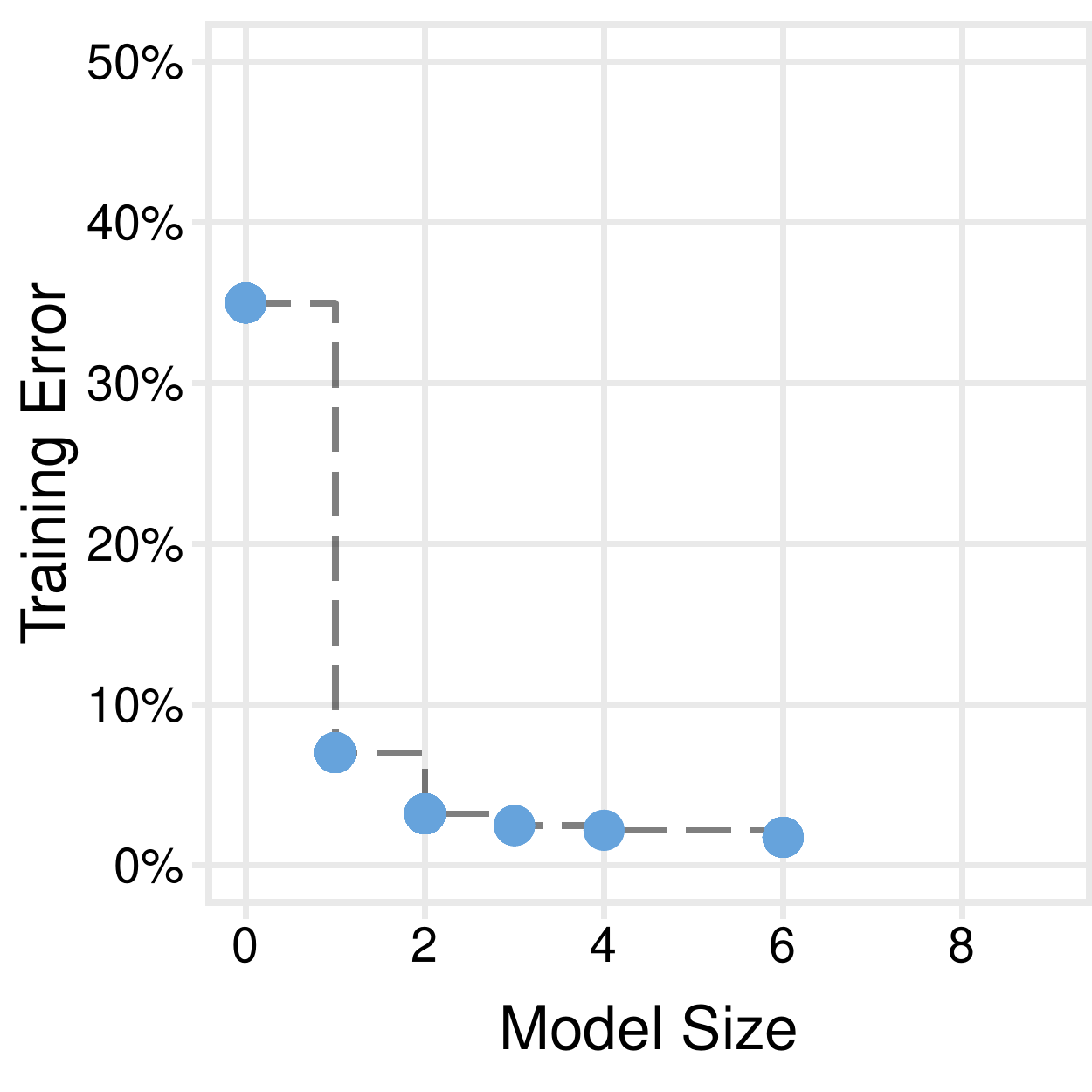}\quad
\includegraphics[width=0.31\textwidth,trim=2mm 2.5mm 0mm 2.5mm,clip=true]{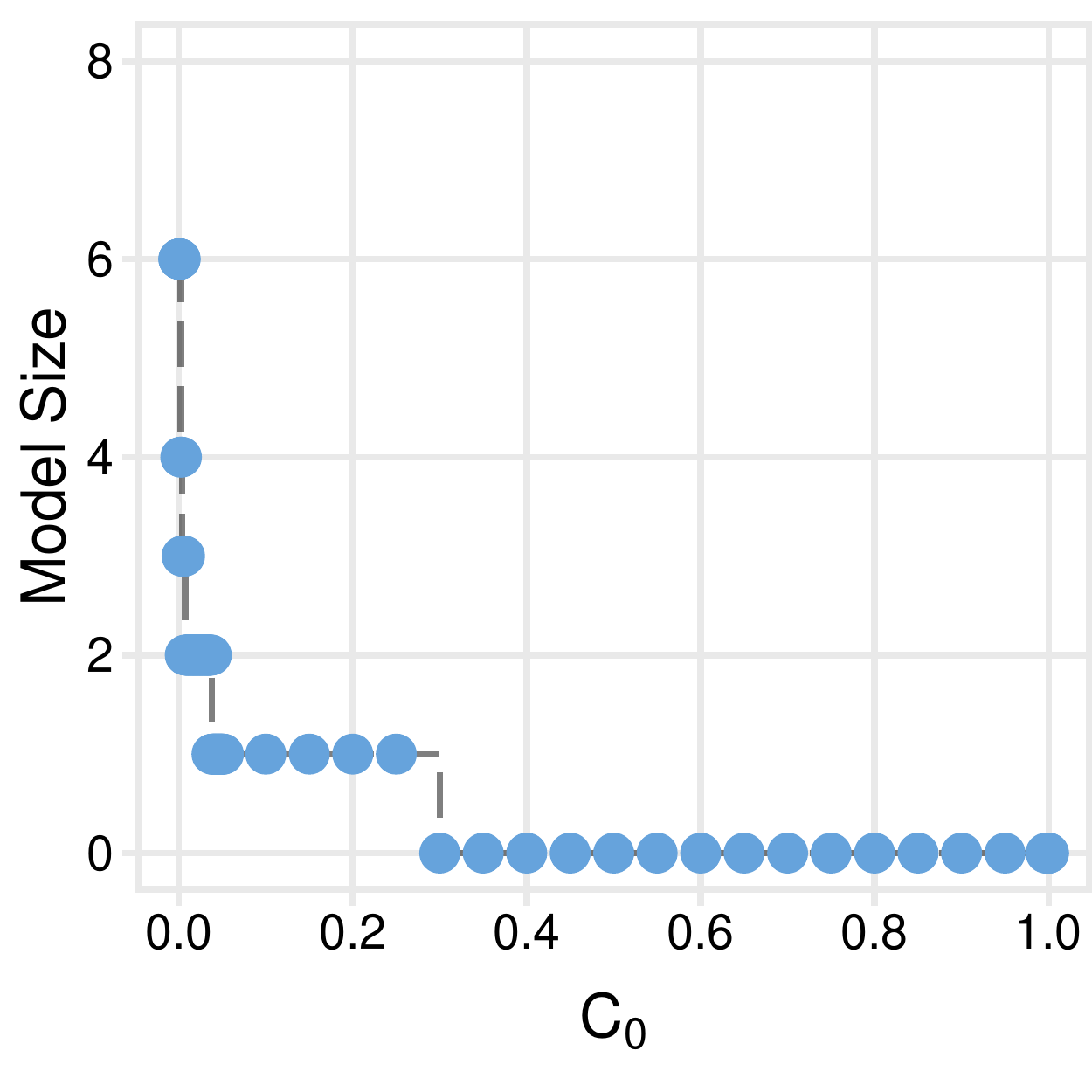}\quad
\includegraphics[width=0.31\textwidth,trim=2mm 2.5mm 0mm 2.5mm,clip=true]{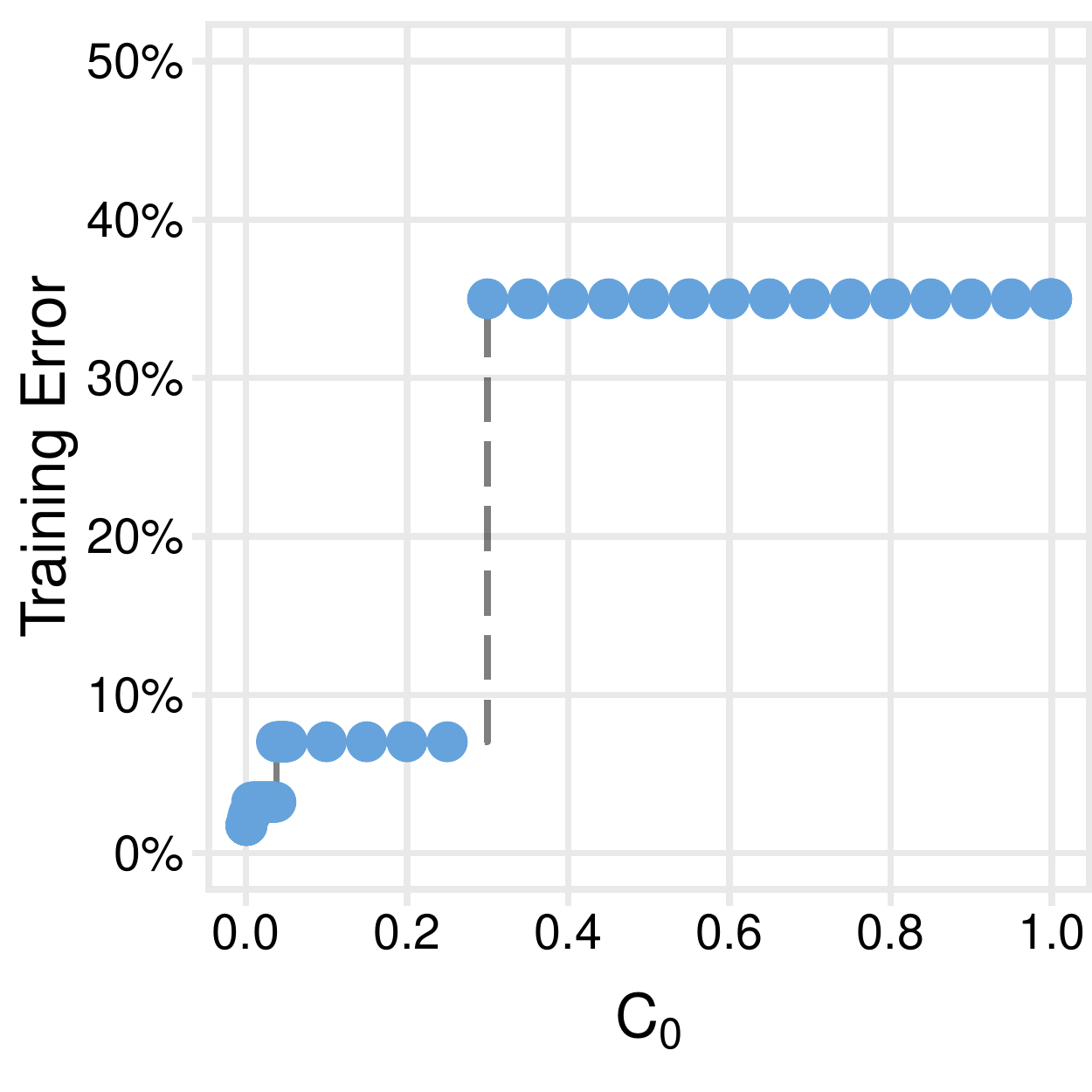}
%
\caption{Training error and model size for linear classifiers trained on the \texttt{breastcancer} dataset for over 200 values of the regularization parameter $C_0$. We restrict $\lambda_j \in \Z\cap [-10,10]$, and regularize with $\IntPen(\lambdab) = \vnorm{\lambdab}_0$ so that $\max{\IntPen(\lambdab)} = P$. 
All possible values of accuracy and interpretability are attained for $C_0 \in [1/NP,1-1/N]$. Setting $C_0 < 1/{NP}$ is guaranteed to yield a model with the most interpretability. Setting $C_0>1-1/N$ is guaranteed to yield a model with the highest accuracy. There are at most $\min({N,P})$ equivalence classes of $C_0$.}
\label{Figure::MeaningfulRegularizationParameter}
\end{figure}

Training models with the 0--1 loss function and a direct interpretability penalty also has the benefit of producing a \textit{meaningful} regularization parameter. When we train models with the 0--1 loss and an $\lzero$-penalty, for example, $C$ represents the number of training examples we would misclassify to change the sparsity of the model by one term. In a more general setting, the regularization parameter, $C$, represents the \textit{price of interpretability} and can be set \textit{a priori} as the maximum training accuracy that we are willing to sacrifice to achieve one unit gain in interpretability in the optimal classifier. Setting $C < \frac{1}{N\max{\IntPen{(\lambdab})}}$ produces a classifier that achieves the highest possible training accuracy. Setting $C > 1-\frac{1}{N}$ produces a classifier that achieves the highest possible interpretability. Thus, we can attain all possible levels of training accuracy and interpretability for our model by constraining the regularization parameter to the interval, $C \in [\frac{1}{N\max\IntPen{(\lambdab)}},1 -\frac{1}{N}]$. Figure \ref{Figure::MeaningfulRegularizationParameter} illustrates this range for the \textds{breastcancer} dataset.

If we require increased scalability, we can replace the 0--1 loss function with any convex loss function, such as the hinge loss or the exponential loss, using a decomposition method (see Section \ref{Sec::Decomposition}). This approach allows us to train models with polynomial running time in $N$, with the same IP solver and with any discrete interpretability penalty and interpretability set. However, it loses the advantages of the 0--1 loss, such as the high degree of accuracy, the robustness to outliers, the meaningful regularization parameter, and the ability to formulate hard constraints on the training error. In light of this, we recommend training models with the 0--1 loss, and only using a different loss if the training process becomes computationally challenging or the final model has to produce conditional probability estimates (in which case we use the logistic loss or the exponential loss).
\subsection{Restricting Coefficients to Any Discrete Set}\label{Sec::AnyDiscreteSet}
We restrict coefficients $\lambdab$ to a \textit{generalized discrete set}, where each coefficient $\lambda_j$ takes one of $K_j$ values from the set $\Lset_j =\{l_{j,1},\ldots, l_{j,K_j}\}$. We do this by defining $K_j$ binary variables, $u_{j,k}\in\B$, and including the following constraints in the IP formulation:
\begin{align*}\centering
&\lambda_j = \sum_{k=1}^{K} l_{j,k} u_{j,k} \textrm{ for all } j\hspace{1cm} \sum_{u=1}^{K} u_{j,k} \leq 1 \textrm{ for all } j.
\end{align*}
Generalized discrete sets can be used, for instance, to restrict coefficient $\lambda_j$ to all possible values that have two significant digits and are between $-9900$ to $9900$ by setting:
\begin{align*}\centering
\Lset = 
\left\{
	\begin{array}{l|c}
	\multirow{4}{*}{$\lambdab \in \mathbb{Z}^P$} &  \lambda_j = d_1 \times 10^{E_1} + d_2 \times 10^{E_2} \, \for \, j = 1,\ldots, P  \\ 
	 & d_1,d_2 \in  \{0, \pm 1,\pm2, \ldots, \pm9\} \\ 
	 & E_1, E_2 \in \{  0, 1, 2, 3 \} \\
	 & E_2=E_1-1 
	 \end{array}
\right\}.
\end{align*}
These sets are especially useful for producing expository models when features have wildly different orders of magnitude. Consider, for instance, a model such as: \textit{predict violent crime in neighborhood if sign[0.0001($\#$residents) -3($\#$parks) +60($\#$thefts$\_$last$\_$year)]$>$0}. Here, using coefficients with one significant digit maintains the expository power of the model, and draws attention to the units of the each feature by clarifying that the values of $residents$ are much larger than those of $parks$. 
\subsection{Incorporating Monotonicity Constraints}\label{Sec::MonotonicityConstraints}
The interpretability of linear models can be significantly improved when the signs of coefficients match the intuition or background knowledge of domain experts \citep{pazzani2001acceptance,verbeke2011building,martens2011performance}. We can train models that include these kind of relationships by using \textit{sign constraints} (also referred to as monotonicity constraints). 

Consider training a model with integer coefficients between $-\Lambda$ and $\Lambda$. In this case, we can restrict coefficients with indices $j \in \J_{pos}$ to be non-negative, coefficients with indices $j \in \J_{neg}$ to be non-positive, and coefficients with indices $j \in \J_{free}$ to take on either sign by defining:
\begin{alignat*}{3}\centering
\Lset_j  &= 
\begin{dcases}
\left\{ \lambda_j \in \Z \cap [0,\Lambda] \right\}             & \text{ if } j \in\J_{pos} \\
\left\{ \lambda_j \in \Z \cap [-\Lambda,0] \right\}             & \text{ if } j \in\J_{neg} \\
\left\{ \lambda_j \in \Z \cap [-\Lambda,\Lambda] \right\}      & \text{ if } j\in \J_{free}.
\end{dcases}
\end{alignat*}
These sets can be added to an IP formulation using lower or upper bound constraints for $\lambda_j$. Sign-constrained formulations can have the side effect of improved computational performance since they narrow down the feasible region of the IP. Correct prior knowledge on the sign of the coefficients may also result in a more accurate predictive model \citep{dawes1979robust}.
\subsection{Incorporating Feature-Based Preferences}

Domain experts sometimes require models that incorporate preferences among different features. Our framework can incorporate such preferences by minimizing a weighted $\lzero$-penalty with a customized regularization parameter for each coefficient ($C_{0,j}$) along with the 0--1 loss. 

Consider a case where we wish for our model to use feature $j$ instead of feature $k$. We can set $C_{0,k} = C_{0,j} + \epsilon$, where $\epsilon > 0$ represents the maximum additional accuracy that we are willing to sacrifice in order to use feature $j$ instead of feature $k$. Thus, setting $C_{0,k} = C_{0,j} + 0.02$ would ensure that we would only be willing to use feature $k$ instead of feature $j$ if it yields an additional 2\% gain in accuracy over feature $k$. 

This approach can also be used to deal with missing data. Consider training a model where feature $j$ contains $M<N$ missing points. Instead of dropping these points, we can impute the values of the $M$ missing examples, and adjust the regularization parameter $C_{0,j}$ so that our model only uses feature $j$ if it yields an additinal gain in accuracy of more than $M$ examples:
\begin{align*}
C_{0,j} = C_0 + \frac{M}{N} .
\end{align*}
The adjustment factor is chosen so that: if $M=0$ then $C_{0,j} = C_0$ and if $M=N$ then $C_{0,j} = 1$ and the coefficient is dropped entirely (see Theorem \ref{Thm::L0Bound}). This adjustment ensures that features with lots of imputed values are more heavily penalized than features with fewer imputed values.
\subsection{Incorporating Feature-Based Constraints}\label{Sec::FeatureBasedConstraints}
The interpretability of classification models is often tied to the composition of input variables \citep{Freitas:2014ic}. Our framework can provide fine-grained control over the choice of input variables of in a model by formulating constraints in terms of discrete indicator variables, $\alpha_j = \indic{\lambda_j \neq 0}$. 

We can use these indicator variables to impose a hard limit on the number of input variables (e.g. 10) in our classification model by adding the following constraint to our IP formulation,
\begin{align*}\centering
\sum_{j = 1}^P{\alpha_j} \leq 10.
\end{align*}
More generally, we can use these variables to fine-tune the composition of input variables in our models. As an example, consider the following constraint, which imposes an ``either-or" condition to ensure that a model will not include both $male$ and $female$ as input variables:
\begin{align*}\centering
\alpha_{male} + \alpha_{female} \leq 1.
\end{align*}
Alternatively, consider the following constraint, which imposes an ``if-then" condition to ensure that a model will only include $hypertension$ and $heart\_attack$ if it also includes $stroke$:
\begin{align*}\centering
\alpha_{heart\_attack} + \alpha_{hypertension} \leq 2 \alpha_{stroke}.
\end{align*}
We can also encode more complicated relationships: we can encode a hierarchical relationship among input variables (a partial order), for instance, by requiring that an input variables in the leaves is only used when all features above it in the hierarchy are also used: 
\begin{align*}\centering
\alpha_{leaf} \leq \alpha_{node} \textrm{ for all nodes above the leaf}.
\end{align*}
%
%
%
\subsection{Training Models for Imbalanced Data}\label{Sec::HandlingImbalancedData}
The vast majority of real-world classification problems are imbalanced. In these problems, training a classifier by maximizing classification accuracy often produces a trivial model (i.e. if the probability of heart attack is 1\%, a classifier that never predicts a heart attack is still 99\% accurate). Handling highly imbalanced data is incredibly difficult for most classification methods: even taking a cost-sensitive approach \citep[see][]{liu2006influence}, it is difficult to produce anything except a model classifier that always predicts either the majority or minority class \citep[see e.g.][]{goh2014box}.

Given $\nplus$ positively-labeled examples from the set $\iplus = \{i:y_i = +1\}$, and $\nminus$ negatively-labeled examples from the set $\iminus = \{i:y_i = -1\}$, the cost-sensitive approach uses a \textit{weighted loss function},
\begin{align*}
\Loss{\lambdab;\data_N} &= \frac{1}{N} \sum_{i\in\iplus} \wplus \Loss{\lambdab;(\xb_i,y_i)} +  \frac{1}{N}\sum_{i\in\I^{-}}  \wminus \Loss{\lambdab;(\xb_i,y_i)}.
\end{align*}
Here, we can adjust the weights $\wplus$ or $\wminus$ to control the accuracy on the positive and negative class, respectively. We assume without loss of generality that $\wplus + \wminus = 1$.

Our framework has several unique benefits when training models for imbalanced problems. When we train models with the weighted 0--1 loss, we can set the values of $\wplus$ and $\wminus$ purposefully. Specifically, we can set $\wplus < \frac{1}{1+\nplus}$ to train a model that classifies all of the negative examples correctly, and set $\wplus > \frac{\nminus}{1+\nminus}$ to train a model that classifies all of the positive examples correctly. Thus, we can train models that attain all possible levels of \textit{sensitivity} (i.e. accuracy on the positive class) and \textit{specificity} (i.e. accuracy on the negative class) by limiting $\wplus \in \left[\frac{1}{1+\nplus}, \frac{\nminus}{1+\nminus} \right]$.

Another benefit is that we can explicitly limit the sensitivity or specificity of our models without tuning. When domain experts specify hard constraints on sensitivity or specificity, we can encode these constraints into the IP, and produce a model in a ``single-shot" procedure that does not require grid search over $\wplus$ and $\wminus$. Consider, for example, a case where we need to train the most accurate model with a maximum error of 20\% on negatively-labeled examples. We can train this model by solving an optimization problem with the form:
\begin{align}
\min_{\lambdab} & \qquad \frac{1}{N} \sum_{i\in\iplus} \wplus \indic{y_i \lambdab^T\xb_i \leq 0} + \frac{1}{N} \sum_{i\in\iminus} \wminus \indic{y_i \lambdab^T\xb_i \leq 0} \notag \\ 
\st & \qquad \frac{1}{N^-}\sum_{i\in\iminus} \indic{y_i \lambdab^T\xb_i \geq 0} \leq 0.20 \label{Con::MaxFPRConstraint}\\
& \qquad \lambdab\in\Lset. \notag
\end{align}
We set $\wplus > \frac{\nminus}{1+\nminus}$ and set $\wminus = 1 - \wplus$ so that the optimization aims to find a classifier that classifies all of the positively-labeled examples correctly, at the expense of misclassifying all of the negatively-labeled examples. Constraint \eqref{Con::MaxFPRConstraint} prevents this from happening,  and limits the error on negatively-labeled examples to 20\%. Thus, the optimal classifier attains the highest accuracy among classifiers with a maximum error of 20\% on negatively-labeled examples.

A similar single-shot procedure can be used for classification problems with an ``intervention budget." These are problems where we need to find a model that attains the highest classification accuracy on a subset of the population. 
Suppose that we had a budget to predict $\hat{y}_i=+1$ at most $25\%$ of the time, because we have the resources to take an action on $25\%$ of the population. We can train this model by solving an optimization problem with form:
\begin{align}
\min_{\lambdab} & \qquad \frac{1}{N} \sum_{i\in\iplus} \wplus \indic{y_i \lambdab^T\xb_i \leq 0} + \frac{1}{N} \sum_{i\in\I^{-}} \wminus \indic{y_i \lambdab^T\xb_i \leq 0} \notag \\ 
\st & \qquad \frac{1}{N}\sum_{i=1}^N \indic{\lambdab^T\xb_i \geq 0} \leq 0.25 \label{Con::Test} \\
& \qquad \lambdab\in\Lset. \notag
\end{align}
Here, constraint \eqref{Con::Test} ensures that any feasible classifier predicts $\hat{y}_i=+1$ at most 25\% of the time. We set $\wplus > \frac{\nminus}{1+\nminus}$ and set $\wminus = 1 - \wplus$ so that the optimization aims to produce a classifier that classifies all of the positively labeled examples accurately. In addition, we set $\wminus = 1 - \wplus$ so that the optimization also aims to classify negatively examples accurately as a secondary objective. Thus, the optimal classifier attains the highest possible training accuracy among classifiers that satisfy the intervention budget.
\clearpage
\section{Models}\label{Sec::Models}
In this section, we present four different kinds of interpretable models that can be produced with our framework. We pair each model with an IP formulation that minimizes the 0--1 loss function. These formulations can be adapted to train models with other loss functions by switching loss constraints (Appendix \ref{Appendix::LossConstraints}) or by using loss decomposition (Section \ref{Sec::Decomposition}).
\subsection{Scoring Systems}\label{Sec::SLIM}
Scoring systems allow users to make quick, hands-on predictions by adding, subtracting and multiplying a few meaningful numbers. These models are in widespread use for assessing the risk of medical outcomes 
%
\citep[e.g.,][]{light1972pleural,ranson1974prognostic,knaus1991apache,le1984simplified,bone1992american,wells1997value,antman2000timi,moreno2005saps}.
Scoring systems are difficult to reproduce with existing methods because they require discrete coefficients. Most popular medical scoring systems are often hand-crafted by domain experts \citep[e.g.,][]{knaus1985apache, gage2001validation} or trained using heuristic procedures \citep[e.g.,][]{le1993new}.
%
%

We can create principled scoring systems by solving an optimization problem of the form:
\begin{align}
\label{Eq::SLIMFormulation}
\begin{split}
\min_{\lambdab} & \qquad \Loss{\lambdab;\data_N} + C_0 \vnorm{\lambdab}_0 + \epsilon \vnorm{\lambdab}_1 \\ 
\st &  \qquad \lambdab \in \Lset. 
\end{split}
\end{align}
We refer to a classifier produced by this problem as a \textit{Supersparse Linear Integer Model} (SLIM), and provide an example for the \textds{breastcancer} dataset in Figure \ref{Fig::SLIMExample}.
\begin{figure}[htbp]
\centering
\small{
\textbf{PREDICT TUMOR IS BENIGN if SCORE $> 17$} \\ 
\vspace{0.25em}
\begin{tabular}{|l l  c | c |}
   \hline
1. & $UniformityOfCellSize$ & $\times$ 4 & $\quad\cdots\cdots$ \\ 
2. & $BareNuclei$ & $\times$ 2  & $+\quad\cdots\cdots$ \\ 
   \hline
 & \small{\textbf{ADD POINTS FROM ROWS 1-2}} & \small{\textbf{SCORE}} & $=\quad\cdots\cdots$ \\ 
   \hline 
\end{tabular}
}
\caption{SLIM scoring system for \textds{breastcancer} when $C_0 = 0.025, \Lset_0 = \Z \cap [\text{\footnotesize{-}}100,100], \text{and} \Lset_j = \Z \cap [\text{\footnotesize{-}}10,10]$. This model has 2 features, which take values between 0--10, and a mean 10-fold CV test error of $3.4 \pm 2.0\%$.}
\label{Fig::SLIMExample}
\end{figure}

SLIM creates scoring systems by restricting coefficients to a small set of bounded integers, such as $\Lset = \{ \lambdab \in \Z^{P+1} \;|\; |\lambda_j| \leq 20 \for j = 0,\ldots, P \}$. Here, the interpretability penalty regularizes the $\lzero$-norm to tune sparsity, and the $\lone$-norm to restrict coefficients to coprime values (i.e. coefficients whose greatest common denominator is 1). The $\epsilon$ is set small enough so that neither training accuracy nor interpretability is influenced by this term.

To illustrate the use of the $\lone$-penalty, consider classifier $\hat{y}=\sign{x_1 + x_2}$. If the objective in \eqref{Eq::SLIMFormulation} minimized only the 0--1 loss and an $\lzero$-penalty, then $\hat{y}=\sign{2 x_1 + 2 x_2}$ would have the same objective value as $\hat{y}=\sign{x_1 + x_2}$ because it makes the same predictions and has the same number of non-zero coefficients. Because the coefficients are restricted to belong to a discrete set, adding a tiny $\lone$-penalty in the objective of \eqref{Eq::SLIMFormulation} yields the classifier with the smallest coefficients, $\hat{y} = \sign{x_1+x_2}$, where the greatest common denominator of the coefficients is 1.

When we train SLIM scoring systems with the 0--1 loss function, the regularization parameter $C_0$ can be set as the maximum accuracy we are willing to sacrifice to remove one feature from the optimal classifier. We can restrict $C_0 \in [\frac{1}{NP},1-\frac{1}{N}]$ as setting $C_0 < \frac{1}{NP}$ is guaranteed to produce a classifier with the highest possible training accuracy while setting $C_0 > 1 - \frac{1}{N}$ is guaranteed to produce a classifier with the highest possible sparsity. Given $C_0$ and $\Lset$, we set $\epsilon <\frac{\min{(\frac{1}{N},C_0})}{\max_{\lambdab\in\Lset}\vnorm{\lambdab}_1}$ so that the maximum value of the $\lone$-penalty $\epsilon \cdot \max_{\lambdab\in\Lset}{\vnorm{\lambdab}_1}$ is smaller than the unit value of accuracy and sparsity in the objective of \eqref{Eq::SLIMFormulation}. This ensures that the $\lone$-penalty is small enough to restrict coefficients to coprime values without affecting accuracy or sparsity.

\clearpage
We can train a SLIM scoring system with the 0--1 loss function by solving the following IP:
\begin{subequations}
\begin{equationarray}{crcl>{\qquad}ll}
\hspace{3cm} \min_{\lambdab,\bf{\psi},\bf{\Phi},\bf{\alpha},\bf{\beta}} &\frac{1}{N}\sum_{i=1}^{N} \loss_i & + & \sum_{j=1}^{P} \IntPen_j  \notag \\
\hspace{3cm} \st          & M_i \loss_i                  & \geq & \gamma -\sum_{j=0}^P y_i \lambda_j x_{i,j}       &\mprange{i}{1}{N} & \mpdes{0--1 loss} \label{Con::SLIMLoss} \\
& \IntPen_j & = & C_0\alpha_j + \epsilon\beta_j &\mprange{j}{1}{P}& \mpdes{int. penalty} \label{Con::SLIMIntPenalty} \\
& -\Lambda_j\alpha_j  & \leq & \lambda_j \leq \Lambda_j\alpha_j  &\mprange{j}{1}{P} & \mpdes{$\lzero$ norm} \label{Con::SLIML0Norm} \\
& -\beta_j &  \leq & \lambda_j  \leq \beta_j &\mprange{j}{1}{P} & \mpdes{$\lone$ norm} \label{Con::SLIML1Norm} \\
& \lambda_j & \in & \Lset_j &  \mprange{j}{0}{P} & \mpdes{int. set} \notag \\ 
& \loss_i & \in & \B &  \mprange{i}{1}{N} & \mpdes{loss variables} \notag  \\
& \IntPen_j  & \in & \R_+  & \mprange{j}{1}{P} & \mpdes{int. penalty variables} \notag \\
& \alpha_j  & \in & \B  & \mprange{j}{1}{P} & \mpdes{$\lzero$ variables} \notag \\
& \beta_j    & \in & \R_+ & \mprange{j}{1}{P} & \mpdes{$\lone$ variables} \notag
\end{equationarray}
\end{subequations}
Here, the constraints in \eqref{Con::SLIMLoss} set the loss variables $\loss_i = \indic{y_i \lambdab^T\xb_i \leq 0}$ to $1$ if a linear classifier with coefficients $\lambdab$ misclassifies example $i$. This is a Big-M formulation for the 0--1 loss that depends on scalar parameters $\gamma$ and $M_i$ (see e.g. \citealt{rubin2009mixed}). The value of $M_i$ represents the ``maximum score when example $i$ is misclassified", and can be set as $M_i = \max_{\lambdab \in \Lset} (\gamma - y_i\lambdab^T\xb_i)$ which is easy to compute since the $\lambda_j$ are restricted to a discrete set. The value of $\gamma$ represents the ``margin" and should technically be set as a lower bound on $y_i\lambdab^T\xb_i$. When the features are binary, $\gamma$ can be set to any value between 0 and 1. In other cases, the lower bound is difficult to calculate exactly, so we set $\gamma=0.1$, which makes an implicit assumption on the values of the features. The constraints in \eqref{Con::SLIMIntPenalty} define the total interpretability penalty for each coefficient as $\IntPen_j = C_0 \alpha_j + \epsilon \beta_j$, where $\alpha_j = \indic{\lambda_j\neq 0}$ is defined by the constraints in \eqref{Con::SLIML0Norm}, and $\beta_j = |\lambda_j|$ is defined by the constraints in \eqref{Con::SLIML1Norm}. We represent the largest absolute value of each coefficient using the parameters $\Lambda_j = \max_{\lambda_j\in\Lset_j} |\lambda_j|$.
\subsection{Personalized Models}\label{Sec::PILM}
A \textit{Personalized Integer Linear Model} (PILM) is a generalization of a Supersparse Linear Integer Model that provides fine-grained soft control over the interpretability of coefficients. To use this model, users define $R+1$ interpretability sets,
\begin{align*}
\Lset^r = \{ l_{r,1},\ldots,l_{r,K_r}\} \for r = 0,\ldots, R,
\end{align*}
as well as a ``personalized" interpretability penalty,
\begin{align*}
\IntPen(\lambda_j) &= 
  \begin{dcases*}
    C_0 & if $\lambda_j \in \Lset^0$ \vspace{-0.1cm}\\
    & $\vdots$ \vspace{-0.1cm} \\
    C_R & if $\lambda_j \in \Lset^R$.\vspace{-0.1cm}
  \end{dcases*}
\end{align*}
These components must be specified so that the penalty regularizes coefficients from less interpretable sets more heavily. This requires that: (i) the interpretability sets, $\Lset^1, \ldots, \Lset^R$ are mutually exclusive; (ii) $\Lset^r$ is more interpretable than $\Lset^{r+1}$; (iii) the regularization parameters are monotonically increasing in $r$, $C_0<C_1<\ldots<C_R$. 

When we train PILM with the 0--1 loss function, we can set the regularization parameters $C_r$ as the minimum gain in training accuracy required to use a coefficient from $\Lset^r$. As an example, consider training a model with the 0--1 loss and the interpretability penalty:
\begin{align*}
\label{Eq::InterpretabiltyExample}
\IntPen(\lambda_j) &= 
  \begin{cases}
    C_0 = 0.00 & \text{if} \quad \lambda_j \in {0}\\
    C_1 = 0.01 & \text{if} \quad \lambda_j \in \pm \{1,\ldots,10\}\\
    C_2 = 0.05 & \text{if} \quad \lambda_j \in \pm \{11,\ldots,100\}.
  \end{cases}
\end{align*}
In this case, the optimal classifier only uses a coefficient from $\Lset^1$ if it yields at least a 1\% gain in training accuracy, and a coefficient from $\Lset^2$ if it yields at least a 5\% gain in training accuracy.
\clearpage
We can train a PILM classifier with the 0--1 loss function by solving the IP:
\begin{subequations}
\begin{equationarray}{crcl>{\hspace{0.1cm}}ll}
\min_{\lambdab,\bf{\psi},\bf{\Phi},\bf{u}} &\frac{1}{N}\sum_{i=1}^{N} \loss_i & + &  \sum_{j=1}^{P} \IntPen_j  \notag \\
\st & M_i\loss_i & \geq & \gamma -\sum_{j=0}^P y_i \lambda_j x_{i,j} & \mprange{i}{1}{N} & \mpdes{0--1 loss} \label{Con::PILMLoss} \\
& \IntPen_j & = & \sum_{r=0}^R \sum_{k=1}^{K_r} C_r u_{j,k,r} & \mprange{j}{1}{P} & \mpdes{int. penalty} \label{Con::PILMIntPenalty} \\
& \lambda_j & = & \sum_{r=0}^R \sum_{k=1}^{K_r} l_{r,k} u_{j,k,r} &\mprange{j}{0}{P} & \mpdes{coefficient values} \label{Con::PILMIntCon} \\
&  1  & = & \sum_{r=0}^R \sum_{k=1}^{K_r} u_{j,k,r} & \mprange{j}{0}{P} & \mpdes{1 int. level per coef.}  \label{Con::PILMOnly1} \\
& \loss_i & \in & \B &  \mprange{i}{1}{N} & \mpdes{loss variables} \notag  \\
& \IntPen_j  & \in & \R_+  & \mprange{j}{1}{P} & \mpdes{int. penalty variables} \notag \\
& u_{j,r,k} & \in & \B &  \mprange{j}{0}{P} \quad \mprange{r}{0}{R} \quad \mprange{k}{1}{K_r} & \mpdes{coef. value variables} \notag 
\end{equationarray}
\end{subequations}
Here, the loss constraints and Big-M parameters in \eqref{Con::PILMLoss} are identical to those from the SLIM IP formulation (see Section \ref{Sec::SLIM}). The $u_{j,k,r}$ are binary indicator variables that are set to 1 if $\lambda_j$ is equal to $l_{k,r}$. Constraints \eqref{Con::PILMOnly1} ensure that each coefficient will use exactly one value from one interpretability set. Constraints \eqref{Con::PILMIntCon} ensure that each coefficient $\lambda_j$ is assigned a value from the appropriate interpretability set, $\Lset^r$, and constraints \label{Con::PILMIntPenalty} ensure that each coefficient $\lambda_j$ is assigned the value specified by the personalized interpretability penalty.
\subsection{Rule-Based Models}
Our framework can also produce rule-based classification models when the training data are composed of \textit{binary rules}. In general, any real-valued feature (e.g. $age$) can be converted into a binary rule by setting a threshold,
\begin{align*}
age \geq 25 = \begin{cases}
1 &  \text{ if } age \geq 25 \\ 
0 & \text{ if } age < 25 .
\end{cases}
\end{align*}
Such thresholds can be set using domain expertise, rule mining or discretization \citep{Liu:2002ty}.

In what follows, we assume that we train our models using training data that contains $T_j$ binary rules $\bm{h}_{j,t} \in \B^N$ for each feature $\xb_j \in \R^N$ in the original data. We make the following assumptions about the conversion process. If $\xb_j$ is a binary variable, then it is left unchanged so that $T_j = 1$ and $\bm{h}_{j,T_j} = \xb_j$. If $\xb_j$ is a categorical variable $\xb_j \in \{1,\ldots,K\}$, then there exists a binary rule for each category so that $T_j=K$ and $\bm{h}_{j,t}=\indic{\xb_j=k}$ for $t = 1,\ldots,K$. If $\xb_j$ is a real variable, then the conversion produces $T_j$ binary rules of the form $\bm{h}_{j,t} = \indic{\xb_j \geq v_{j,t}}$ where $v_{j,t}$ denotes the $t^\text{th}$ threshold for feature $j$. Note that while there exists an infinite number of thresholds for a real-valued feature, we need to consider at most $N-1$ thresholds in practice (i.e. one threshold placed each pair of adjacent values, $x_{(i),j}<v_{j,t}<x_{(i+1),j}$); using additional thresholds will produce the same set of binary rules and the same rule-based model.

We do not extract binary rules for the intercept term so that $\lambda_0 \in \Lset_0$. We also use the same notation for coefficients of binary rules as we do for regular features, $\lambda_{j,t} \in \Lset_j \text{ for } j = 1,\ldots,P$. Thus, rule-based models from our framework have the form:
\begin{align}
y = \sign{ \lambda_0 + \sum_{j=1}^P \sum_{t=1}^{T_j} \lambda_{j,t}h_{j,t}}.  
\end{align}
\subsubsection{M-of-N Rule Tables}\label{Sec::MofNRuleTables}
\textit{M-of-N rule tables} are rule-based models that make predictions as follows: given a set of $N$ rules, predict $\hat{y} = +1$ if at least M of them are true. These models have the major benefit that they do not require the user to compute a mathematical expression \citep[see][]{Freitas:2014ic}. M-of-N rule tables were originally proposed as auxiliary models that could be extracted from neural nets \citep{Towell:1993tx}.

We can use our framework to produce fully optimized M-of-N rule tables as follows:
\begin{align*}
\hspace{5cm} \min_{\lambdab} & \qquad \Loss{\lambdab;\data_N} + C_0 \vnorm{\lambdab}_0 & \notag \\
\st & \qquad \lambda_0 \in \Z \cap [-P,0] & \notag \\ 
& \qquad \lambda_{j,t} \in \B &     \mprange{j}{1}{P} \quad \mprange{t}{1}{T_j}. 
\end{align*}
Here, we can achieve exact $\lzero$-regularization using an $\lone$-penalty since $\vnorm{\lambda_{j,t}}_0 = \vnorm{\lambda_{j,t}}_1$ when $\lambda_{j,t} \in \B$. When we use the 0--1 loss, the regularization parameter $C_0$ can be set as the maximum sacrifice in training accuracy to remove each rule from the optimal table. The coefficients from this optimization problem yield an M-of-N rule table with $M=\lambda_0 + 1$ and $N= \sum_{j=1}^P\sum_{t=1}^{T_j} \lambda_{j,t}$. We provide an example for the \texttt{breastcancer} dataset in Figure \ref{Fig::MNRuleExample}.
\begin{figure}[htbp]
\centering{
\small{
\begin{tabular}{c}
  \bfcell{c}{PREDICT TUMOR IS BENIGN IF \\IF AT LEAST 5 OF THE FOLLOWING 8 RULES ARE TRUE}\\ 
  \toprule
  $UniformityOfCellSize\geq3$ \\ 
  $UniformityOfCellShape\geq3$ \\ 
  $MarginalAdhesion\geq3$ \\ 
  $SingleEpithelialCellSize\geq3$ \\ 
  $BareNuclei\geq3$ \\ 
  $BlandChromatin\geq3$ \\ 
  $NormalNucleoli\geq3$ \\ 
  $Mitoses\geq3$
  \end{tabular}
}
}
\caption{M-of-N rule table for the \texttt{breastcancer} dataset for $C_0 = 0.9/NP$. This model has 8 rules and a mean 10-fold CV test error of $4.8 \pm 2.5\%$. We trained this model with binary rules that we created by setting a threshold for each feature at 3.}
\label{Fig::MNRuleExample}
\end{figure}

We can train an M-of-N rule table with the 0--1 loss function by solving the following IP:
\begin{subequations}
\begin{equationarray}{crcl>{\hspace{0.1cm}}ll}
\min_{\lambdab,\bf{\psi},\bf{\Phi}} &\frac{1}{N}\sum_{i=1}^{N} \loss_i & + & \sum_{j=1}^{P} \IntPen_j  \notag \\
\st          & M_i \loss_i                  & \geq & \gamma -\sum_{j=0}^P\sum_{t=1}^{T_j} y_i \lambda_{j,t} h_{i,j,t}       &\mprange{i}{1}{N} & \mpdes{0--1 loss} \label{Con::MNLoss} \\
& \IntPen_{j,t} & = & C_1 \lambda_{j,t} &\mprange{j}{1}{P} \quad \mprange{t}{1}{T_j} & \mpdes{int. penalty} \label{Con::MNIntPenalty} \\
& \lambda_0 & \in &  \Z \cap [-P,0] &  & \mpdes{intercept values} \notag \\ 
& \lambda_{j,t} & \in & \{0,1\} &  \mprange{j}{1}{P} \quad \mprange{t}{1}{T_j} & \mpdes{coefficient values} \notag \\
& \loss_i & \in & \B &  \mprange{i}{1}{N} & \mpdes{0--1 loss indicators} \notag  \\
& \IntPen_{j,t}  & \in & \R_+  & \mprange{j}{1}{P} \quad \mprange{t}{1}{T_j} & \mpdes{int. penalty values} \notag 
\end{equationarray}
\end{subequations}
Here, the loss constraints and Big-M parameters in \eqref{Con::MNLoss} are identical to those from the SLIM IP formulation (see Section \ref{Sec::SLIM}). Constraints \eqref{Con::MNIntPenalty} define the interpretability penalty variables, $\IntPen_{j,t}$ as the value of the $\lone$-penalty using the fact that $\vnorm{\lambda_{j,t}}_0 = \vnorm{\lambda_{j,t}}_1 = \lambda_{j,t}$ when $\lambda_{j,t} \in \B$.
\subsubsection{Threshold-Rule Models}\label{Sec::TILM}
A \textit{Threshold-Rule Integer Linear Model} (TILM) is a scoring system where the input variables are thresholded versions of the original feature set (i.e. decision stumps). These models are well-suited to problems where the outcome has a non-linear relationship with real-valued features. As an example, consider the SAPS II medical scoring system of \citealt{le1993new}, which assesses the mortality of patients in intensive care using thresholds on real-valued features such as $blood\_pressure>200$ and $heart\_rate<40$. TILM scoring systems optimize the binarization of real-valued features by using feature selection on a large (potentially exhaustive) pool of binary rules for each real-valued feature. \citet{carrizosa2010binarized} and \citet{goh2014box} take different but related approaches for constructing classifiers with binary threshold rules. 

We train TILM scoring systems using an optimization problem of the form:
\begin{align}
\label{Eq::TILMFormulation}
\begin{split}
\min_{\lambdab} & \qquad \Loss{\lambdab;\data_N} + C_f \cdot \text{Features} + C_t \cdot \text{Rules per Feature} + \epsilon \vnorm{\lambdab}_1 \\ 
\st & \qquad \lambdab \in \Lset, \\
    & \qquad \sum_{t=1}^{T_j} \indic{\lambda_{j,t}\neq 0} \leq R_{max} \text{ for } j = 1,\ldots,P,  \\ 
    & \qquad \sign{\lambda_{j,1}} = \sign{\lambda_{j,2}} =\ldots = \sign{\lambda_{j,T_j}} \text{ for } j = 1,\ldots,P.
\end{split}
\end{align}
TILM uses an fine-grained interpretability penalty that includes terms for the number of rules used in the classifier as well as the number of features associated with these rules. The small $\lone$-penalty in the objective restricts coefficients to coprime values as in Section \ref{Sec::SLIM}. Here, $C_f$ tunes the number of features used in the model, $C_t$ tunes the number of rules per feature, and $\epsilon$ is set to a small value to produce coprime coefficients. TILM includes hard constraints to limit the number of binary rules per feature to $R_{max}$ (e.g. $R_{max} = 3$), and to ensure that the coefficients for binary rules from a single feature agree in sign (this improves the interpretability of the model by ensuring that each feature maintains a strictly monotonically increasing or decreasing relationship with the outcome).

We train TILM scoring systems with the 0--1 loss by solving the following IP formulation:
\begin{subequations}
\begin{equationarray}{crcl>{\hspace{0.1cm}}ll}
\min_{\lambdab,\bf{\psi},\bf{\Phi},\bf{\tau},\bf{\nu},\bf{\delta}} &\frac{1}{N}\sum_{i=1}^{N} \loss_i & + & \sum_{j=1}^{P} \IntPen_j  \notag \\
\st          & M_i \loss_i                  & \geq & \gamma -\sum_{j=0}^P\sum_{t=1}^{T_j} y_i \lambda_{j,t} h_{i,j,t} &\mprange{i}{1}{N} & \mpdes{0--1 loss} \label{Con::TILMLoss} \\
& \IntPen_j & = & C_f\nu_j + C_t \tau_j + \epsilon \sum_{t=1}^{T_j}{\beta_{j,t}} & \mprange{j}{1}{P} & \mpdes{int. penalty} \label{Con::TILMIntPenalty} \\
& T_j \nu_j           & = & \sum_{t=1}^{T_j} \alpha_{j,t}      &\mprange{j}{1}{P} & \mpdes{feature use} \label{Con::TILMFeatures}\\ 
& \tau_j              & = & \sum_{t=1}^{T_j} \alpha_{j,t}-1    &\mprange{j}{1}{P} & \mpdes{threshold/feature} \label{Con::TILMThresholdPerFeature} \\
&-\Lambda_j\alpha_{j,t}    & \leq & \lambda_{j,t} \leq \Lambda_j\alpha_{j,t}  &\mprange{j}{1}{P} \quad \mprange{t}{1}{T_j} & \mpdes{$\lzero$ norm} \label{Con::TILML0NormLower} \\
&-\beta_{j,t}              & \leq & \lambda_{j,t}  \leq \beta_{j,t}  &\mprange{j}{1}{P} \quad \mprange{t}{1}{T_j} & \mpdes{$\lone$ norm} \label{Con::TILML1NormUpper} \\
& \tau_j              & \leq & R_{max} + 1  &\mprange{j}{1}{P} & \mpdes{max thresholds} \label{Con::TILMMaxThresholds} \\
&-\Lambda_j (1-\delta_{j})    & \leq & \lambda_{j,t} \leq \Lambda_j\delta_{j}  &\mprange{j}{1}{P} \quad \mprange{t}{1}{T_j} & \mpdes{$\lzero$ norm} \label{Con::TILMSigns} \\
& \lambda_{j,t} & \in & \Lset_j &  \mprange{j}{0}{P} \quad \mprange{t}{1}{T_j} & \mpdes{coefficient values} \notag \\ 
& \loss_i & \in & \B &  \mprange{i}{1}{N} & \mpdes{loss variables} \notag  \\
& \IntPen_j  & \in & \R_+  & \mprange{j}{1}{P} & \mpdes{int. penalty variables} \notag \\
& \alpha_j  & \in & \B  & \mprange{j}{1}{P} & \mpdes{$\lzero$ variables} \notag \\
& \beta_j    & \in & \R_+ & \mprange{j}{1}{P} & \mpdes{$\lone$ variables} \notag \\
& \nu_j    & \in & \B & \mprange{j}{1}{P} & \mpdes{feature use indicators} \notag \\ 
& \tau_j    & \in & \Z_+ & \mprange{j}{1}{P} & \mpdes{threshold/feature variables} \notag \\ 
& \delta_j    & \in & \B & \mprange{j}{1}{P} & \mpdes{sign indicators} \notag
\end{equationarray}
\end{subequations}
Here, the loss constraints and Big-M parameters in \eqref{Con::MNLoss} are identical to those from the SLIM IP formulation (see Section \ref{Sec::SLIM}). The interpretability penalty for each coefficient, $\IntPen_j$, is set as $C_f\nu_j + C_t \tau_j + \epsilon \sum_{t=1}^{T_j}{\beta_{j,t}}$ in constraints \eqref{Con::TILMIntPenalty}. The variables used in the interpretability penalty include: $\nu_j$, which indicate that we use a non-zero coefficient for a binary rule from feature $j$; $\tau_j$, which counts the number of additional binary rules we use from feature $j$; and $\beta_{j,t} = |\lambda_{j,t}|$. The values of $\nu_j$ and $\tau_j$ are derived from the variable $\alpha_{j,t} = \indic{\lambda_{j,t}\neq0}$ in constraints \eqref{Con::TILMFeatures} and \eqref{Con::TILMThresholdPerFeature}. Constraints \eqref{Con::TILMMaxThresholds} limit the total number of binary rules associated with feature $j$ to $\R_{max}$. Constraints \eqref{Con::TILMSigns} ensure that all of the coefficients of binary rules from a feature $j$ agree in sign; these constraints depend on the variables $\delta_j$, which are set to 1 when $\lambda_{j,t} \geq 0$, and to 0 when $\lambda_{j,t} \leq 0$ (the value of $\delta_j$ does not matter if some or all of the coefficients $\lambda_{j,t}$ are all 0 ).
\clearpage
\section{Methods to Enhance Scalability}\label{Sec::Methods}
In this section, we present two methods to enhance the scalability of our framework, which we refer to as \textit{loss decomposition} and \textit{data reduction}.
%
%

\subsection{Loss Decomposition}\label{Sec::Decomposition}
Consider a generic optimization problem from our framework, $\OriginalP$,
\begin{align*}
\begin{split}
\min_{\lambdab} & \qquad \Loss{\lambdab;\data_N} + C \cdot \IntPen(\lambdab) \\ 
\st & \qquad \lambdab \in \Lset.
\end{split}
\end{align*}
%
%
Usually, we would solve $\OriginalP$ by formulating an IP that uses $N$ variables and $N$ constraints to represent the \textit{individual losses}, $\Loss{\lambdab;(\xb_i,y_i)} \, i = 1,\ldots, N$. This approach does not scale well and may result in intractable formulations that exceed memory limits for large datasets. The quantity of interest, however, is not the individual losses but the \textit{aggregate loss}, $$\Loss{\lambdab;\data_N} = \frac{1}{N}\sum_{i=1}^N\Loss{\lambdab;(\xb_i,y_i)}.$$ 

Decomposition methods, also known as cutting-plane or localization methods, are a popular class of techniques to solve large-scale optimization problems (see \citealt{boyd2004convex} and \citealt{joachims2006training}, 2009 for recent applications). The main benefit in applying these methods in our framework is that we can delegate all data-related computation to an \textit{oracle function}. The oracle function is stand-alone function that is called by the IP solver to compute the individual losses, $\Loss{\lambdab;(\xb_i,y_i)} \,\mprange{i}{1}{N}$, and return information about the aggregate loss, $\Loss{\lambdab;\data_N}$. In this setup, the IP solver \textit{queries} the oracle function to obtain information about the aggregate loss at different values of $\lambdab$. Thus, the IP solver handles a \textit{proxy problem}, $\ProxyP{}$, that can drop the $N$ variables and $N$ constraints used to compute the individual losses in the original optimization problem, $\OriginalP$.

Decomposition methods use an iterative algorithm that queries the oracle function to build a piecewise linear approximation of the aggregate loss function in $\ProxyP{}$. With each iteration, the piecewise linear approximation of the aggregate loss improves, and the solution to $\ProxyP{}$ converges to the solution of $\OriginalP$. In this way, these methods allows us to obtain the solution for arbitrarily large instances of $\OriginalP$, since all of the computation for the individual losses is done by the oracle function, which can accomodate distributed computation and generally scales with the same time complexity as matrix-vector multiplication, O$(N^2)$. In addition, these methods allow us to train models with \textit{any} convex loss function (i.e. including non-linear functions) using an IP solver because the IP solver repeatedly solves $\ProxyP{}$, which contains a piecewise linear approximation of the loss function.
\subsubsection{Benders' Decomposition}
\begin{figure}[htbp]
\centering
\includegraphics[width=0.45\textwidth,trim=17.5mm 10mm 12.5mm 10mm,clip=true,page=2]{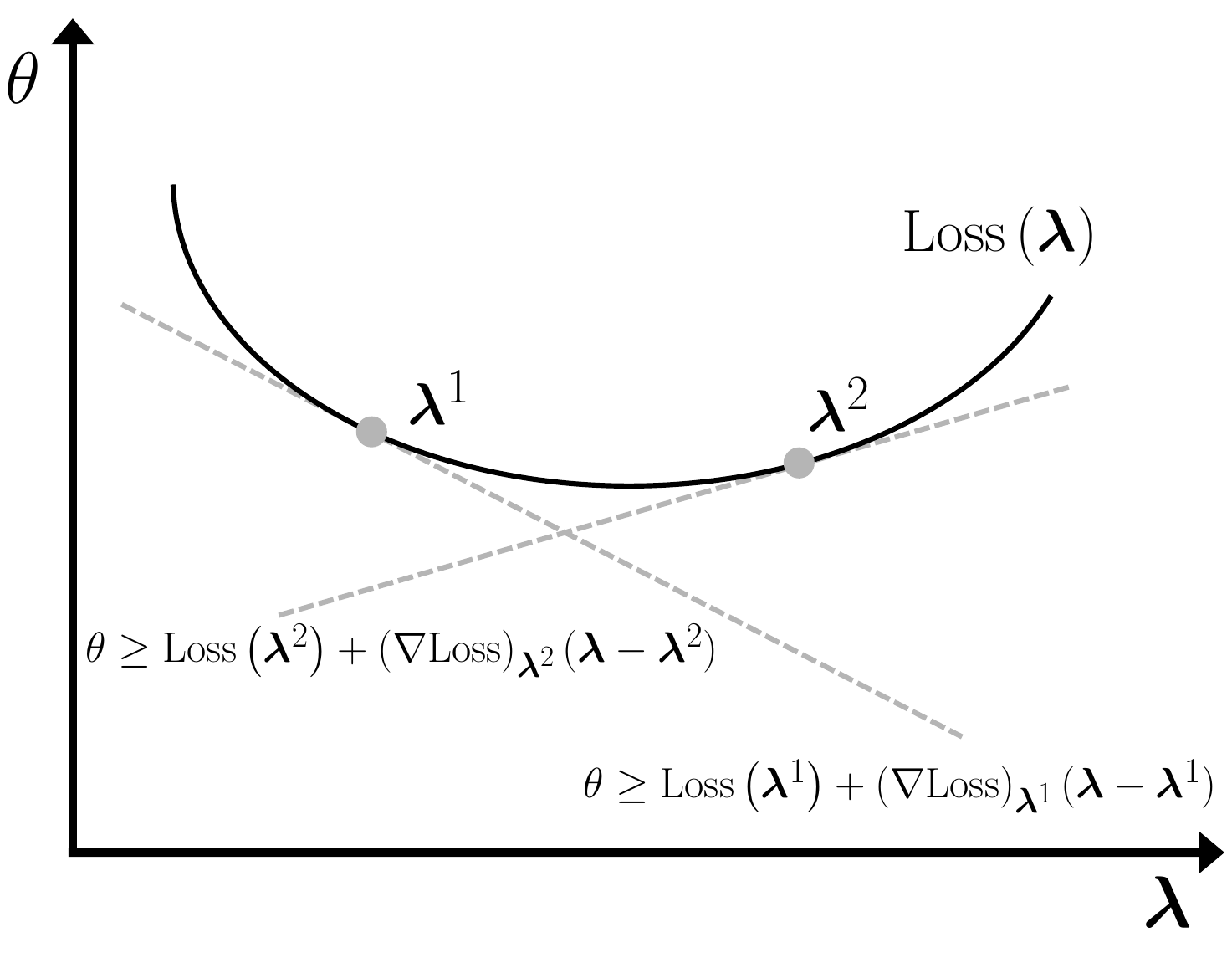} 
\includegraphics[width=0.45\textwidth,trim=17.5mm 10mm 12.5mm 10mm,clip=true,page=3]{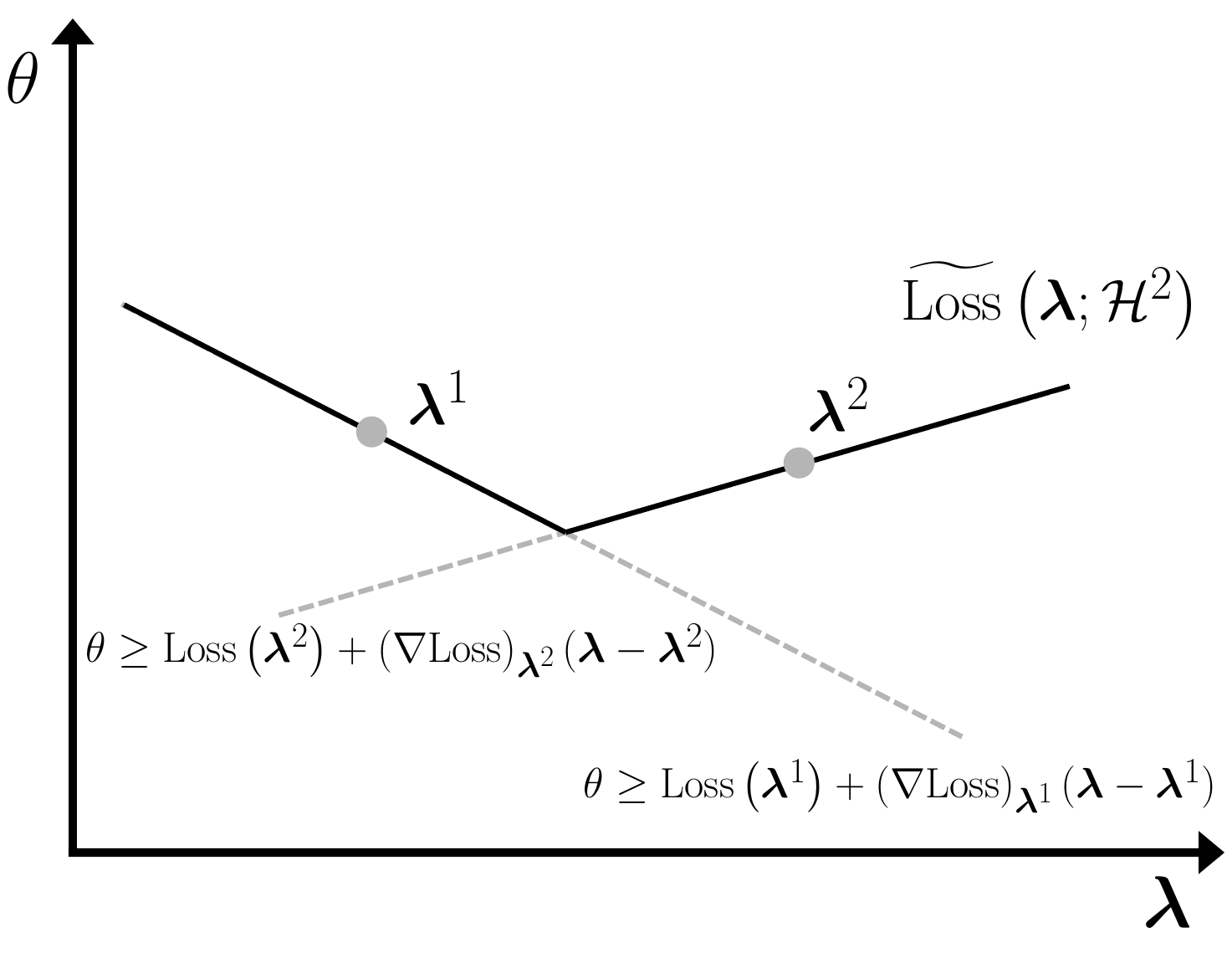} 
\caption{Building a piecewise linear approximation to a convex function using cutting planes. The figure on the left depicts the aggregate loss function $\Loss{\lambdab}$ (black) with cutting planes at the points $\lambdab^1$ and $\lambdab^2$ (grey). The figure on the right depicts the piecewise linear approximation of the aggregate loss function $\ApproxLoss{\lambdab;\CutPoints{2}}$ (black), which is formed as the smallest value of $\theta$ that lies above the collection of cutting planes, $\CutPoints{2}$.}
\label{Fig::ApproxLossFunction}
\end{figure}

We present a popular decomposition algorithm, known as Benders' decomposition, in Algorithm \ref{Alg::BendersGeneric}. 
%
%
This algorithm is initialized with a proxy problem $\ProxyP{0}$ that represents the aggregate loss using the variable $\theta \in \R$:
\begin{equationarray*}{clcl}
\min_{\lambdab,\theta} & \theta & + & C \cdot \IntPen(\lambdab) \\
\st & \lambdab & \in & \Lset \\
    & \theta & \in & \R. 
\end{equationarray*}
On the $k^\text{th}$ iteration, the algorithm solves $\ProxyP{k}$ to obtain the solution $\lambdab^{k}$. Next, it queries the oracle function to obtain a \textit{cutting plane} to the aggregate loss function at $\lambdab^{k}$. A cutting plane at the point $\lambdab^{k}$ is a supporting hyperplane to the aggregate loss function at $\lambdab^{k}$ with the form:
\begin{align}
\theta \geq \Loss{\lambdab^k} + \dLoss{\lambdab^k}(\lambdab - \lambdab^k).\label{Eq::SupportingHyperplane}
\end{align}
Here, $\Loss{\lambdab^k} \in \R$ and $\dLoss{\lambdab^k} \in \R^P$ are fixed quantities that denote the value and subgradient of the aggregate loss at $\lambdab^k$, respectively (note that we drop $\data_N$ for clarity). The algorithm adds the cutting plane in \eqref{Eq::SupportingHyperplane} as a constraint to $\ProxyP{k}$ to yield the proxy problem $\ProxyP{k+1}$,
\clearpage
\begin{equationarray*}{clclcl}
\min_{\lambdab,\theta} & \theta & + & C \cdot \IntPen(\lambdab) & &  \notag \\ 
\st & \lambdab & \in & \Lset & & \notag \\ 
    & \theta & \in & \R & & \notag \\ 
    & \theta & \geq &\Loss{\lambdab^1} & + & \dLoss{\lambdab^1}(\lambdab^1)(\lambdab - \lambdab^1) \notag \\ 
    & & \vdots & \notag \\ 
    & \theta & \geq & \Loss{\lambdab^k} & + & \dLoss{\lambdab^k}(\lambdab^k)(\lambdab - \lambdab^k). \notag
 \end{equationarray*}
In Figure \ref{Fig::ApproxLossFunction}, we show how the algorithm uses a collection of $k$ cutting planes, $\CutPoints{k}$ to create a piecewise linear approximation of the aggregate loss function, $\ApproxLoss{\lambdab;\CutPoints{k}}$. This figure also illustrates why decomposition requires a convex loss function (i.e. the convexity of the loss guarantees that the cutting plane approximation underestimates the true aggregate loss function).

In practice, the piecewise linear approximation of the aggregate loss in $\ProxyP{k}$ improves monotonically with each iteration, and the solution to $\ProxyP{k}$ converges to the solution of $\OriginalP$ (see \citealt{floudas1995nonlinear}). We can detect convergence when the solution to $\ProxyP{k}$ does not change across multiple iterations (see Appendix \ref{Appendix::ProgressOfDecomposition}) or by comparing upper and lower bounds for objective value of $\OriginalP$. 

To describe these bounds, let us denote the objective function of $\OriginalP$ as $Z(\lambdab)$ and denote an optimal solution as $\lambdab^* \in \argmin Z(\lambdab)$. In addition, let us denote the objective function of $\ProxyP{k}$ as $\tilde{Z}(\lambdab;\CutPoints{k})$ and denote an optimal solution as $\tilde{\lambdab}^k \in \argmin \tilde{Z}(\lambdab;\CutPoints{k})$. To obtain a lower bound to $Z(\lambdab^*)$, notice that a piecewise linear approximation of a convex loss function underestimates the true value for all $\lambdab \in \Lset \subseteq \R^P$. This implies:
\begin{alignat}{2} 
\ApproxLoss{\lambdab;\CutPoints{k}} & \leq  \Loss{\lambdab}  & \qquad \forall \lambdab \in \Lset, \nonumber \\
\min_{\lambdab\in\Lset} ~ \ApproxLoss{\lambdab;\CutPoints{k}} +  C \cdot \IntPen(\lambdab) & \leq \min_{\lambdab\in\Lset} ~ \Loss{\lambdab}  +  C \cdot \IntPen(\lambdab), & \label{Eq::DecompEq1} \\ 
\tilde{Z}(\lambdab;\CutPoints{k}) & \leq Z(\lambdab). & \nonumber
\end{alignat}
To obtain an upper bound to $Z(\lambdab^*)$, notice that a point that is feasible for $\ProxyP{k}$ is  also feasible for $\OriginalP$ since both $\ProxyP{k}$ and $\OriginalP$ constrain $\lambdab \in \Lset$ (all other constraints in $\ProxyP{k}$ and $\OriginalP$ are related to the loss function and do not affect feasible values of $\lambdab$). This implies:
\begin{align}
Z(\lambdab^*) \leq Z(\lambdab^k). \label{Eq::DecompEq2}
\end{align}
Since $\lambdab^k$ is the minimizer of $\tilde{Z}(\lambdab;\CutPoints{k})$, we have that $$\tilde{Z}(\lambdab^{k};\CutPoints{k}) \leq \tilde{Z}(\lambdab^*;\CutPoints{k}).$$ We can now combine the inequality in \eqref{Eq::DecompEq1} for $\lambdab = \lambdab^*$ with the inequality in \eqref{Eq::DecompEq2} to see that $$\tilde{Z}(\lambdab^{k};\CutPoints{k}) \leq \tilde{Z}(\lambdab^*;\CutPoints{k}) \leq Z(\lambdab^*) \leq Z(\lambdab^k).$$ Thus, we have derived a lower bound, $LB^k = \tilde{Z}(\lambdab^{k};\CutPoints{k})$, and an upper bound, $UB^k = Z(\lambdab^k)$ for $Z(\lambdab^*)$. These bounds provide a guarantee on the optimality of the solution $\lambdab^k$ at iteration $k$.
%
%
\begin{algorithm}
\caption{Benders' decomposition}
\begin{algorithmic}
  \REQUIRE $\delta > 0$, tolerance gap between upper and lower bound
  \REQUIRE $\OriginalP$, original optimization problem with objective function, $Z(\lambdab) = \Loss{\lambdab} + C\cdot\IntPen(\lambdab)$
  \REQUIRE $\ProxyP{0}$, initial proxy problem with objective, $\tilde{Z}(\lambdab;\CutPoints{0}) = \theta + C\cdot\IntPen(\lambdab)$ 
  \REQUIRE oracle function to compute $\Loss{\lambdab}$ and $\dLoss{\lambdab}$ for any $\lambdab \in \Lset$
  \STATE \textbf{Initialize:} $k \longleftarrow 0$, $UB^k \longleftarrow \infty$, $LB^k \longleftarrow 0$
  \WHILE {$UB^k - LB^k < \delta$}
    \STATE Solve $\ProxyP{k}$ to obtain $\lambdab^k$
    \STATE Query the oracle function to obtain $\Loss{\lambdab^k}$ and $\dLoss{\lambdab^k}$
    \STATE Obtain $\ProxyP{k+1}$ by adding the cut, $\theta \geq \Loss{\lambdab^k} +\dLoss{\lambdab^k}(\lambdab - \lambdab^k)$ to $\ProxyP{k}$
    \STATE $UB^{k+1} \longleftarrow \max(UB^{k},Z(\lambdab^k))$
    \STATE $LB^{k+1} \longleftarrow \tilde{Z}(\lambdab^k;\CutPoints{k})$
    \STATE $k \longleftarrow k + 1$
  \ENDWHILE
  \ENSURE $\lambdab^k$, $\delta$-optimal solution for $\OriginalP$
\end{algorithmic}
\end{algorithm}\label{Alg::BendersGeneric}
\subsubsection{Trade-offs Between Accuracy and Scalability}
Loss decomposition requires the use of a convex loss function, and involves trade-offs between accuracy and computation. In theory, we know that classifiers that minimize the 0--1 loss attain the lowest possible training accuracy (by definition). In practice, however, we know that this approach may be intractable for large datasets. Here, we illustrate these tradeoffs using a controlled experiment where we compare the accuracy of classifiers produced with different loss functions as we increase the size of the training data.

We considered a basic setup where the interpretability penalty includes only an $\lzero$-penalty and the interpretability set restricts coefficients to integers between -10 and 10. With this setup, we trained one set of classifiers with the logistic loss using the Benders' decomposition in Algorithm \ref{Alg::BendersGeneric}. These classifiers represent the solution to the following mixed-integer non-linear program:
\begin{align}
\label{Eq::DecompositionSLIMLog}
\begin{split}
\min_{\lambdab} & \qquad \LogLoss{\lambdab} + C_0 \vnorm{\lambdab}_0 \\ 
\st &  \qquad \lambdab \in \Z^{P+1}\cap[-10,10]^{P+1}.
\end{split}
\end{align}
As a comparison, we also trained a set of classifiers with the 0--1 loss by directly optimizing the IP:
\begin{align}
\label{Eq::DecompositionSLIM01}
\begin{split}
\min_{\lambdab} & \qquad \ZeroOneLoss{\lambdab} + C_0 \vnorm{\lambdab}_0 \\
\st &  \qquad \lambdab \in \Z^{P+1} \cap[-10,10]^{P+1}.
\end{split}
\end{align}
In both cases, we set the value of $C_0$ to $\frac{0.9}{N}$ -- small enough so that the optimal classifiers for problem \eqref{Eq::DecompositionSLIM01} would attain the highest possible training accuracy.

We trained a classifier using each loss function for 17 datasets with the number of examples ranging between $N=50$ to $N=10\,000\,000$. We generated training data from two $5$-dimensional Gaussian distributions with means $\mu_{+1}=(2,2,2,2,2)$ and $\mu_{-1}=(0,0,0,0,0,0)$ and a unit covariance matrix $\Sigma_{_+1} = \Sigma_{-1} = I$. Thus, we were able to maintain the same level of difficulty for each $N$ and also determine that the optimal classifier was linear with coefficients $\lambdab^*=(-10,2,2,2,2,2) \in \Lset$. For each $N$, we trained a classifier with the logistic loss by running Benders' decomposition until convergence. We then ran the IP solver for the time that it took Benders' decomposition to converge. Thus, the classifiers we trained with the logistic loss represent optimal solutions for problem \eqref{Eq::DecompositionSLIMLog} while the classifiers we trained with the 0--1 loss represent the best feasible solution to \eqref{Eq::DecompositionSLIM01} that were produced within a severe time restriction.

\begin{figure}[htbp]
\centering
\includegraphics[width=0.45\textwidth,trim=2.5mm 2.5mm 0mm 2.5mm,clip=true]{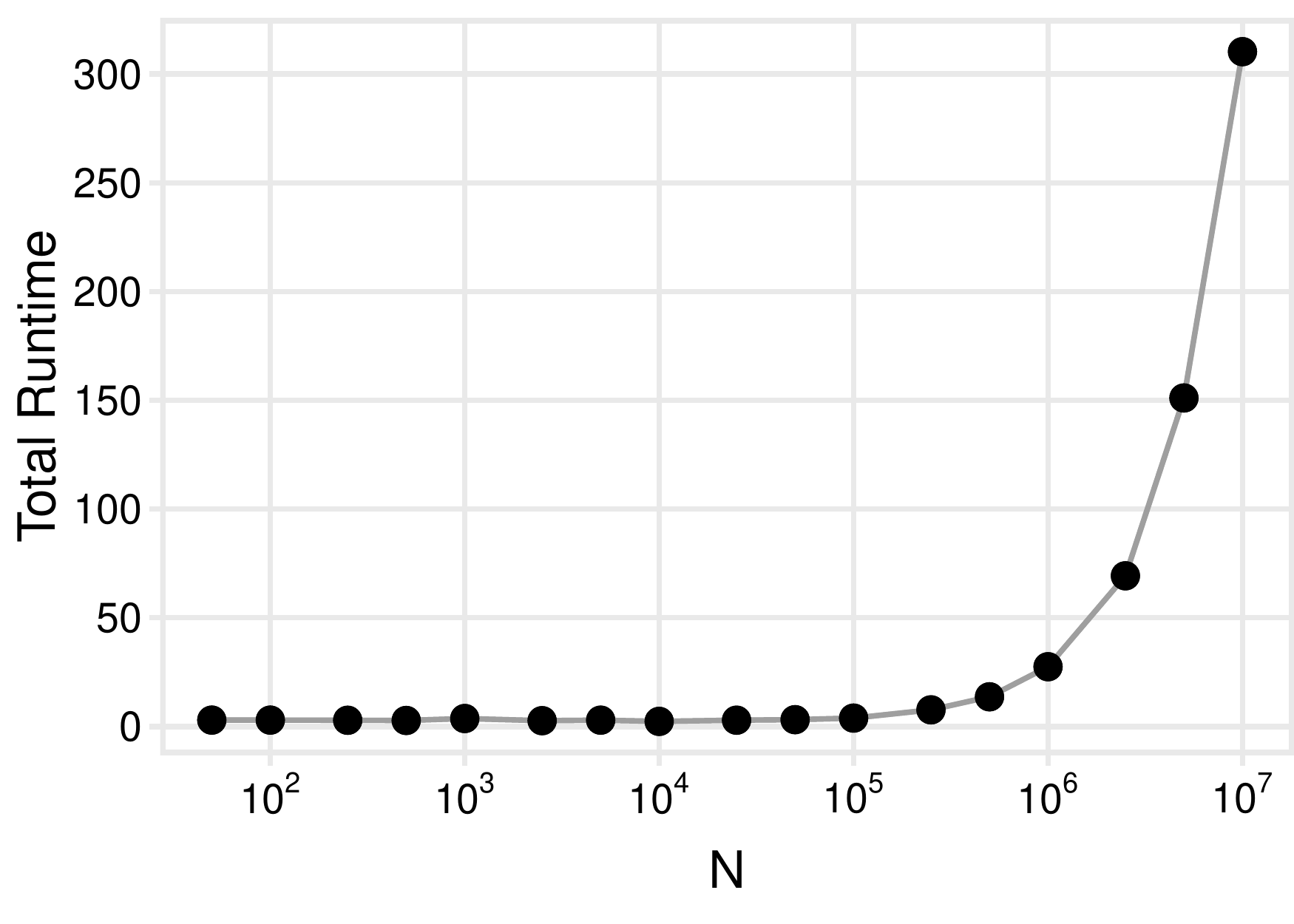}
\caption{Time required to train a classifier with the logistic loss and Benders' decomposition on simulated datasets with $N=50$ to $N = 10000000$. This setup can optimize a model with $N = 10000000$ data points in 310.3 seconds. We include an iteration-by-iteration overview of the algorithm for the $N = 10000000$ case in Appendix \ref{Appendix::ProgressOfDecomposition}.}
\label{Fig::DecompositionRuntime}
\end{figure}

\begin{figure}[htbp]
\centering
\includegraphics[width=0.45\textwidth,trim=2.5mm 2.5mm 0mm 2.5mm,clip=true]{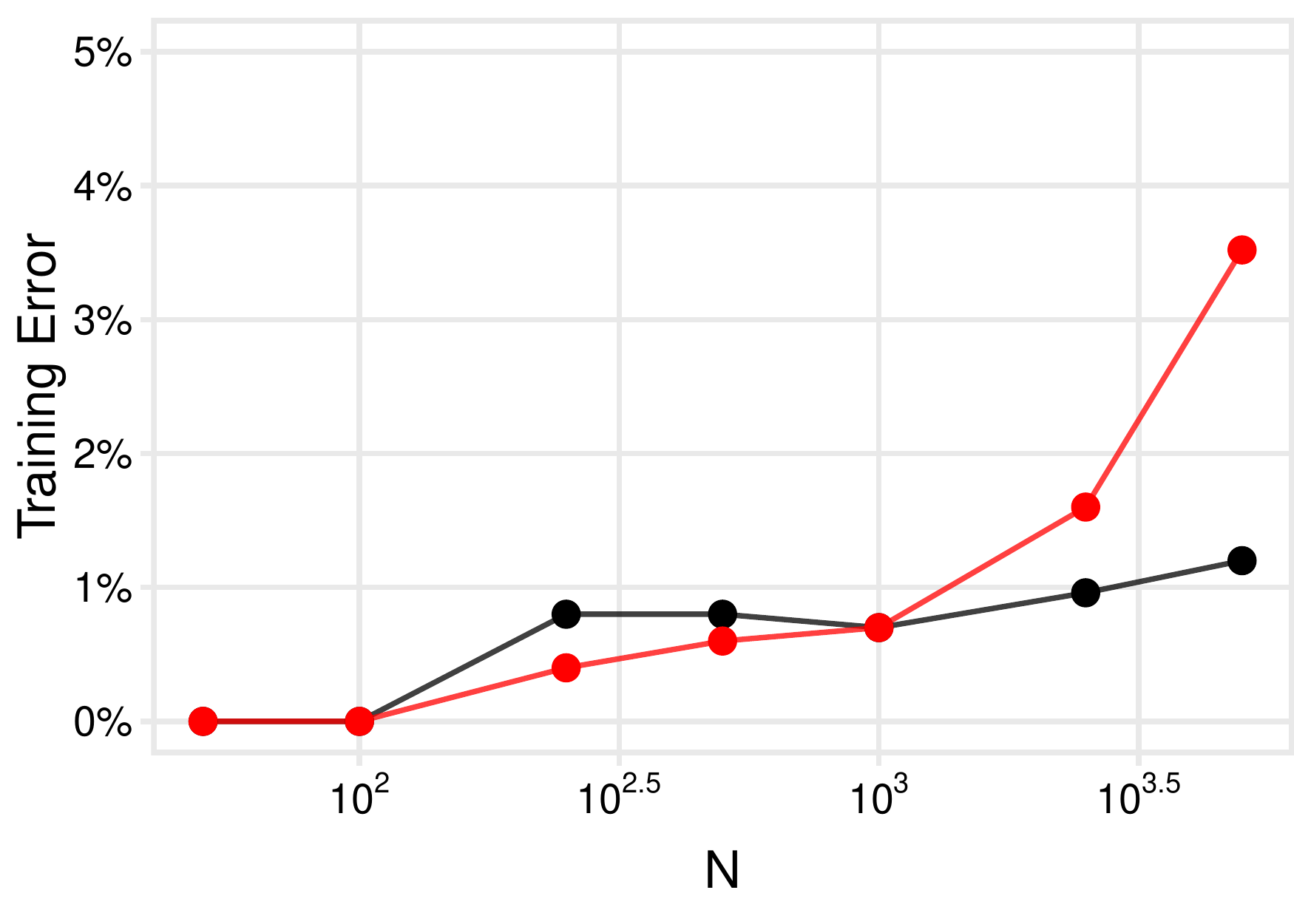}
\includegraphics[width=0.45\textwidth,trim=2.5mm 2.5mm 0mm 2.5mm,clip=true]{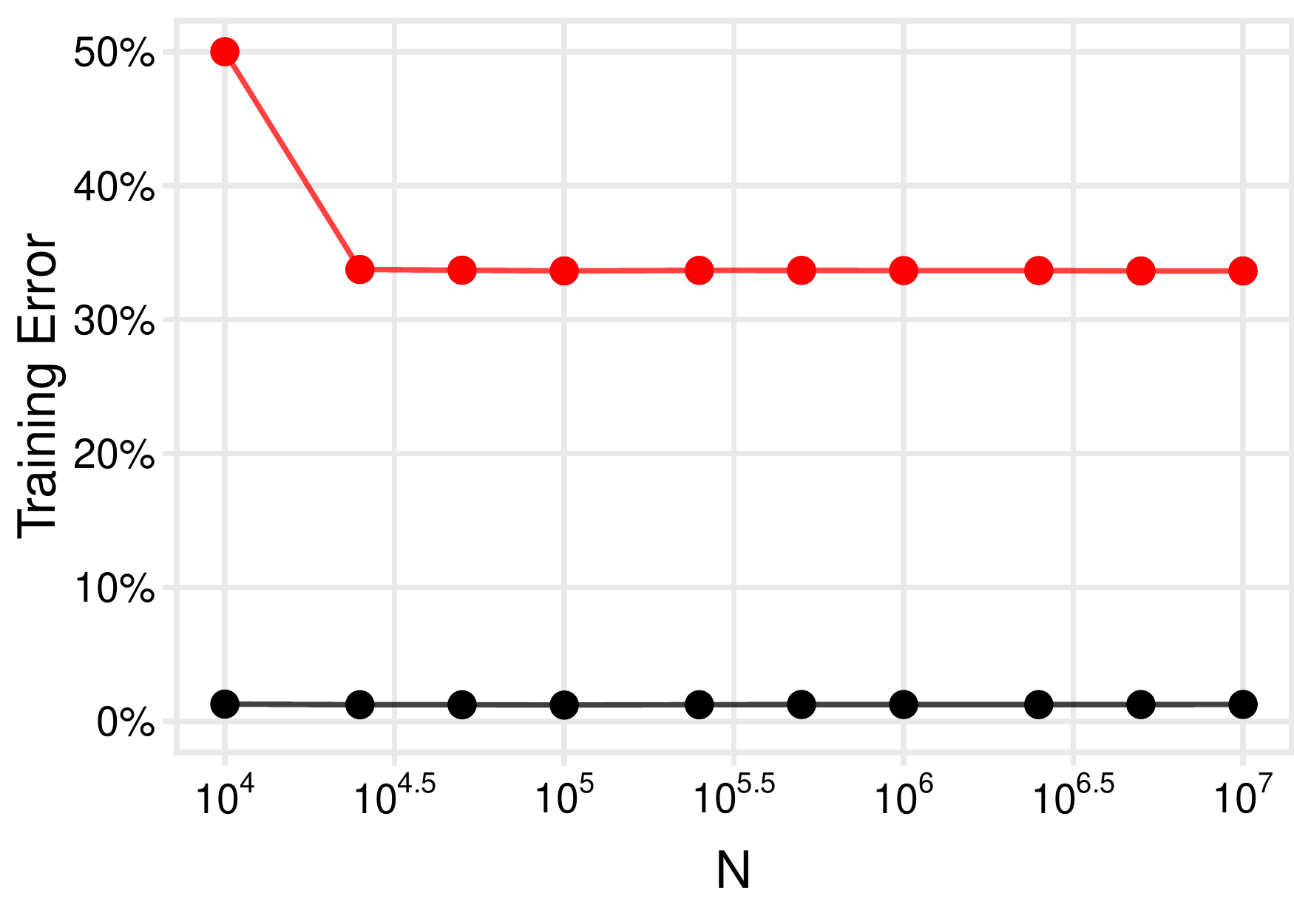}
\caption{Training error of classifiers trained with the logistic loss (black) and the 0--1 loss (red) on datasets with $N<10000$ points (left) and $N\geq10000$ points (right). For each $N$, we trained the optimal classifier with the logistic loss by running Benders' decomposition until convergence. We then trained a classifier with the 0--1 loss by solving an IP for an equivalent amount of time. Thus, the classifiers trained with the 0--1 loss correspond to the best feasible solution for the IP obtained in the time that it took Benders' decomposition to converge. Here, the limited training time affects the accuracy of classifiers trained with the 0--1 loss starting at $N=5000$ points; classifiers trained with the 0--1 loss achieve very poor accuracy for $N\geq 10000$ due to the severe time restriction.}
\label{Fig::DecompositionTrainingError}
\end{figure}

We summarize the results of this experiment in Figures \ref{Fig::DecompositionRuntime} and \ref{Fig::DecompositionTrainingError}. These results were produced using simple implementations with default settings in MATLAB 2014a and the CPLEX 12.6 API on 2.6 GHZ machine with 16GB RAM. Figure \ref{Fig::DecompositionRuntime} shows the time required to train classifiers with the logistic loss using Benders decomposition for different $N$ (i.e. the runtime for both methods). In this case, the oracle function computes cutting planes using matrix-vector multiplication in MATLAB and scales with O($N^2$). As a result, the algorithm produces the optimal classifier for a dataset with $N = 10\,000\,000$ points in 310.3 seconds. We provide an detailed overview of the algorithm for this case in Appendix \ref{Appendix::ProgressOfDecomposition}. Figure \ref{Fig::DecompositionTrainingError} shows the training error for classifiers trained with the logistic loss and the 0--1 loss for different $N$. Here, we see that when we impose a limit on computation, there is a threshold above which classifiers trained with the 0--1 loss achieve higher training error than the classifiers trained with the logistic loss. In this case, classifiers trained with the 0--1 loss have higher training error for datasets with $N\geq5000$. As a reminder, these results do not imply we cannot train classifiers for $N\geq10000$ with the 0--1 loss, as we imposed a severe time limit on the IP solver for the purposes of experimentation, as shown in Figure \ref{Fig::DecompositionRuntime}.
\subsubsection{Discussion}
Loss decomposition involves trade-offs between accuracy, scalability and flexibility. On one hand, the method allows us to efficiently train classifiers with any convex loss function using a IP solver, which may be necessary when we need to train models that can produce conditional probability estimates (i.e. using the logistic loss). On the other hand, the method requires us to forfeit key in minimizing the 0--1 loss, such as the robustness to outliers and the ability to formulate hard constraints on accuracy. Even so, loss decomposition allows us to benefit from scalability while maintaining substantial control over the interpretability by means a discrete interpretability penalty and discrete interpretability set. 

Loss decomposition is well-suited to train models on large-scale datasets because it confines all data-related computation to an oracle function, which can compute cutting planes for popular loss functions with polynomial running time in $N$ (as shown in Appendix \ref{Table::LossFunctions}, cutting planes can be computed using simple operations such as matrix-vector multiplication). Note that loss decomposition does not scale polynomially in $P$: increasing $P$ requires the approximation of a high-dimensional loss function which involves an exponential number of cutting planes. In practice, we can improve the baseline performance of loss decomposition on datasets with large $P$ by using a \textit{one-tree implementation} \citep{Bai2009,NaoumSawaya:2010ky}. We can also substantially improve scalability in $P$ by adding cutting planes at geometrically significant points of the feasible region of $\ProxyP{k}$, such as its center of gravity \citep{Newman:1965de,levin1965algorithm}, its Chebyshev center \citep{Elzinga:1975eq}, or its analytic center \citep{atkinson1995cutting,goffin2002convex}.
\subsection{Data Reduction}\label{Sec::Reduction}
Data reduction is a procedure for filtering training data when we train models using a robust non-convex loss function, such as the 0--1 loss. Given initial training data, $\data_N = (\xb_i,y_i)_{i=1}^N$, and a proxy to our original optimization problem, data reduction solves $N+1$ variants of the proxy problem to identify examples whose class can be determined ahead of time. These examples are then removed to produce \textit{reduced} training data, $\data_M \subseteq \data_N$. The computational gain associated with data reduction comes from training models with $\data_M$, which requires us to solve an instance of our original optimization problem with $N-M$ fewer loss constraints.

We provide an overview of data reduction in Algorithm \ref{Alg::DataReduction}. To explain how the algorithm works, let us consider an optimization problem from our framework, expressed in terms of classifier functions:
\begin{align}
\label{Eq::ReductionOriginal}
\min_{f\in\F} ~ Z(f;\data_N).
\end{align}
Here, $Z(f;\data_N) = \Loss{\lambdab;\data_N}+C\cdot \IntPen(\lambdab)$ and $\F = \{f:\X\to\Y \;|\; f(\xb) = \sign{\lambdab^T\xb} \text{ and } \lambdab \in \Lset \}$. Data reduction filters the training data by solving a convex proxy:
\begin{align}
\label{Eq::ReductionProxy}
\min_{f\in\tilde{\F}} ~ \tilde{Z}(f;\data_N).
\end{align}
Here, the objective function of the proxy problem, $\tilde{Z}:\tilde{\F}\rightarrow\R$, is chosen as a convex approximation of the objective function to the original problem, $Z:\F\rightarrow\R$. Similarly, the set of feasible classifiers of the proxy problem, $\tilde{\F}$, is chosen as a convex approximation of the set of feasible classifiers of the original optimization problem, $\F$. We assume, without loss of generality, that $\F \subseteq \tilde{\F}$. 

Data reduction works with any proxy problem so long as we can hypothesize that the $\varepsilon$-level set of the proxy problem contains the set of optimizers to the original problem. That is, we can use any feasible set $\tilde{\F}$ and any objective function $\tilde{Z}:\tilde{\F}\rightarrow\R$, as long as we can specify a value of $\varepsilon$ that is large enough for the following \textit{level set condition} to hold:
\begin{align}
\hspace{0.3\textwidth} & \tilde{Z}(f^*) \leq  \tilde{Z}(\tilde{f}^*) + \varepsilon & \forall f^* \in \F^* \text{ and } \tilde{f}^* \in \tilde{\F}^*. \label{Eq::ReductionLevelSet}
\end{align}
Here, $f^*$ denotes an optimal classifier to the original problem from the set $\F^* = \argmin_{f\in\F} Z(f)$ and $\tilde{f}^*$ denotes an optimal classifier to the proxy problem from the set $\tilde{\F}^* = \argmin_{f\in\tilde{\F}} \tilde{Z}(f)$. The width of the the level set, $\varepsilon$ is related to the amount of data that will be filtered: if $\varepsilon$ is chosen too large, the method will not filter very many examples and will be less helpful for reducing computation (see Figure \ref{Fig::BankruptcyDataReduction}). In what follows, we often refer to the value $\tilde{Z}(\tilde{f}^*) + \varepsilon$ as the \textit{upper bound on the objective value of all classifiers in the $\varepsilon$-level set}. 

\begin{figure}[htbp]
\centering
\includegraphics[width=0.45\textwidth,trim=5mm 20mm 5mm 20mm,clip=true]{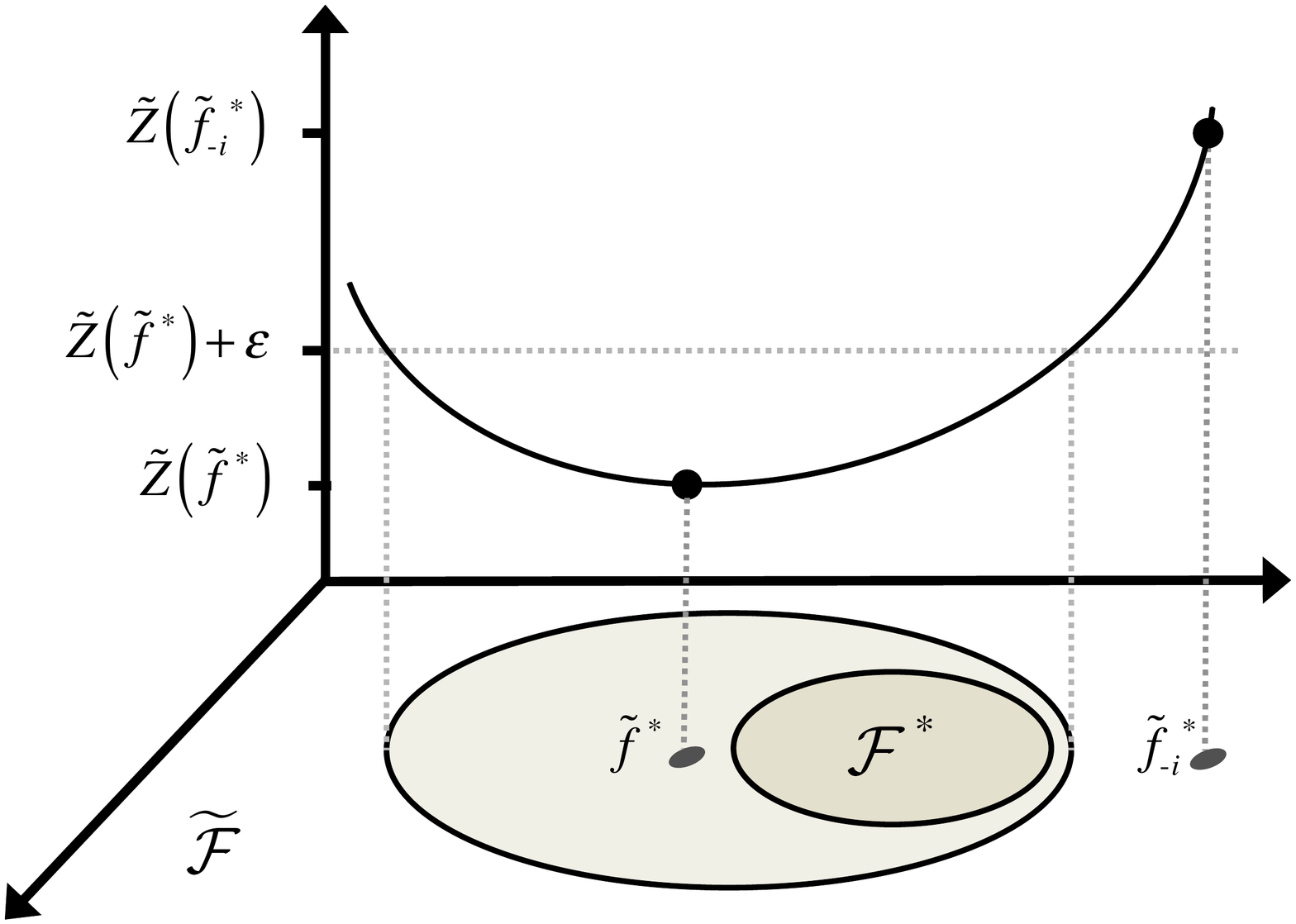}
\caption{We initialize data reduction with $\varepsilon$ large enough so that $\tilde{Z}(f^*)<\tilde{Z}(\tilde{f}^*)+\varepsilon$ for all $f^* \in \F^*$ and all $\tilde{f}^* \in \tilde{\F}^*$. Here, $f^*$ denotes an optimal solution to the original optimization problem from the set of optimal solutions $\F^*$, and $\tilde{f}^*$ denotes a solution to the proxy optimization problem from the set of optimal solutions $\tilde{\F}^*$. Data reduction trains an classifier $\tilde{f}^*_{\text{-}i}$ for each example in the initial training data, $\data_N$, by solving a variant of the proxy problem that forces $\tilde{f}^*_{\text{-}i}$ to label example $i$ in a different way than $\tilde{f}^*$. Here, $\tilde{Z}(\tilde{f}^*_{\text{-}i})>\tilde{Z}(\tilde{f}^*)+\varepsilon$. Thus, we know the predicted sign of example $i$ and do not include it in the reduced training data, $\data_M$.}
\label{Fig::Reduction}
\end{figure}

In the first stage of data reduction, we solve the convex proxy in order to (i) compute the upper bound on the objective value of all classifiers in the $\varepsilon$-level set, $\tilde{Z}(\tilde{f}^*)+\varepsilon$, and (ii) to identify a set of baseline labels, $\tilde{y}_i = \sign{\tilde{f}^*(\xb_i)}.$ In the second stage of data reduction, we solve a variant of the convex proxy for each of the $N$ examples. Here, the $i^\text{th}$ variant of the convex proxy for contains an additional constraint that forces example $i$ to be classified as $-\tilde{y}_i$:
\begin{align}
\label{Eq::DataReductionConvexVariant}
\min_{\substack{f\in\tilde{\F}\\ \tilde{y}_i f(\xb_i) < 0}} ~ \tilde{Z}(f).
\end{align}
We denote the optimal classifier obtained by solving the $i^\text{th}$ variant problem as $\tilde{f}^*_{\text{-}i}$. If the optimal value of the $i^\text{th}$ variant problem exceeds the upper bound of classifiers in the $\varepsilon$-level set (i.e. $\tilde{Z}(\tilde{f}^*_{\text{-}i})>\tilde{Z}(\tilde{f}^*)+\varepsilon$) then no classifier in the $\varepsilon$-level set can label point $i$ as $-\tilde{y}_i$. In other words, all classifiers in the $\varepsilon$-level set must label this point as $\tilde{y}_i$. Since the $\varepsilon$-level set contains the optimal classifiers to the original problem by the assumption in \eqref{Eq::ReductionLevelSet}, we can remove example $i$ from the reduced dataset $\data_M$ because we know that any optimal classifier to the original problem will label this point as $\tilde{y}_i$. We illustrate this situation in Figure \ref{Fig::Reduction}, 

In Theorem \ref{Thm::DataReduction}, we prove that we obtain the same set of optimal classifiers if we train a model with the original data, $\data_N$, or the reduced data, $\data_M$. In Theorem \ref{Thm::ReductionSufficientConditions}, we provide a set of conditions that are sufficient for a proxy to the 0--1 loss function to satisfies the level set condition from \eqref{Eq::ReductionLevelSet} for any given $\varepsilon$. These conditions can be used to determine a value of $\varepsilon$ that is large enough for the level set condition to hold for any loss function chosen to approximate the 0--1 loss function. Alternatively, these conditions can also be used to craft a proxy loss function that fulfills the level-set condition for a fixed value of $\varepsilon$.
\begin{algorithm}
\caption{Data Reduction from $\data_N$ to $\data_M$\label{Alg::DataReduction}}
\begin{algorithmic}
  \REQUIRE $\data_N = (\xb_i,y_i)_{i=1}^N$, initial training data \\
  \REQUIRE $\min \, \tilde{Z}(f;\data_N) \textrm{ s.t. }  f \in \tilde{\F}$, convex proxy problem trained with $\data_N$ \\
  \REQUIRE $\varepsilon$, width of the convex proxy level set
  \STATE \textbf{Initialize:} $\data_M \longleftarrow \emptyset$
  \STATE $\tilde{f}^* \longleftarrow \argmin_f \tilde{Z}(f;\data_N)$ 
  \FOR {$i = 1,\ldots,N$}
    \STATE $\tilde{y}_i \longleftarrow \sign{\tilde{f}^*(\xb_i)}$
    \STATE $\tilde{f}^*_{\text{-}i} \longleftarrow \argmin \tilde{Z}(f;\data_N) \; \st \; f \in \tilde{\F}, \; \tilde{y}_i f(\xb_i) < 0$ 
           \IF {$\tilde{Z}(\tilde{f}^*_{\text{-}i};\data_N) \leq \tilde{Z}(\tilde{f}^*;\data_N) + \varepsilon$}
               \STATE $\data_M \longleftarrow \data_M \cup (\xb_i,y_i)$
            \ENDIF
            \ENDFOR
  \ENSURE $\data_M$, reduced training data
\end{algorithmic}
\end{algorithm}
\FloatBarrier
\begin{thm}[Equivalence of the Reduced Data]\label{Thm::DataReduction}
Consider an optimization problem used to train a classifier $f \in \F$ with data $\data_N$,
$$\min_{f \in \F} Z(f;\data_N),$$
as well as a proxy optimization problem used to train a classifier $f \in \tilde{\F}$ with data $\data_N$,
$$\min_{f \in \tilde{\F}} \tilde{Z}(f;\data_N).$$
Let $f^*$ denote an optimal classifier to the original problem from the set of optimal classifiers $\F^* = \argmin_{f\in\F} Z(f;\data_N)$, and let $\tilde{f}$ denote a classifier to the proxy problem from the set of optimal classifiers $\tilde{\F}^* = \argmin_{f\in\tilde{\F}}  \tilde{Z}(f;\data_N)$.

If we choose a value of $\varepsilon>0$ large enough so that
\begin{align}
\label{Eq::ReductionAssumption}
\centering
\tilde{Z}(f^*;\data_N) &\leq \tilde{Z}(\tilde{f}^*;\data_N) + \varepsilon \quad \forall f^* \in \F^* \text{ and } \tilde{f}^* \in \tilde{\F}^*, 
\end{align}
then data reduction (Algorithm \ref{Alg::DataReduction}) will output a reduced dataset $\data_M \subseteq \data_N$ such that
\begin{align}\label{Eq::ReductionWTS}
\argmin_{f\in\F} Z(f;\data_N) = \argmin_{f\in\F} Z(f;\data_M).
\end{align}
\end{thm}
\proof{Proof:} See Appendix \ref{Appendix::Proofs}. \endproof
\newcommand{\z}[2]{Z_{#1}\left(#2\right)}
\newcommand{\Zmip}[1]{\z{01}{#1}}
\newcommand{\Zcvx}[1]{\z{\psi}{#1}}
\newcommand{\lmip}[0]{\lambdab^*_{01}}
\newcommand{\lcvx}[0]{\lambdab^*_{\psi}}
\begin{thm}[Sufficient Conditions to Satisfy the Level Set Condition]\label{Thm::ReductionSufficientConditions}
Consider an optimization problem from our framework, $$\min_{\lambdab \in \R^P} \Zmip{\lambdab},$$ where the objective $Z_{01}: \R^P\rightarrow\R$ contains the 0--1 loss function. In addition, consider a proxy optimization problem used in a reduction method, $$\min_{\lambdab \in \R^P} \Zcvx{\lambdab},$$ where the objective contains a proxy loss function, $\psi:\R^P\rightarrow\R$.  Let $\lmip \in \argmin_{\lambdab \in \R^P} \Zmip{\lambdab}$ denote an optimizer of the 0--1 loss function, and $\lcvx \in \argmin_{\lambdab \in \R^P} \Zcvx{\lambdab}$ denote an optimizer of the proxy loss function. 

If the proxy loss function, $\psi$, satisfies the following properties:
  \begin{enumerate}[label=\textnormal\emph{\Roman*.}]
   \item \textit{Upper Bound on the 0--1 Loss:} $\Zmip{\lambdab} \leq \Zcvx{\lambdab} \qquad \forall \lambdab \in \R^P$
   \item \textit{Lipschitz Near $\lmip$:} $\| \lambdab - \lcvx \| < A \implies \Zcvx{\lambdab} - \Zcvx{\lcvx} < L \| \lambdab - \lcvx \| $
   \item \textit{Curvature Near $\lcvx$:} $\| \lambdab - \lcvx \| > C_{\lambdab} \implies \Zcvx{\lambdab} - \Zcvx{\lcvx} > C_{\psi}$
   \item \textit{Closeness of Values Near $\lmip$:} $| \Zcvx{\lmip} - \Zmip{\lmip}| < \varepsilon $
  \end{enumerate}
then it will also satisfy a level-set condition required for reduction: $$\Zcvx{\lmip} \leq \Zcvx{\lcvx}+\varepsilon$$ for all $\lmip \in \argmin_{\lambdab \in \R^P} \Zmip{\lambdab}$  and $\lcvx \in \argmin_{\lambdab \in \R^P} \Zcvx{\lambdab}$, whenever $\varepsilon = LC_{\lambdab}$ obeys $C_{\psi} > 2\varepsilon$.
\end{thm}
\proof{Proof:} See Appendix \ref{Appendix::Proofs}. \endproof
\subsubsection{Off-the-Shelf Implementation}\label{Sec::ReductionDemo}
Data reduction can be applied to any optimization problem from our framework by using its convex relaxation as the proxy problem. This ``off-the-shelf" approach avoids the intricacies in finding a suitable proxy problem for data reduction.

Given an IP from our framework, $\min Z(f) \;\;\st\;\; f \in \F,$ let us denote the convex relaxation of this IP as $\min Z(f) \;\;\st\;\; f \in \tilde\F.$ We assume that $\F$ is a discrete set, and that $\tilde{\F}$ is the convex hull of this set. 
When we use the convex relaxation to the IP as the proxy problem, we can determine a value of $\varepsilon$ that is large enough to satisfy the level set condition from \eqref{Eq::ReductionLevelSet} using feasible solutions to the IP. To see this, let $\hat{f} \in \F$ denote a feasible solution to the IP, let $f^* \in \argmin_{f\in\F}Z(f)$ denote an optimal solution to the IP, and let $\tilde{f}^* \in \argmin_{f\in\tilde{\F}} Z(f)$ denote an optimal solution to the convex relaxation of the IP, and note that:
\begin{align}
Z(\tilde{f}^*) \leq Z(f^*) \leq Z(\hat{f}) \label{Eq::RelaxFeas}.
\end{align}
Here the inequality $Z(\tilde{f}^*) \leq Z(f^*)$ follows from the fact that $\F \subseteq \tilde{\F}$; and the inequality $Z(f^*) \leq  Z(\hat{f})$ follows from the fact that $f^*$ is the optimal solution to the IP while $\hat{f}$ is a feasible solution to the IP.
Thus, we can use any feasible solution to the IP, $\hat{f} \in \F$, to determine the following value of $\varepsilon$ that satisfies the level set condition in \eqref{Eq::ReductionLevelSet}:
\begin{align}
\varepsilon(\hat{f}) = Z(\hat{f}) - Z(\tilde{f}^*).
\end{align}

\begin{figure}[htbp]
\centering
\includegraphics[width=0.45\textwidth,trim=0mm 2.5mm 0mm 2.5mm,clip=true]{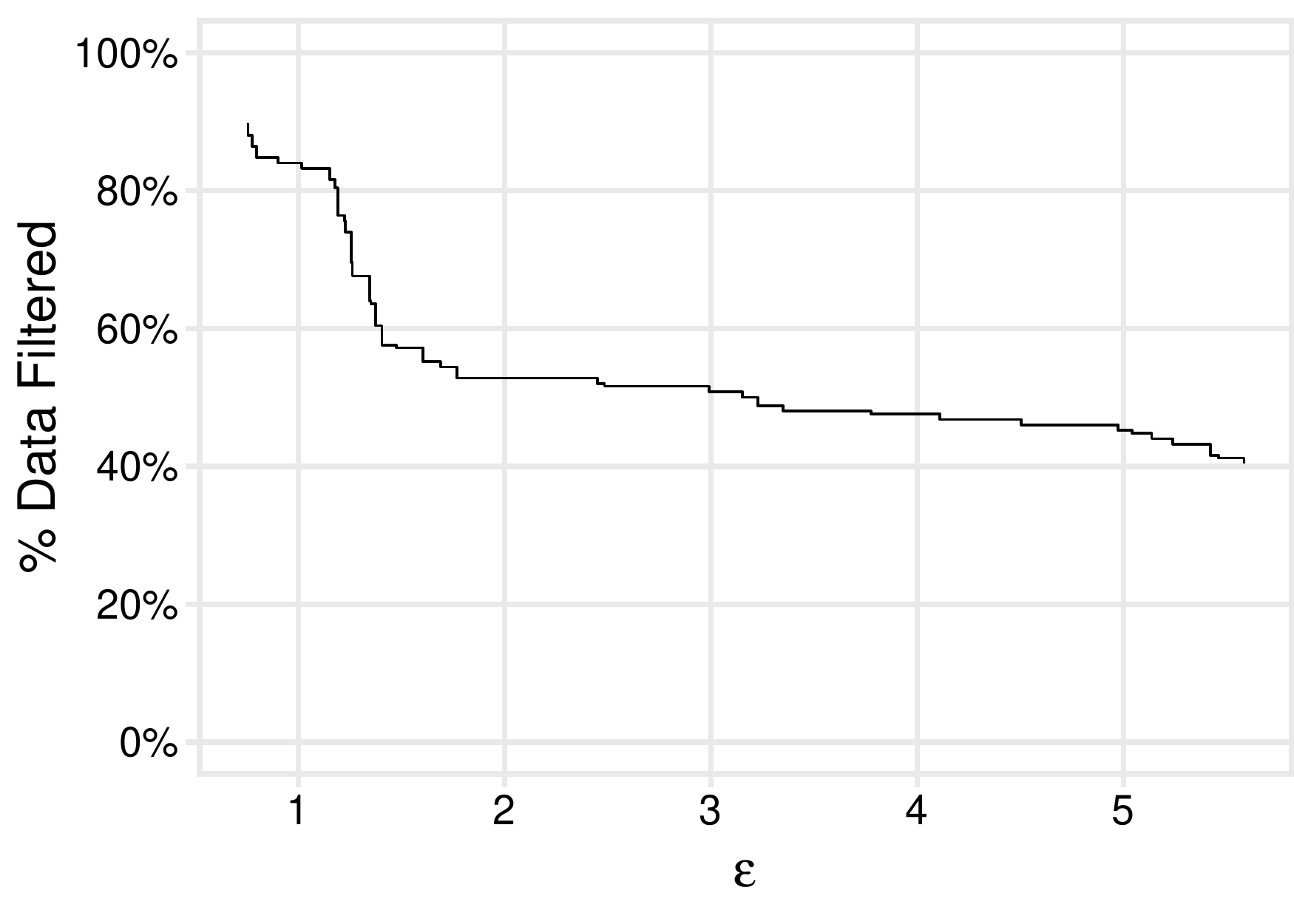}
\caption{Proportion of the data filtered as a function of the width of the level set, $\varepsilon$. For each $\varepsilon$, we were able to train a second classifier that attained the same objective value using only the reduced data.}
\label{Fig::BankruptcyDataReduction}
\end{figure}

In Figure \ref{Fig::BankruptcyDataReduction}, we demonstrate this approach by applying data reduction on a simple optimization problem used to train a SLIM scoring system on the \textds{bankruptcy} dataset for a range of different values for the width of the level set parameter, $\varepsilon$. Specifically we show the proportion of data that was filtered by data reduction as we increased the width of the level set, $\varepsilon$, from its smallest possible value, $\varepsilon_{\min}$, to the largest possible value we would use in practice, $\varepsilon_{\max}$. In this case, our original optimization problem was an instance of the SLIM IP from Section \ref{Sec::SLIM} with  $\Lset_j = \Z\cap[-10,10]$, $C_0 = 0.01$, $\epsilon = 0.9/NP$, $\gamma=0.1$ and $M_i = \max_{\lambdab \in \Lset} (\gamma - y_i \lambdab^T x_{i,j})$. The proxy problem was a convex relaxation of this IP. Here, $\varepsilon_{\min}$ corresponds the value of $\varepsilon$ that we computed using the best feasible solution to the IP (i.e. the optimal solution to the IP) as: $$\varepsilon_{\min} = \tilde{Z}(f^*) -  \tilde{Z}(\tilde{f}^*),$$ and $\varepsilon_{\max}$ corresponds to the value of $\varepsilon$ that was computed using a trivial feasible solution to the IP that users could guess without any computation (i.e. a linear classifier with $\lambdab = 0$): $$\varepsilon_{\max} = \tilde{Z}(0) -  \tilde{Z}(\tilde{f}^*).$$ 

The results in Figure \ref{Fig::BankruptcyDataReduction} show that the proportion of data that is filtered decreases as we increase the width of the level set parameter, $\varepsilon$. These results also that a large amount of data can be filtered by using high quality feasible solutions. Keep in mind that for each value of $\varepsilon$ in Figure \ref{Fig::BankruptcyDataReduction}, we were able to train a model that attained the same objective value using only the reduced data.
%
%
%
\subsubsection{Discussion}
Unlike loss decomposition, data reduction can decrease the computation when training any model from our framework, including models that we train with the 0--1 loss function. Data reduction can be used in a preliminary procedure before the training process, as an iterative procedure that is called by the IP solver during the training process, or even in conjunction with a screening test for the convex problem (see e.g. \citealt{wang2013lasso,liu2013safe}). In situations where it may be difficult to find a suitable proxy problem, data reduction can easily be applied by using the convex relaxation of the original optimization as the proxy. This ``off-the-shelf" approach is similar to branch-and-bound, since it makes use of the convex relaxation and proceeds by imposing conditions on the feasible region. Even so, data reduction is fundamentally different from branch-and-bound. In particular, data reduction aims to reduce the feasible region of the IP by imposing individual constraints on the the way that each point in the data is labeled. In contrast, branch-and-bound algorithms aim to solve the problem to optimality, by imposing multiple constraints on all variables in the problem. In doing so, branch-and-bound algorithms may fail to effectively exploit the structure of the problem.
\clearpage
\section{Discretization Bounds and Generalization Bounds}\label{Sec::Theory}
In this section, we present new bounds on the accuracy of linear classifiers with discrete coefficients.
\subsection{Discretization Bounds}
\label{Sec::DiscretizationBounds}
Our first result shows that we can always craft a discrete set so that the training accuracy of a linear classifier with discrete coefficients $\lambdab \in \Lset$ (e.g. SLIM) is no worse than the training accuracy of a linear classifier with real-valued coefficients $\rhob \in \R^P$ (e.g. SVM).
\begin{thm}[Minimum Margin Resolution Bound]\label{Thm::MinMarginBound}
Let $\rhob = (\rho_1,\ldots,\rho_P) \in \R^P$ denote the coefficients of a linear classifier trained with any method using data $\data_N = (\xb_i,y_i)_{i=1}^N$. Let $X_{\max} = \max_i \|\xb_i\|_2$ denote the largest magnitude of any training example, and let $\gamma_{\min} = \min_i \frac{|\rhob^T\xb_i|}{\vnorm{\rhob}_2}$ denote the minimum margin achieved by any training example. 

Consider training a linear classifier with discrete coefficients $\lambdab = (\lambda_1,\ldots,\lambda_P)$ from the set:
\begin{align*}
\Lset = \left\{\lambdab \in \Z^P ~\Big|~ |\lambda_j| \leq \Lambda \; \for j = 1,\ldots,P \right\}.
\end{align*}
If we choose $\Lambda$ such that
\begin{align}
\Lambda &> \frac{X_{\max}\sqrt{P}}{2 \gamma_{\min}}, \label{Eq::MinMarginLambda}
\end{align}
then there exists a $\lambdab\in \Lset$ such that the 0--1 loss of $\lambdab$ is less than or equal to the 0--1 loss of $\rhob$: 
\begin{align}
\sum_{i=1}^N \indic{y_i\lambdab^T\xb_i\leq 0} \leq \sum_{i=1}^N \indic{y_i \rhob^T \xb_i \leq 0}. \label{Eq::MinMarginStatement}
\end{align}
\end{thm}
\proof{Proof:} See Appendix \ref{Appendix::Proofs}. \endproof

\noindent Theorem \ref{Thm::MinMarginBound} can be used to choose the resolution parameter $\Lambda$ so that the discrete set $\Lset$ is guaranteed to contain a classifier that attains the same 0--1 loss as $\rhob$. The classifiers produced from the rounding procedure often attain a lower value of the 0--1 loss than $\indic{y_i\rhob^T\xb_i\leq 0}$ because our training process optimizes the 0--1 loss directly. The rounding procedure alters many coefficients simultaneously, where each small change influences accuracy; thus, it is very easy for a rounding procedure to choose a non-optimal solution.

The following corollary produces additional discretization bounds by considering progressively larger values of the margin. These bounds can be used to relate the resolution parameter to a worst-case guarantee on training accuracy.
\begin{corollary}[$k^\text{th}$ Margin Resolution Bound]\label{Thm::MinMarginBoundCorollary}
Let $\rhob = (\rho_1,\ldots,\rho_P) \in \R^P$ denote the coefficients of a linear classifier trained with data $\data_N = (\xb_i,y_i)_{i=1}^N$. Let $(k)$ denote the training example with the $k^{\textrm{th}}$ smallest margin, so that 
$\gamma_{(k)}:=\frac{|\rhob^T\xb_{(k)}|}{\vnorm{\rhob}_2}$ are the margins in increasing order. Let $\I_{(k)}$ denote the indices of training examples $i$ with $\frac{|\rhob^T\xb_i|}{\vnorm{\rhob}_2} \leq \gamma_{(k)}$, and let $X_{(k)} = \max_{i \not\in \I_{(k)}} \|\xb_i\|_2$ denote the largest magnitude of any training example $\xb_i \in \data_N$ for $i \not\in \I_{(k)}$.

Consider training a linear classifier with discrete coefficients $\lambdab = (\lambda_1,\ldots,\lambda_P)$ from the set:
\begin{align*}
\Lset = \left\{\lambdab \in \Z^P ~\Big|~ |\lambda_j| \leq \Lambda \; \for j = 1,\ldots,P \right\}. \\ 
\intertext{If we choose $\Lambda$ such that:}
\Lambda > \frac{X_{(k)}\sqrt{P}}{2 \gamma_{(k)}}, \hspace{8em}\\
\intertext{then there exists $\lambdab \in \Lset$ such that the 0--1 loss of $\lambdab$ and the 0--1 loss of $\rhob$ differ by at most $k-1$:}
\sum_{i=1}^N \indic{y_i\lambdab^T\xb_i\leq 0}  - \sum_{i=1}^N \indic{y_i \rhob^T \xb_i \leq 0} \leq  k - 1. 
\end{align*}
\end{corollary}
\proof{Proof of Corollary \ref{Thm::MinMarginBoundCorollary}:}
The proof follows by applying Theorem \ref{Thm::MinMarginBound} to a dataset that does not contain any of the examples $i \in \I_{(k)}$, that is $\data_N \backslash \I_{(k)}$. \endproof
We have now shown that good discretized solutions exist and can be constructed easily. This motivates that optimal discretized solutions, which by definition are better than rounded solutions, will also be good relative to the best non-discretized solution.
\subsection{Generalization Bounds}\label{Sec::GeneralizationBounds}
According to the principle of structural risk minimization \citep{vapnik1998statistical}, fitting a classifier from a simpler class of models may lead to an improved guarantee on predictive accuracy. Consider training a classifier $f:\X\rightarrow\Y$ with data $\data_N = (\xb_i,y_i)_{i=1}^N$, where $\xb_i \in \X \subseteq \R^P$ and $y_i \in \Y = \{-1,1\}$. In what follows, we provide uniform generalization guarantees on the predictive accuracy of all functions, $f \in \F$. These guarantees bound the true risk, $R^\text{true}(f) = \mathbb{E}_{\X,\Y} \indic{f(\xb) \neq y},$ by the empirical risk, $R^\text{emp}(f) = \frac{1}{N}\sum_{i=1}^N \indic{f(\xb_i) \neq y_i},$ and other quantities important to the learning process.
\begin{thm}[Occam's Razor Bound for Linear Classifiers with Discrete Coefficients]\label{Thm::NormalBound}
Let $\F$ denote the set of linear classifiers with coefficients $\lambdab \in \Lset$:
\begin{align*}
\F = \left\{f :\X\to\Y \;\big|\; f(\xb)=\sign{\lambdab^T\xb} \textnormal{ and } \lambdab \in \Lset \right\}.
\end{align*}
For every $\delta > 0,$ with probability at least $1-\delta$, every classifier $f\in\F$ obeys:
\begin{align*} 
R^\textnormal{true}(f) \leq R^\textnormal{emp}(f) + \sqrt{\frac{\log(|\Lset|) - \log(\delta)}{2N}}.
\end{align*}
\end{thm}
The proof of Theorem \ref{Thm::NormalBound} uses Hoeffding's inequality for a single function, $f$, combined with the union bound over all functions $f \in \F$. The result that more restrictive hypothesis spaces can lead to better generalization provides motivation for using discrete models without necessarily expecting a loss in predictive accuracy. As the amount of data $N$ increases, the bound indicates that we can include more functions in the set $\Lset$. When a large amount of data are available, we can reduce the empirical error by using, for instance, one more significant digit for each coefficient $\lambda_j$.

One notable benefit of our framework is that we can improve the generalization bound from Theorem \ref{Thm::NormalBound} by excluding suboptimal models from the hypothesis space \textit{a priori}. When we train discrete linear classifiers using an optimization problem whose objective minimizes a loss function and regularizes the $\lzero$-norm, for instance, we can bound the number of features in the classifier using the value of the $\lzero$-penalty, $C_0$. In Theorem \ref{Thm::L0Bound}, we use this principle to tighten the generalization bound from Theorem \ref{Thm::NormalBound} for a case where we train linear classifiers with an optimization problem that minimizes the 0--1 loss function, regularizes the $\lzero$-norm, and restricts coefficients to a set of bounded integers (e.g. SLIM, TILM, and M-of-N rule tables).
\begin{thm}[Generalization of Linear Classifiers with $\lzero$-Regularization]\label{Thm::L0Bound}
\hspace{5em}
Let $\F$ denote the set of linear classifiers with coefficients $\lambdab \in \Lset$ such that 
\begin{align*}
\F = \Big\{f :\X\to\Y \;\big|\; f(\xb)=\sign{\lambdab^T\xb} \textnormal{ and } 
\lambdab \in \argmin_{\lambda \in \Lset} 
Z(\lambdab;\data_N) \textnormal{ for some }
\{\X\times\Y\}^N \Big\} 
\end{align*}
where
\begin{align*}
Z(\lambdab;\data_N) = \frac{1}{N} \sum_{i=1}^N \indic{y_i \lambdab^T \xb_i \leq 0} + C_0\vnorm{\lambdab}_0.
\end{align*}
For every $\delta > 0,$ with probability at least $1-\delta$, every classifier $f\in\F$ obeys:
\begin{align*} 
R^{\textnormal{true}}(f) &\leq R^{\textnormal{emp}}(f) + \sqrt{\frac{\log(|\mathcal{H}_{P,C_0}|) - \log(\delta)}{2N}}.
\end{align*}
where
\begin{align*} 
\mathcal{H}_{P,C_0} &= \Bigg\{\lambdab \in \Lset \;\big|\; \vnorm{\lambdab}_0 \leq \left\lfloor \frac{1}{C_0} \right\rfloor \Bigg\}.
\end{align*}
\end{thm}
\proof{Proof:} See Appendix \ref{Appendix::Proofs}. \endproof

\noindent This theorem states that the class of minimizers of $Z(\lambdab;\data_N)$ for any dataset $\data_N$ is bounded in size through regularization coefficient $C_0$. Namely, each minimizer obeys $\|\lambdab\|_0\leq \left\lfloor \frac{1}{C_0} \right\rfloor $. This translates into a better risk bound.

In Theorem \ref{Thm::CoprimeBound}, we present a generalization guarantee that uses this idea to relate discrete classification and number theory. The theorem applies to cases where we optimize a scale-invariant objective function over a set of bounded integer vectors, and use a small $\lone$-penalty to restrict coefficients to coprime integers (e.g. SLIM). Here, we can refine the hypothesis space to only include $P$-dimensional coprime integer vectors bounded by $\Lambda$, and express the generalization bound from  Theorem \ref{Thm::NormalBound} in terms of the $P$-dimensional Farey points of level $\Lambda$ \citep[][see e.g.]{2012arXiv1207.0954M}.
\begin{thm}[Generalization of Discrete Linear Classifiers with Coprime Coefficients]\label{Thm::CoprimeBound}
Let $\F$ denote the set of linear classifiers with coprime integer coefficients, $\lambdab$, bounded by $\Lambda$:
\begin{align*}
\F &= \left\{f :\X\to\Y \;\big|\; f(\xb)=\sign{\lambdab^T\xb} \textnormal{ and } \lambdab \in \Lset \right\},\\
\Lset &= \{\lambdab \in \mathbb{\hat{Z}}^P : |\lambda_j| \leq \Lambda \textnormal{ for } j=1,\ldots,P\},\\
\hat{\Z}^P &= \left\{\bm{z}\in\Z^P:\textnormal{gcd}(\bm{z}) = 1\right\}.
\end{align*}
For every $\delta > 0,$ with probability at least $1-\delta$, every classifier $f\in\F$ obeys:
\begin{align*} 
R^{\textnormal{true}}(f) &\leq R^{\textnormal{emp}}(f) + \sqrt{\frac{\log(|\mathcal{C}_{P,\Lambda}|) - \log(\delta)}{2N}}, \\
\intertext{where $\mathcal{C}_{P,\Lambda}$ denotes the set of Farey points of level $\Lambda$:}
\mathcal{C}_{P,\Lambda} &= \left\{ \frac{\lambdab}{q} \in [0,1)^P: (\lambdab,q) \in \mathbb{\hat{Z}}^{P+1} \text{ and } 1 \leq q \leq \Lambda \right\}.
\end{align*}
\end{thm}
%
%
The proof involves a counting argument over coprime integer vectors, using the definition of Farey points from number theory.

\begin{figure}[htbp]
\centering
\includegraphics[width=0.425\textwidth,keepaspectratio=true,trim=0mm 0mm 0mm 14mm,clip=true]{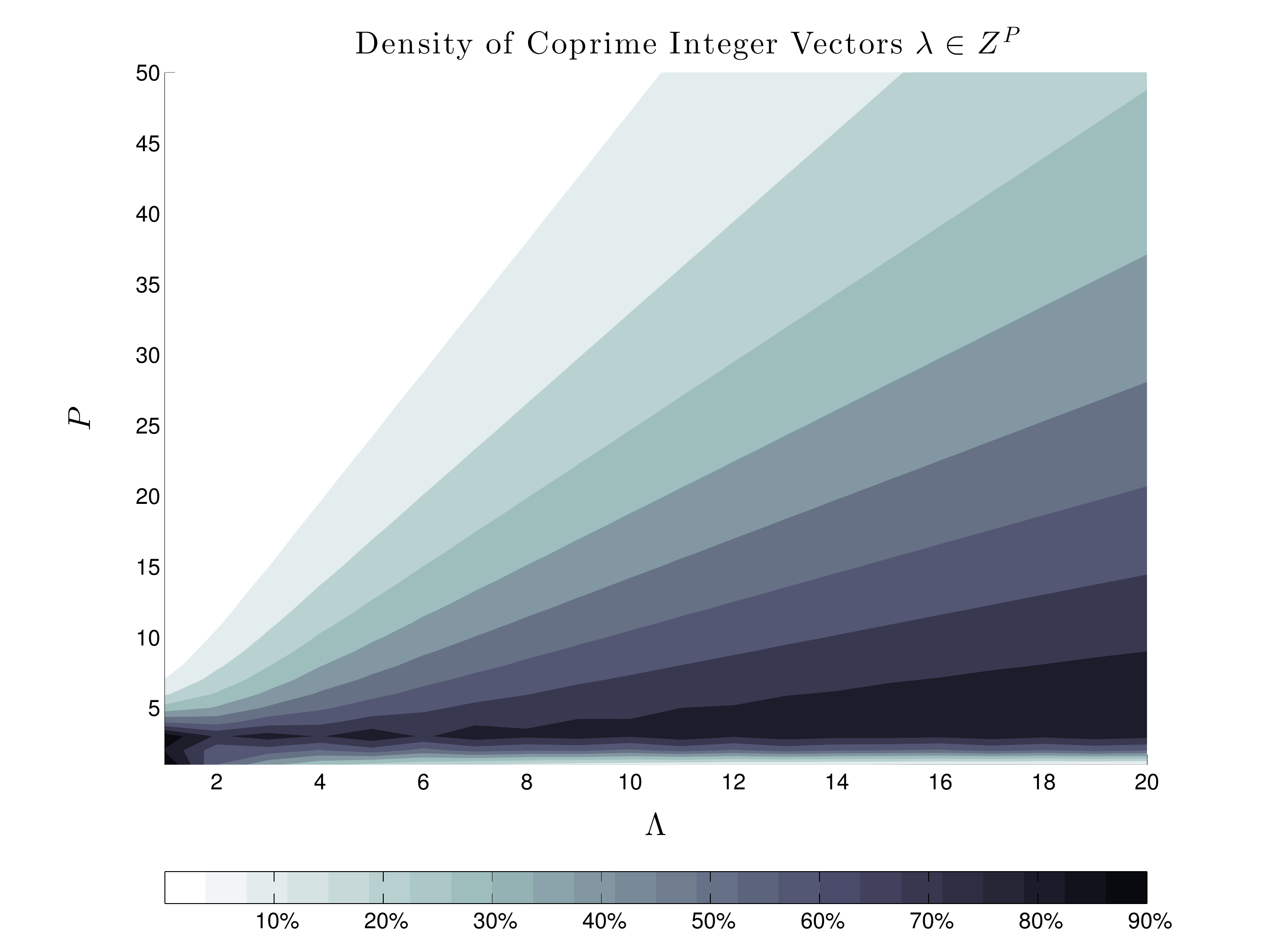}
\includegraphics[width=0.425\textwidth,keepaspectratio=true,trim=0mm 0mm 0mm 20mm,clip=true]{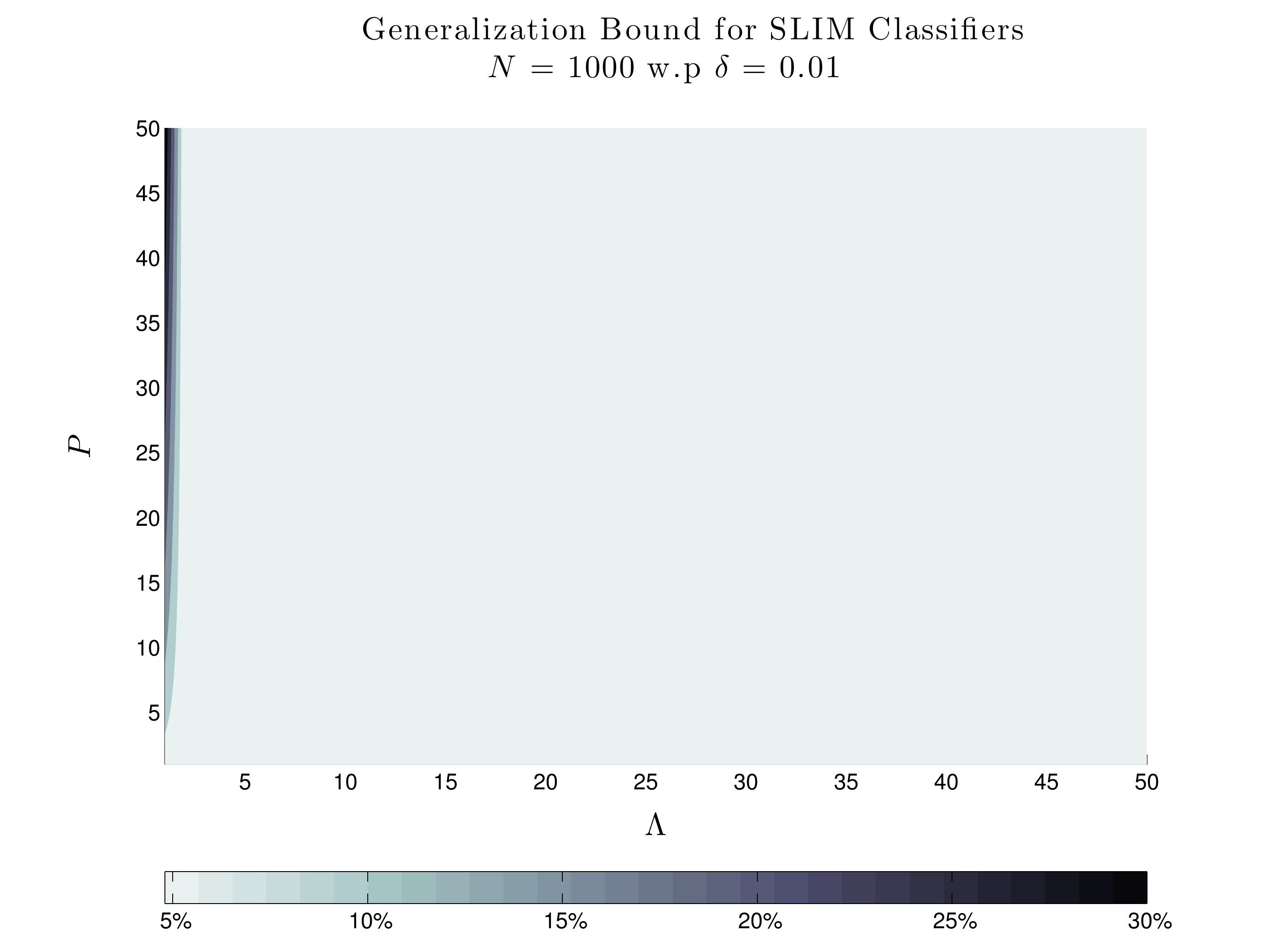}
\caption{Relative density of coprime integer vectors in $\Z^P$ (left), and the relative improvement in the generalization bound due to the use of coprime coefficients for $\delta=0.01$ (right).}
\label{Fig::CoprimeFigure}
\end{figure}

In Figure \ref{Fig::CoprimeFigure}, we plot the relative density of coprime integer vectors in $\Z^P$ bounded by $\Lambda$ (i.e. $|\mathcal{C}_{P,\Lambda}|/(2\Lambda+1)^P$) 
and the relative improvement in the generalization bound due to the use of coprime coefficients. This shows that using coprime coefficients can significantly reduce the number of classifiers based on the dimensionality of the data and the value of $\Lambda$. The corresponding improvement in the generalization bound is usually small, but may be large when the data are high dimensional and $\Lambda$ is small (e.g. M-of-N rules and SLIM with very small coefficients).
\FloatBarrier
\clearpage
\section{Application to Sleep Apnea Diagnosis}\label{Sec::Demonstrations}
In this section, we present a case study where we used our framework to build a clinical tool for sleep apnea diagnosis. Our goal is to demonstrate the flexibility and performance our framework in comparison to existing methods on a real-world problem that requires a tailored prediction model.
\subsection{Data Overview and Model Requirements}
The data for this study was provided to us as part of an ongoing collaboration with physicians at the Massachusetts General Hospital Sleep Laboratory. It contains $N = 1922$ records of patients and $P = 112$ binary features related to their health and sleep habits. The classification task is to identify a patient with some form of sleep apnea, where $y_i=+1$ if patient $i$ has obstructive sleep apnea or upper airway resistance syndrome. There is significant class imbalance, with Pr$(y_i=+1) = 76.9\%$. 

To ensure that we would produce a model that would be used and accepted in a clinical context, our collaborators also provided us with the following list of \textit{model requirements}:
\begin{enumerate}[leftmargin=0.45cm,topsep=0pt,parsep=0pt]

\item \textit{Limited FPR:} The model had to achieve the highest possible true positive rate (TPR) while maintaining a maximum false positive rate (FPR) of 20\%. This would ensure that the model could diagnose as many cases of sleep apnea as possible but limit the number of faulty diagnoses. 

\item \textit{Limited Model Size:} The model had to be transparent and use at most 10 features. This would ensure that the model was simple enough to be explained and understood by other physicians in a short period of time.

\item \textit{Sign Constraints:} The model had to obey established relationships between well-known risk factors and the incidence of sleep apnea (i.e. it could not suggest that a patient with hypertension had a lower risk of sleep apnea since hypertension is a well-known risk factor for sleep apnea).

\end{enumerate}
\subsection{Training Setup and Model Selection}
We built an appropriate model by training a SLIM scoring system with integer coefficients between -20 and 20. Our framework let us easily address all three model requirements using  \textbf{one} instance of the free parameters, as follows:
\begin{enumerate}[leftmargin=0.45cm,topsep=2pt,parsep=2pt]

\item We added a hard constraint on the 0--1 loss to limit the maximum FPR at 20\%. We then set $\wplus = \nminus/(1+\nminus)$ to guarantee that the optimization process would yield a classifier with the highest possible TPR with a maximum FPR less than 20\% (see Section \ref{Sec::HandlingImbalancedData}).

\item We added a hard constraint on the $\lzero$-norm to limit the maximum number of features to 10 (see Section \ref{Sec::FeatureBasedConstraints}). We then set $C_0 = 0.9\wplus/NP$ to guarantee that the the optimization process would yield a classifier that did not sacrifice accuracy for sparsity (see Section \ref{Sec::Framework}). 

\item We added sign constraints to ensure that our classifier would not violate established relationships between features and outcomes (see Section \ref{Sec::MonotonicityConstraints}).

\end{enumerate}
With this setup, we trained 10 models with subsets of the data to assess predictive accuracy through 10-fold cross validation (CV), and 1 final model with all of data to hand over to our collaborators. We solved each IP for 3 hours, in parallel, on 12-core 2.7GHZ machine with 48GB RAM. This required 3 hours of total computing time.

As a comparison, we trained models using 8 state-of-the-art classification methods summarized in Table \ref{Table::SleepApneaSetup}. We dealt with the class imbalance by using a weighted loss function where we varied the sensitivity parameter, $\wplus$, across its full range (see Section \ref{Sec::HandlingImbalancedData}) and sought to address the remaining requirements by extensively exploring different settings and free parameters.

Model selection was inherently difficult for the baseline methods as their performance varied jointly over $\wplus$ and other free parameters. To choose the best model that satisfied all of the model requirements without mixing training and testing data, we proceeded by: (i) dropping any instance of the free parameters where any model requirements were violated for at least one of the 10 folds; (ii) choosing the model that maximized the mean 10-fold CV test TPR among remaining instances. 
%
%
%
%

\begin{table}[htbp]
\scriptsize
\centering
\begin{tabular}{lccl}
\toprule 

\textbf{Method} & \bfcell{c}{Controls} & \bfcell{c}{\# Instances} & \bfcell{l}{Free Parameter Grid} \\ 

\midrule

CART & \cell{c}{Max FPR\\Model Size} & 39 &  

\cell{l}{
39 values of $\wplus \in (0.025,0.05,\ldots,0.975)$
} \\

\midrule

C5.0T & \cell{c}{Max FPR} & 39 &

\cell{l}{
39 values of $\wplus \in (0.025,0.05,\ldots,0.975)$
} \\ 

\midrule

C5.0R & \cell{c}{Max FPR\\Model Size} & 39 &

\cell{l}{
39 values of $\wplus \in (0.025,0.05,\ldots,0.975)$
} \\ 

\midrule

Lasso & \cell{c}{Max FPR\\Model Size\\Sign} & 39000 & 

\cell{l}{
39 values of $\wplus \in (0.025,0.05,\ldots,0.975)$ \\
$\times$ 1000 values of $\lambda$ chosen by \pkg{glmnet}
} \\ 

\midrule

Ridge & \cell{c}{Max FPR\\Model Size\\Sign} & 39000 & 

\cell{l}{
39 values of $\wplus \in (0.025,0.05,\ldots,0.975)$ \\
$\times$ 1000 values of $\lambda$ chosen by \pkg{glmnet}
} \\ 

\midrule

E.Net & \cell{c}{Max FPR\\Model Size\\Sign}  & 975000 &

\cell{l}{
39 values of $\wplus \in (0.025,0.05,\ldots,0.975)$ \\
$\times$ 1000 values of $\lambda$ chosen by \pkg{glmnet}\\
$\times$ 19 values of $\alpha \in (0.05,0.10,\ldots,0.95)$
} \\ 

\midrule

SVM Lin.  & \cell{c}{Max FPR} &  975 & 

\cell{l}{
39 values of $\wplus \in (0.025,0.05,\ldots,0.975)$ \\
$\times$ 25 values of $C \in \{10^t | t = (-3,-2.75\ldots,2.75,3)\}$
} \\ 

\midrule

SVM RBF & \cell{c}{Max FPR} & 975 & 

\cell{l}{
39 values of $\wplus \in (0.025,0.05,\ldots,0.975)$ \\
$\times$ 25 values of $C \in \{10^t | t = (-3,-2.75\ldots,2.75,3)\}$
} \\ 

\midrule

SLIM & \cell{c}{Max FPR\\Model Size\\Sign} & 1 &

\cell{l}{
$\wplus = \nminus/(1+\nminus)$,  $C_0 = 0.9/NP$, \\
$\lambda_0 \in \{-20,\ldots,20\}$, $\lambda_j \in \{-20,\ldots,20\}$} \\ 

\bottomrule 
\end{tabular}
\caption{Overview of the training setup for all methods. Each instance represents a unique combination of free parameters. We list model requirements that each method should be able to handle under the controls column.}
\label{Table::SleepApneaSetup}
\end{table}
\subsection{Discussion of Results}
In what follows, we provide a separate discussions of the flexibility, performance and interpretability of models produced by all 9 classification methods. We provide a summary of the performance of all methods in Table \ref{Table::DemoTableMaxFPR}, and a summary of their flexibility in Table \ref{Table::DemoFlexibilityTable}.
\begin{table}[ht]
\centering
{\scriptsize
\begin{tabular}{lccccccccc}
  \toprule
 & & \multicolumn{2}{c}{\textbf{REQUIREMENTS}} & \multicolumn{1}{c}{\textbf{OBJECTIVE}} & \multicolumn{5}{c}{\textbf{OTHER INFORMATION}} \\ \cmidrule(lr){3-4}\cmidrule(lr){5-5}\cmidrule(lr){6-10}
\bfcell{c}{Method} & \bfcell{c}{Requirements\\Satisfied} & \bfcell{c}{Train\\FPR} & \bfcell{c}{Model\\Size} & \bfcell{c}{Test\\TPR} & \bfcell{c}{Test\\FPR} & \bfcell{c}{Train\\TPR} & \bfcell{c}{Final\\Train\\TPR} & \bfcell{c}{Final\\Train\\FPR} & \bfcell{c}{Final\\Model\\Size} \\ 
  
\toprule

SLIM & \scriptsize{\cell{c}{Max FPR\\Model Size\\Sign}} & \cell{c}{\scriptsize{19.9$\%$}\\\tiny{19.8 - 20.0$\%$}} & \cell{c}{\scriptsize{10}\\\tiny{10 - 10}} & \cell{c}{\scriptsize{61.7$\%$}\\\tiny{56.6 - 66.0$\%$}} & \cell{c}{\scriptsize{25.1$\%$}\\\tiny{12.5 - 34.8$\%$}} & \cell{c}{\scriptsize{63.6$\%$}\\\tiny{57.5 - 69.8$\%$}} & \cell{c}{\scriptsize{66.6$\%$}\\\tiny{-}} & \cell{c}{\scriptsize{19.8$\%$}\\\tiny{-}} & \cell{c}{\scriptsize{10}\\\tiny{-}} \\ 
   
\midrule

Lasso & \scriptsize{\cell{c}{Max FPR\\Model Size\\Sign}} & \cell{c}{\scriptsize{10.8$\%$}\\\tiny{9.5 - 14.4$\%$}} & \cell{c}{\scriptsize{9}\\\tiny{9 - 10}} & \cell{c}{\scriptsize{47.5$\%$}\\\tiny{38.8 - 60.7$\%$}} & \cell{c}{\scriptsize{11.7$\%$}\\\tiny{5.0 - 20.8$\%$}} & \cell{c}{\scriptsize{48.0$\%$}\\\tiny{46.0 - 53.8$\%$}} & \cell{c}{\scriptsize{47.1$\%$}\\\tiny{-}} & \cell{c}{\scriptsize{10.1$\%$}\\\tiny{-}} & \cell{c}{\scriptsize{10}\\\tiny{-}} \\ 
   
\midrule

E. Net & \scriptsize{\cell{c}{Max FPR\\Model Size}} & \cell{c}{\scriptsize{15.1$\%$}\\\tiny{10.6 - 19.5$\%$}} & \cell{c}{\scriptsize{9}\\\tiny{9 - 9}} & \cell{c}{\scriptsize{50.6$\%$}\\\tiny{42.8 - 62.8$\%$}} & \cell{c}{\scriptsize{16.2$\%$}\\\tiny{8.2 - 24.1$\%$}} & \cell{c}{\scriptsize{51.7$\%$}\\\tiny{44.6 - 56.0$\%$}} & \cell{c}{\scriptsize{53.5$\%$}\\\tiny{-}} & \cell{c}{\scriptsize{16.2$\%$}\\\tiny{-}} & \cell{c}{\scriptsize{9}\\\tiny{-}} \\ 
   
\midrule

Ridge & \scriptsize{\cell{c}{Max FPR}} & \cell{c}{\scriptsize{19.2$\%$}\\\tiny{18.6 - 20.0$\%$}} & \cell{c}{\scriptsize{110}\\\tiny{110 - 110}} & \cell{c}{\scriptsize{68.7$\%$}\\\tiny{63.2 - 74.1$\%$}} & \cell{c}{\scriptsize{21.1$\%$}\\\tiny{12.5 - 31.2$\%$}} & \cell{c}{\scriptsize{69.3$\%$}\\\tiny{67.3 - 71.2$\%$}} & \cell{c}{\scriptsize{69.4$\%$}\\\tiny{-}} & \cell{c}{\scriptsize{19.1$\%$}\\\tiny{-}} & \cell{c}{\scriptsize{110}\\\tiny{-}} \\ 
   
\midrule

SVM Lin. & \scriptsize{\cell{c}{Max FPR}} & \cell{c}{\scriptsize{18.5$\%$}\\\tiny{17.3 - 19.6$\%$}} & \cell{c}{\scriptsize{111}\\\tiny{111 - 111}} & \cell{c}{\scriptsize{70.4$\%$}\\\tiny{65.8 - 76.6$\%$}} & \cell{c}{\scriptsize{27.6$\%$}\\\tiny{12.5 - 41.7$\%$}} & \cell{c}{\scriptsize{73.3$\%$}\\\tiny{72.5 - 74.2$\%$}} & \cell{c}{\scriptsize{73.3$\%$}\\\tiny{-}} & \cell{c}{\scriptsize{18.5$\%$}\\\tiny{-}} & \cell{c}{\scriptsize{111}\\\tiny{-}} \\ 
   
\midrule

SVM RBF & \scriptsize{\cell{c}{Max FPR}} & \cell{c}{\scriptsize{16.1$\%$}\\\tiny{14.6 - 19.2$\%$}} & \cell{c}{\scriptsize{NA}\\\tiny{NA - NA}} & \cell{c}{\scriptsize{88.4$\%$}\\\tiny{84.9 - 91.6$\%$}} & \cell{c}{\scriptsize{57.7$\%$}\\\tiny{50.0 - 70.8$\%$}} & \cell{c}{\scriptsize{99.3$\%$}\\\tiny{98.9 - 99.5$\%$}} & \cell{c}{\scriptsize{99.1$\%$}\\\tiny{-}} & \cell{c}{\scriptsize{16.9$\%$}\\\tiny{-}} & \cell{c}{\scriptsize{NA}\\\tiny{-}} \\ 
   
\midrule

C5.0R & \scriptsize{\cell{c}{-}} & \cell{c}{\scriptsize{28.1$\%$}\\\tiny{17.3 - 36.4$\%$}} & \cell{c}{\scriptsize{27}\\\tiny{17 - 43}} & \cell{c}{\scriptsize{81.4$\%$}\\\tiny{73.7 - 89.9$\%$}} & \cell{c}{\scriptsize{43.9$\%$}\\\tiny{25.0 - 57.1$\%$}} & \cell{c}{\scriptsize{86.1$\%$}\\\tiny{80.9 - 89.9$\%$}} & \cell{c}{\scriptsize{84.5$\%$}\\\tiny{-}} & \cell{c}{\scriptsize{31.5$\%$}\\\tiny{-}} & \cell{c}{\scriptsize{20}\\\tiny{-}} \\ 
   
\midrule

C5.0T & \scriptsize{\cell{c}{-}} & \cell{c}{\scriptsize{21.5$\%$}\\\tiny{13.6 - 32.4$\%$}} & \cell{c}{\scriptsize{78}\\\tiny{48 - 98}} & \cell{c}{\scriptsize{78.0$\%$}\\\tiny{73.1 - 84.8$\%$}} & \cell{c}{\scriptsize{47.8$\%$}\\\tiny{36.7 - 57.1$\%$}} & \cell{c}{\scriptsize{87.6$\%$}\\\tiny{80.7 - 92.7$\%$}} & \cell{c}{\scriptsize{85.3$\%$}\\\tiny{-}} & \cell{c}{\scriptsize{29.5$\%$}\\\tiny{-}} & \cell{c}{\scriptsize{41}\\\tiny{-}} \\ 
   
\midrule

CART & \scriptsize{\cell{c}{-}} & \cell{c}{\scriptsize{52.3$\%$}\\\tiny{41.8 - 63.6$\%$}} & \cell{c}{\scriptsize{12}\\\tiny{9 - 16}} & \cell{c}{\scriptsize{88.6$\%$}\\\tiny{83.2 - 94.0$\%$}} & \cell{c}{\scriptsize{57.7$\%$}\\\tiny{42.5 - 71.4$\%$}} & \cell{c}{\scriptsize{91.4$\%$}\\\tiny{89.4 - 95.4$\%$}} & \cell{c}{\scriptsize{96.5$\%$}\\\tiny{-}} & \cell{c}{\scriptsize{76.8$\%$}\\\tiny{-}} & \cell{c}{\scriptsize{8}\\\tiny{-}} \\ 
   
\bottomrule
\end{tabular}
}
\caption{TPR, FPR and model size for all methods. We report the 10-fold CV mean for TPR and FPR and the 10-fold CV median for the model size; the ranges in each cell represent the 10-fold CV minimum and maximum.}
\label{Table::DemoTableMaxFPR}
\end{table}

\FloatBarrier
\vspace{0.20cm}
\noindent\textbf{On the Flexibility of Baseline Methods}
\vspace{0.10cm}

\begin{figure}[htbp]
\centering
\includegraphics[width=0.9\fwidth]{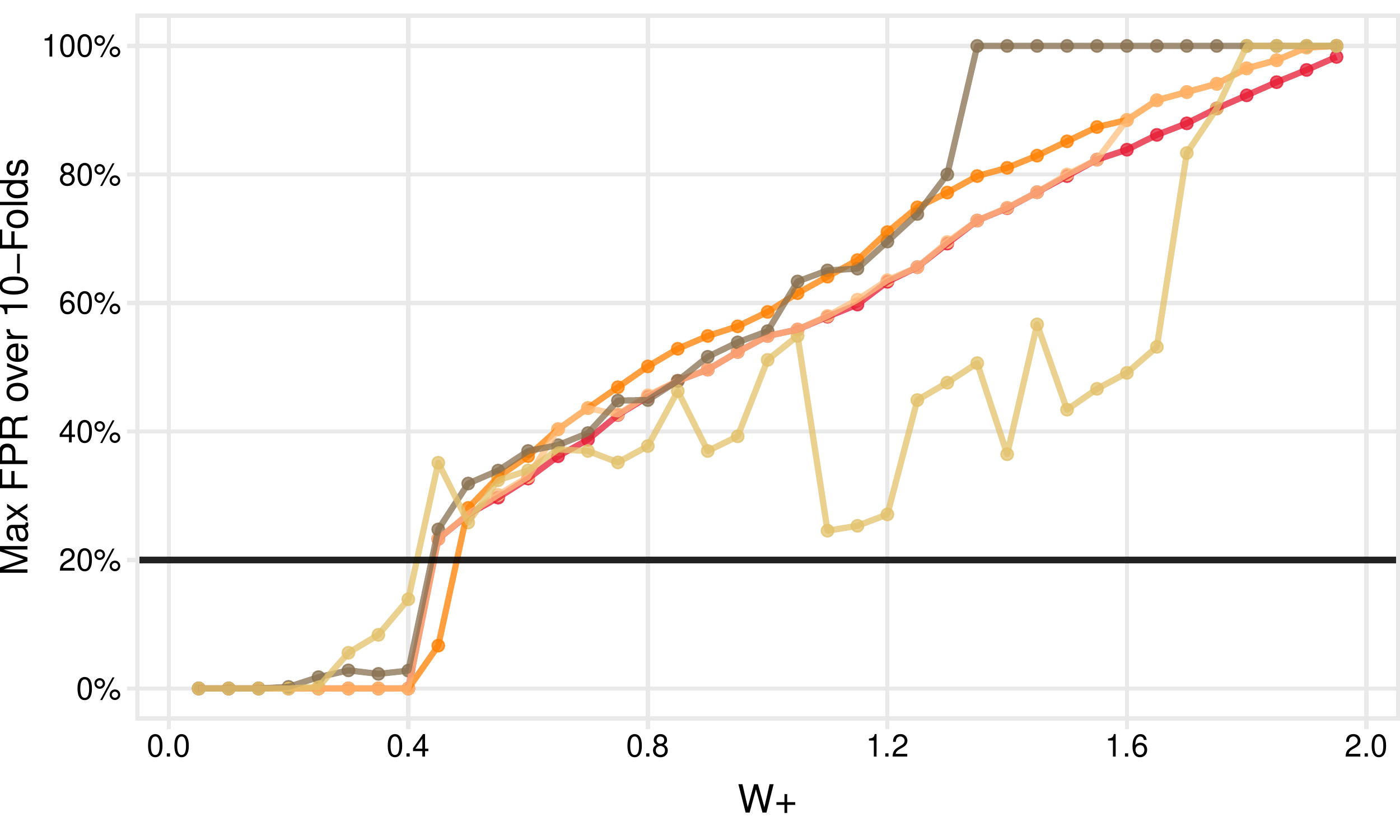} 
\includegraphics[trim=1in 0in 0in 0in,clip,width=0.81\fwidth,]{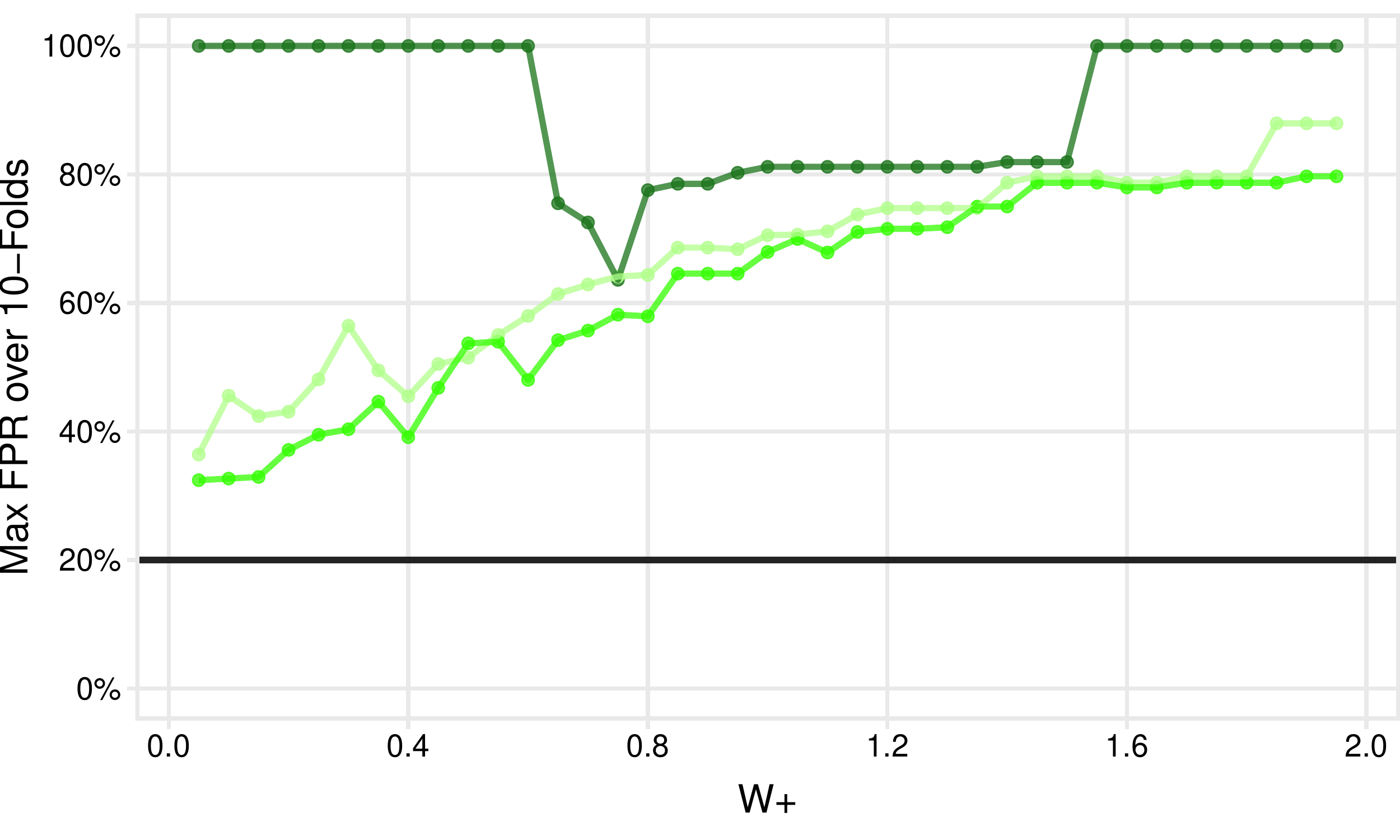} 
\includegraphics[trim=9in 0in 0in 0in,clip,width=0.1\fwidth,]{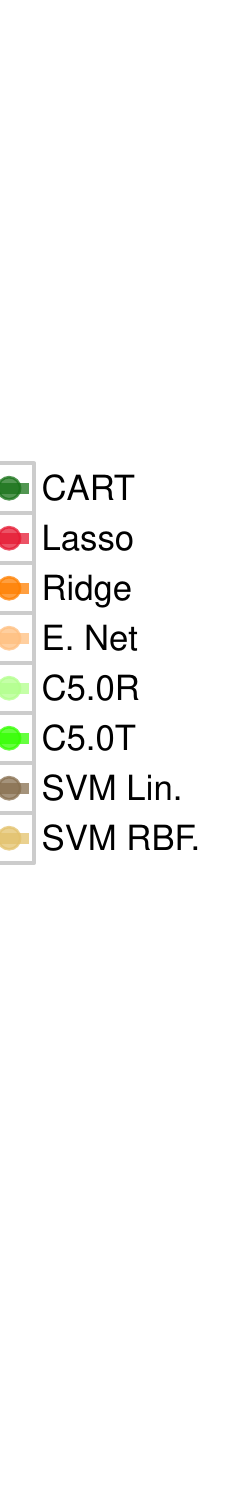}
\caption{10-fold CV max FPR for classifiers produced using the baseline methods across the full range of $\wplus$. For each method and value of $\wplus$, we plot the classifier that has the smallest 10-fold CV max FPR. The figure on the right highlights methods that cannot produce a model with a FPR less than 20\% for any value of $\wplus$. It shows that CART, C5.0, C5.0T cannot be tuned to satisfy the max FPR $\leq 20 \%$ requirement.}
\label{Fig::WRangeGraph}
\end{figure}

\noindent Among the 9 classification methods that we used in this study, only SLIM and Lasso were able to satisfy the three model requirements given to us by physicians. Tree and rule-based methods such as CART, C5.0 Tree and C5.0 Rule were unable to produce a model with a maximum FPR of 20\% (see Figure \ref{Fig::WRangeGraph}). Methods that used $\ltwo$-regularization such as SVM Lin., SVM RBF and Ridge were unable to achieve the required level of sparsity. E. Net, which uses both $\lone$- and $\ltwo$-regularization, was able to achieve the required level of sensitivity and sparsity, but unable to maintain them after the inclusion of sign constraints (possibly due to a numerical issue in the \pkg{glmnet} package).

We did not expect all methods to satisfy all of the model requirements as all methods were not designed to produce tailored models. To be clear, we expected: all methods to be able to satisfy the max FPR requirement; methods with sparsity controls to fulfill the max FPR and model size requirements (Lasso, E.Net, CART, C5.0T and C5.0R); and methods with sparsity and monotonicity controls to satisfy the max FPR, model size and sign constraint requirements (Lasso and E.Net). Even so, we included all methods in our comparison to highlight the following important but often overlooked points. 

State-of-the-art methods for predictive modeling do not:
\begin{itemize}[leftmargin=0.45cm,topsep=2pt,parsep=2pt]

\item Accommodate reasonable constraints that are crucial models to be used and accepted. There is simply no mechanism in most implementations to adjust important model qualities. That is, there is no mechanism to control sparsity in C5.0T, and no mechanism to incorporate sign constraints in SVM. Incorporating reasonable constraints is a difficult process, and results in a poor trade-off with accuracy when it is possible.

\item Have controls that work correctly. Even when a method can accommodate reasonable constraints by allowing us to set parameters, these controls are indirect and do not always allow us to incorporate multiple constraints simultaneously. Finding a feasible model requires a tuning process that involves grid search over a large free parameter grid. Even after extensive tuning, however, it is possible to never find a model that can satisfy the model constraints (e.g. CART, C5.0R, C5.0T for the max FPR requirement as shown in Figure \ref{Fig::WRangeGraph}). 

\item  Allow tuning to be portable when the training set changes. Consider a standard approach for model selection where we choose free parameters so as to maximize predictive accuracy. In this case, we would train models on several folds for each free instance of the parameters, choose an instance of the free parameters that obeys all model requirements while maximizing predictive accuracy, and then train a final model using this instance. Unfortunately, there is no guarantee that the final model we produce using the selected parameters will obey all model requirements. In contrast, the models from our framework have such a guarantee as we can encode hard constraints in the interpretability set.

\end{itemize}
\begin{table}[htbp]
\centering
{\scriptsize
\begin{tabular}{lccc}
  \toprule
\bfcell{c}{Method} & \bfcell{c}{\% Instances w.\\Acceptable FPR} & \bfcell{c}{\% Instances w.\\Acceptable FPR\\\& Model Size} & \bfcell{c}{\% Instances w.\\Acceptable FPR\\Model Size \& Signs} \\ 
     \toprule SLIM & 100.0\% & 100.0\% & 100.0\% \\ 
   \midrule Lasso & 21.0\% & 12.8\% & 12.8\% \\ 
   \midrule E. Net & 18.3\% & 1.8\% & 0.0\% \\ 
   \midrule Ridge & 29.3\% & 0.0\% & 0.0\% \\ 
   \midrule SVM Lin. & 19.6\% & 0.0\% & 0.0\% \\ 
   \midrule SVM RBF. & 36.6\% & 0.0\% & 0.0\% \\ 
   \midrule C5.0R  & 0.0\% & 0.0\% & 0.0\% \\ 
   \midrule C5.0T & 0.0\% & 0.0\% & 0.0\% \\ 
   \midrule CART & 0.0\% & 0.0\% & 0.0\% \\ 
   \bottomrule \end{tabular}
}
\caption{Summary of the proportion of instances that fulfilled each model requirement. Each instance represents a unique combination of free parameters for a given method.}
\label{Table::DemoFlexibilityTable}
\end{table}
\FloatBarrier
\vspace{0.20cm}
\noindent\textbf{On the Sensitivity of Acceptable Models}
\vspace{0.10cm}

\begin{figure}
\centering
\includegraphics[width=0.5\textwidth,trim=0mm 4mm 0mm 0mm,clip]{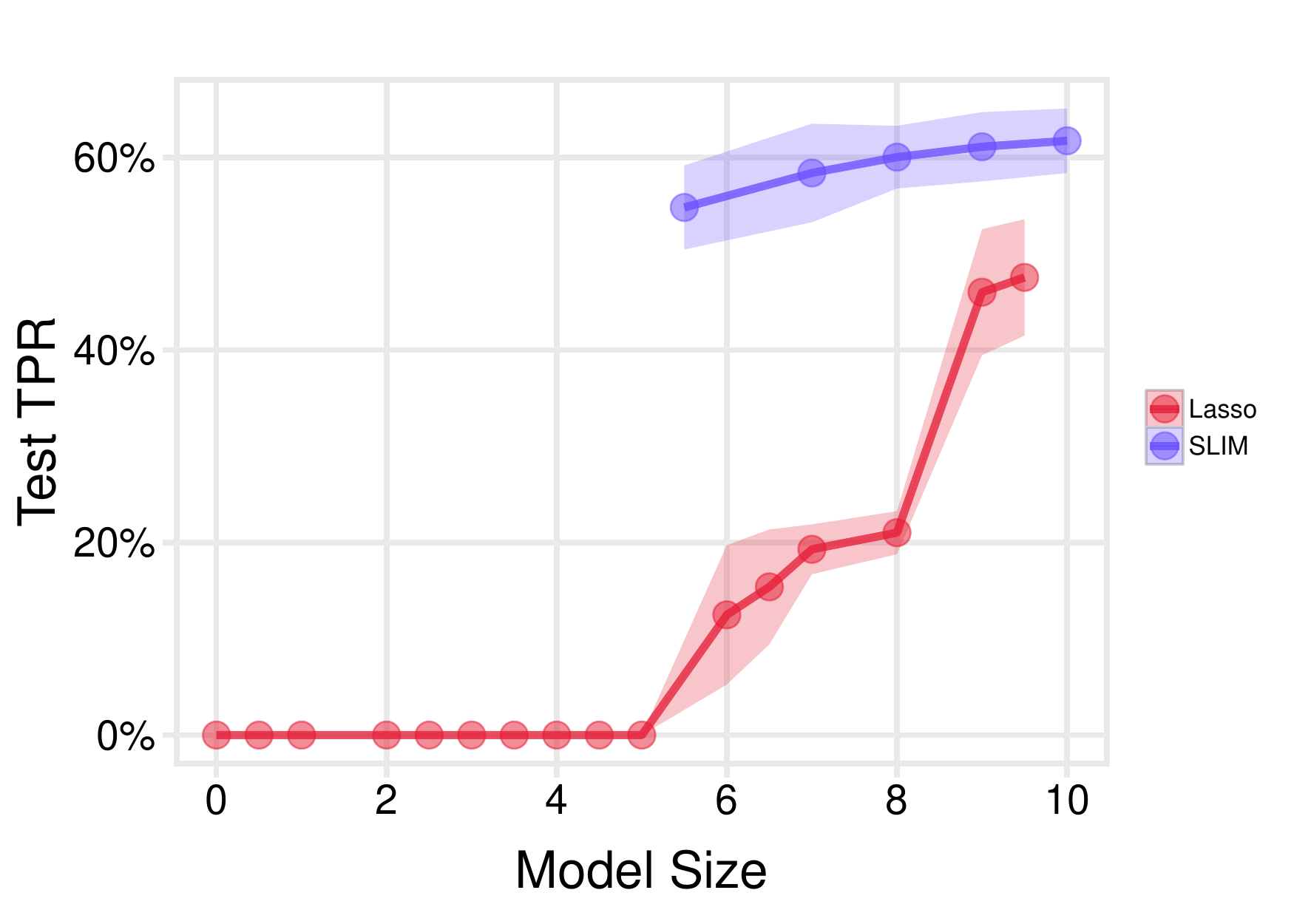}
\caption{Sensitivity of models satisfied all three requirements. Models produced by SLIM dominate those produced by Lasso across the full regularization path.}
\label{Figure::SleepApneaRegPath}
\end{figure}

\noindent Among the two methods that produced acceptable models, the SLIM model had significantly higher sensitivity than the Lasso model -- a result that we expected given that SLIM minimizes the 0--1 loss and an $\lzero$-penalty while Lasso minimizes convex surrogates of these quantities. To show that this result held true across the entire regularization path of the Lasso model, we have plotted the sensitivity and sparsity of SLIM models trained for $C_0 = (0.01,0.08,0.07,0.05)$ in Figure \ref{Figure::SleepApneaRegPath}.

The benefits of avoiding approximations are also clear when, for instance, we compare the performance of the SLIM model and the Ridge model in Table \ref{Table::DemoTableMaxFPR}. Here, both the SLIM model and the Ridge model attain similar levels of sensitivity even as SLIM is fitting linear models from a far smaller hypothesis space (i.e. linear classifiers with 10 features and integer coefficients vs. linear classifiers with 112 features and real coefficients).

\vspace{0.20cm}
\noindent\textbf{On the Interpretability of Final Models}
\vspace{0.10cm}

\noindent We present the score functions of the final SLIM and Lasso models in Figure \ref{Fig::SleepApneaModels} (we include the best models from methods that were unable to fulfill all three model requirements in Appendix \ref{Appendix::SleepApneaModels}). Both SLIM and Lasso produced models that aligned with the domain knowledge of our collaborators: they complied with sign constraints and also included large coefficients for well-known risk factors such as $bmi$, $sex$, age and $hypertension$. Our collaborators commented on the interpretability benefits of using integer coefficients, as it made it easier for them to understand the SLIM model through two distinct mechanisms: first, by clearly exposing relationships between the features and the outcome; second, by making it easier for others to validate our tool by making predictions for hypothetical examples without a calculator.
\begin{figure}[htbp]
\centering
\begin{tabularx}{5in}{lX}
SLIM & \begin{tabular}{p{1mm}lp{1mm}lp{1mm}l}
   & $18 ~age\geq 60$ & $\scriptsize{+}$ & $10 ~bmi\geq25$ & $\scriptsize{+}$ & $10 ~bmi \geq 40$ \\ 
   $\scriptsize{+}$ & $8 ~snoring$ & $\scriptsize{+}$ & $6 ~hypertension$ & $\scriptsize{+}$ & $4 ~bronchitis$ \\ 
   $\scriptsize{+}$ & $2 ~cataplexy$ & $\scriptsize{+}$ & $2 ~dozes\_off\_watching\_TV$ & $\scriptsize{+}$ & $2 ~retired$ \\
   $\scriptsize{-}$ & $8 ~female$ & $\scriptsize{-}$ & $19$ &  & \\ 
  \end{tabular} \\ 
\vspace{0.5em} & \\ 
Lasso & \begin{tabular}{p{1mm}lp{1mm}lp{1mm}l}
  & $0.52 ~male$ & $\scriptsize{+}$ & $0.37 ~hypertension$ & $\scriptsize{+}$ & $0.30 ~age \geq 60$ \\
  $\scriptsize{+}$ & $0.22 ~snoring$ & $\scriptsize{+}$ & $0.20 ~snoring\_2$ & $\scriptsize{+}$ & $0.15 ~bmi \geq 30$ \\ 
  $\scriptsize{+}$ & $0.01 ~stopbreathing$ & $\scriptsize{+}$ & $8.49\times10^{-14} ~bmi \geq 25$ & $\scriptsize{-}$ & $0.19 ~bmi < 25$ \\ 
  $\scriptsize{-}$ & $8.07 \times 10^{-14}~female$ & $\scriptsize{-}$ & $1.05$ &  &  \\ 
\end{tabular}
\end{tabularx}
\caption{Score functions of the final SLIM and Lasso models, which satisfied all of the model requirements. The Lasso coefficients for $female$ and $bmi\geq25$ are effectively 0. To eliminate them, we would need to set a threshold for the coefficients in an arbitrary way. Note that $snoring = 1$ if a patient was referred to the sleep clinic due to snoring, and $snoring\_2 = 1$ if a patient believes that snoring is the symptom of obstructive sleep apnea; $snoring$ and $snoring_2$ are highly correlated, but not identical.}
\label{Fig::SleepApneaModels}
\end{figure}

\begin{table}[htbp]
\centering
\small{\textbf{PREDICT PATIENT HAS OSA OR UARS if SCORE $> 19$}\vspace{0.5em}}\\
\begin{tabular}{|l l  c | c |}
   \hline
  1. & age $\geq$ 60 & 18 points & $\quad\cdots\cdots$ \\ 
  2. & body mass index $\geq$ 25 & 10 points & $+\quad\cdots\cdots$ \\ 
  3. & body mass index $\geq$ 40 & 10 points & $+\quad\cdots\cdots$ \\ 
  4. & snoring & 8 points & $+\quad\cdots\cdots$ \\ 
  5. & hypertension & 6 points & $+\quad\cdots\cdots$ \\ 
  6. & bronchitis & 4 points & $+\quad\cdots\cdots$ \\ 
  7. & cataplexy & 2 points & $+\quad\cdots\cdots$ \\ 
  8. & dozes off while watching TV & 2 points & $+\quad\cdots\cdots$ \\ 
  9. & retired & 2 points & $+\quad\cdots\cdots$ \\ 
  10. & female & -8 points & $+\quad\cdots\cdots$ \\ 
   \hline
 & \small{\textbf{ADD POINTS FROM ROWS 1-10}} & \small{\textbf{SCORE }} & $=\quad\cdots\cdots$ \\ 
   \hline
\end{tabular}
\caption{SLIM scoring system trained to diagnose sleep apnea. This model achieves a TPR of 61.7\%, obeys all model requirements, and was trained using a single instance of the free parameters.}
\label{Fig::SleepApneaModels}
\end{table}
\FloatBarrier
%
%
%
%
%
%
%
%
%
\clearpage
\section{Numerical Experiments}\label{Sec::NumericalExperiments}
We present numerical experiments that compare the accuracy and sparsity of 10 classification methods on 8 popular classification datasets. Our goal is to illustrate the off-the-shelf performance of models from our framework, provide empirical evidence on the performance of discrete linear classifiers, and show that we can train accurate models by solving IPs in a matter of minutes.
\subsection{Experimental Setup}\label{Sec::ExperimentalSetup}
\textit{Datasets}: We ran numerical experiments using various popular datasets from the UCI Machine Learning Repository \citep{Bache+Lichman:2013}, summarized in Table \ref{Table::ExperimentalDatasets}. We chose these datasets to allow a comparison with other works, and explore the performance of each method as we varied the size and nature of the training data. We processed each dataset by binarizing all categorical features as some real-valued features. We trained all methods using the same processed dataset, except for MN Rules, where we needed to binarize all of the features (we include the number of binary rules as $P_{rules}$ in Table \ref{Table::ExperimentalDatasets}; note that for each binary rule in the MN Rules datasets, $\bm{h}_{j,t}$, we also included the opposite of that binary rule, $1-\bm{h}_{j,t}$). For the purposes of reproducibility, we plan to include all of our datasets and finalized code in the supplementary materials.
%
%
\begin{table}[htbp]
\scriptsize
\centering
\begin{tabular}{lcccl}
\toprule
\textbf{Dataset} & $N$ & $P$ &$P_{rules}$ & \textbf{Classification Task}\\
\toprule \texttt{adult} & 32561 & 36 & 70 & predict if a U.S. resident is earning more than $\$50000$ \\
\midrule \texttt{breastcancer} & 683 & 9 & 36 & detect breast cancer using a biopsy \\ 
\midrule \texttt{bankruptcy} & 250 & 6 & 18 & predict if a firm will go bankrupt \\ 
\midrule \texttt{haberman} & 306 & 3 & 8 & predict the 5-year survival of patients after breast cancer surgery \\ 
\midrule \texttt{heart} & 303 & 32 & 26 & identify patients with an elevated risk of heart disease \\
\midrule \texttt{mammo} & 961 & 12 & 62 & detect breast cancer using a mammogram \\
\midrule \texttt{mushroom} & 8124 & 113 & 218 &  predict if a mushroom is poisonous \\ 
\midrule \texttt{spambase} & 4601 & 57 & 114 & predict if an e-mail is spam or not \\ 
\bottomrule
\end{tabular}
\caption{Overview of all datasets used in the numerical experiments.}
\label{Table::ExperimentalDatasets}
\end{table}
\newline
\noindent \textit{Methods}: We summarize the training setup for each method in Table \ref{Table::ExperimentalMethods}. We trained SLIM and MN Rules using the CPLEX 12.6 API in MATLAB and 8 state-of-the-art baseline methods using packages in R 3.0.2 \citep{Rcitation}. We note that we trained SLIM and MN Rules because they were well-suited as off-the-shelf classifiers, unlike PILM and TILM which are designed for data-dependent settings and interpretability functions. 

For each method, each dataset, and each unique combination of free parameters, we trained 10 models using subsets of the data to assess predictive accuracy via 10-fold cross-validation (CV), and 1 final model using all of the data to assess interpretability. We trained SLIM and MN Rules models by solving a total of $6 \times 11$ IPs (6 values of $C_0$, 11 training runs per $C_0$). We allocated at most 10 minutes of computing time to solve each IP, and solved 12 IPs at a time, in parallel, on a 12-core 2.7 GHZ processor with 48 GB RAM. Thus, it took at most 1 hour of computing time to train SLIM or MN rules for each dataset. We aimed to compare the performance of our methods against the best possible performance of the baseline methods, and therefore ran the baseline methods without time constraints using a large grid of free parameters. 

We restricted the hypothesis spaces for SLIM and MN Rules to highlight the impact of using restricting coefficients to a small discrete set. In general, the set of coefficients for SLIM was roughly 10 times larger than the set of coefficients for MN Rules. For SLIM, we used an $\Lset$ set that contained 20 non-zero integer coefficients for each feature, $\Lset_j=\{-10,\ldots,10\}$. For MN Rules, we used an $\Lset$ that contained 1 non-zero coefficient for each binary rule -- this was equivalent to training a model with 2 non-zero coefficients $\Lset_j=\{-1,0,1\}$ since we trained these models on a dataset that also contained the opposite of each rule. The only exception for this was \texttt{haberman}, where we used a scaled set of coefficients for the $age$ variable from the set $\Lset_j = \{-1.0,-0.9,\ldots,1.0\}$ and refined set of coefficients for the intercept $\Lset_0 = \{-100,-99.9,-99.8,\ldots,100\}$.

Since the \texttt{adult} and \texttt{haberman} datasets were imbalanced, we trained all methods for these datasets with a weighted loss function where we set $\wplus = \nminus/N$ and $\wminus = \nplus/N$ so that the classifier would attain roughly the same degree of accuracy on both positive and negative examples.
\begin{table}[htbp]
\scriptsize
\centering
\begin{tabular}{llcl}

\toprule 
\textbf{Method} & \textbf{Acronym} & \textbf{Software} & \textbf{\cell{l}{Settings and Free Parameters}} \\ 

\toprule

CART Decision Trees & CART & \pkg{rpart} \citep{the2012rpart} & default settings \\ 

\midrule

C5.0 Decision Trees & C5.0T & \pkg{c50} \citep{kuhn2012c50} & default settings  \\ 

\midrule

C5.0 Decision Rules  & C5.0R & \pkg{c50}  \citep{kuhn2012c50} & default settings \\ 

\midrule

LARS Lasso, Binomial Link & Lasso & \pkg{glmnet} \citep{friedman2010glmnet} & 1000 values of $\lambda$ chosen by \pkg{glmnet} \\

\midrule

LARS Ridge, Binomial Link & Ridge & \pkg{glmnet} \citep{friedman2010glmnet} & 1000 values of $\lambda$ chosen by \pkg{glmnet} \\ 

\midrule

LARS Elastic Net, Binomial Link & E.Net & \pkg{glmnet} \citep{friedman2010glmnet} &

\cell{l}{
1000 values of $\lambda$ chosen by \pkg{glmnet}\\
$\times$ 19 values of $\alpha \in (0.05,0.10,\ldots,0.95)$
} \\ 

\midrule

SVM, Linear Kernel & SVM Lin. &  \pkg{e1071} \citep{meyer2012e1071} & 25 values of $C \in \{10^t | t = (-3,-2.75\ldots,2.75,3)\}$ \\ 

\midrule

SVM, RBF Kernel & SVM RBF & \pkg{e1071} \citep{meyer2012e1071} & 25 values of $C \in \{10^t | t = (-3,-2.75\ldots,2.75,3)\}$ \\ 

\midrule

SLIM Scoring System & SLIM  & \pkg{CPLEX 12.6} & 

\cell{l}{
$C_0 \in \{0.01, 0.075, 0.05, 0.025, 0.001, \frac{0.9}{NP} \}$ \\
$\lambda_j \in \{-10,\ldots,10\}$; $\lambda_0 \in \{-100,\ldots,100\}$ 
} \\ 

\midrule

M-of-N Rule Tables & MN Rules & \pkg{CPLEX 12.6} & 

\cell{l}{
$C_0 \in \{0.01, 0.075, 0.05, 0.025, 0.001, \frac{0.9}{NP} \}$ \\
$\lambda_j \in \{0,1\}$; $\lambda_0 \in \{-P,\ldots,0\}$ 
} \\ 

\bottomrule 
\end{tabular}
\caption{Training setup for classification methods used for the numerical experiments.}
\label{Table::ExperimentalMethods}
\end{table}
\subsection{Results}\label{Sec::NumericalExperimentsResults}
We summarize the results of our experiments in Table \ref{Table::ExpResultsTable}. We report the sparsity of models using a metric we call \textit{model size}. We define model size as it pertains to the interpretability of different models. Model size represents the number of coefficients for linear classifiers (Lasso, Ridge, E.Net, MN Rules, SLIM, SVM Lin.), the number of leaves for decision tree classifiers (C5.0T, CART), and the number of rules for rule-based classifiers (C5.0R). For completeness, we set the model size for black-box models (SVM RBF) to the number of features since model size does not relate to interpretability for these methods. 

We plot a visual representation of the results of Table \ref{Table::ExpResultsTable} in the plots on the left side of Figures \ref{Fig::ExpPlots1}--\ref{Fig::ExpPlots2}. These plots highlight the accuracy and sparsity of all methods on each dataset separately. In a given figure, we plot a point for each method corresponding to the mean 10-fold CV test error and the median 10-fold CV model size. We surround this point with a box to highlight the variation in accuracy and sparsity for each algorithm. In this case, the box ranges over the 10-fold CV standard deviation in test error and the 10-fold CV min/max of model sizes. When a method shows no variation in model size over the 10 folds, we plot a vertical line rather than a box (i.e. no horizontal variation). When two methods produce models with the same model size (e.g. Lasso, Ridge and E. Net on \textds{breastcancer}) the boxes or lines will also coincide.

We include the regularization paths for linear models such as SLIM, MN Rules and Lasso in the plots on the right side of Figures \ref{Fig::ExpPlots1}--\ref{Fig::ExpPlots2}. These plots show the test error achieved at different levels of sparsity. Note that the regularization path for SLIM and MN Rules always includes the most accurate model that can be produced by the hypothesis space but that this model does not always use all of the features in the dataset (e.g. the most accurate SLIM model for \textds{bankruptcy} uses 3 out of the 6 features). This is actually due to the $\Lset$ set restriction: if the $\Lset$ were relaxed, the method would most likely use more coefficients to attain a higher training accuracy.
\begin{table}[htbp]
\scriptsize
\centering
\resizebox{\textwidth}{!} {
\setlength{\tabcolsep}{1.5pt}
\begin{tabular}{>{\scriptsize}l>{\scriptsize}c>{\scriptsize}lcccccccccc}
   \toprule 
   \bf{Dataset}&\bf{Details}&\bf{Metric}&\bf{Lasso}&\bf{Ridge}&\bf{E. Net}&\bf{C5.0R}&\bf{C5.0T}&\bf{CART}&\bf{SVM Lin.}&\bf{SVM RBF}&\bf{MN Rules}&\bf{SLIM} \\ \toprule 
   
    \texttt{adult} & \ddcell{$N$&32561\\$P$&36\\$\Pr(y\text{=+1})$&24\%\\$\Pr(y\text{=-1})$&76\%} & \bfcell{l}{w. test error\\w. train error\\model size\\model range} & \cell{c}{17.3 $\pm$ 0.9$\%$\\17.2 $\pm$ 0.1$\%$\\14\\13 - 14} & \cell{c}{17.6 $\pm$ 0.9$\%$\\17.6 $\pm$ 0.1$\%$\\36\\36 - 36} & \cell{c}{17.4 $\pm$ 0.9$\%$\\17.4 $\pm$ 0.1$\%$\\17\\16 - 18} & \cell{c}{26.4 $\pm$ 1.8$\%$\\25.3 $\pm$ 0.4$\%$\\41\\38 - 46} & \cell{c}{26.3 $\pm$ 1.4$\%$\\24.9 $\pm$ 0.4$\%$\\84\\78 - 99} & \cell{c}{75.9 $\pm$ 0.0$\%$\\75.9 $\pm$ 0.0$\%$\\4\\4 - 4} & \cell{c}{16.8 $\pm$ 0.8$\%$\\16.7 $\pm$ 0.1$\%$\\36\\36 - 36} & \cell{c}{16.3 $\pm$ 0.5$\%$\\16.3 $\pm$ 0.1$\%$\\36\\36 - 36} & \cell{c}{19.2 $\pm$ 1.0$\%$\\19.2 $\pm$ 0.1$\%$\\4\\4 - 4} & \cell{c}{17.4 $\pm$ 1.4$\%$\\17.5 $\pm$ 1.2$\%$\\19\\7 - 26} \\ 

   \midrule \texttt{breastcancer} & \ddcell{$N$&683\\$P$&9\\$\Pr(y\text{=+1})$&35\%\\$\Pr(y\text{=-1})$&65\%} & \bfcell{l}{test error\\train error\\model size\\model range} & \cell{c}{3.4 $\pm$ 2.2$\%$\\2.9 $\pm$ 0.3$\%$\\9\\8 - 9} & \cell{c}{3.4 $\pm$ 1.7$\%$\\3.0 $\pm$ 0.3$\%$\\9\\9 - 9} & \cell{c}{3.1 $\pm$ 2.1$\%$\\2.8 $\pm$ 0.3$\%$\\9\\9 - 9} & \cell{c}{4.3 $\pm$ 3.3$\%$\\2.1 $\pm$ 0.3$\%$\\7\\6 - 9} & \cell{c}{5.3 $\pm$ 3.4$\%$\\1.6 $\pm$ 0.4$\%$\\13\\7 - 16} & \cell{c}{5.6 $\pm$ 1.9$\%$\\3.6 $\pm$ 0.3$\%$\\4\\3 - 7} & \cell{c}{3.1 $\pm$ 2.0$\%$\\2.7 $\pm$ 0.2$\%$\\9\\9 - 9} & \cell{c}{3.5 $\pm$ 2.5$\%$\\0.3 $\pm$ 0.1$\%$\\9\\9 - 9} & \cell{c}{4.8 $\pm$ 2.5$\%$\\4.1 $\pm$ 0.2$\%$\\8\\7 - 8} & \cell{c}{3.4 $\pm$ 2.0$\%$\\3.2 $\pm$ 0.2$\%$\\2\\2 - 2} \\ 

   \midrule \texttt{bankruptcy} & \ddcell{$N$&250\\$P$&6\\$\Pr(y\text{=+1})$&57\%\\$\Pr(y\text{=-1})$&43\%} & \bfcell{l}{test error\\train error\\model size\\model range} & \cell{c}{0.0 $\pm$ 0.0$\%$\\0.0 $\pm$ 0.0$\%$\\3\\3 - 3} & \cell{c}{0.4 $\pm$ 1.3$\%$\\0.4 $\pm$ 0.1$\%$\\6\\6 - 6} & \cell{c}{0.0 $\pm$ 0.0$\%$\\0.4 $\pm$ 0.7$\%$\\3\\3 - 3} & \cell{c}{0.8 $\pm$ 1.7$\%$\\0.4 $\pm$ 0.2$\%$\\4\\4 - 4} & \cell{c}{0.8 $\pm$ 1.7$\%$\\0.4 $\pm$ 0.2$\%$\\4\\4 - 4} & \cell{c}{1.6 $\pm$ 2.8$\%$\\1.6 $\pm$ 0.3$\%$\\2\\2 - 2} & \cell{c}{0.4 $\pm$ 1.3$\%$\\0.4 $\pm$ 0.1$\%$\\6\\6 - 6} & \cell{c}{0.4 $\pm$ 1.3$\%$\\0.4 $\pm$ 0.1$\%$\\6\\6 - 6} & \cell{c}{1.6 $\pm$ 2.8$\%$\\1.6 $\pm$ 0.3$\%$\\3\\3 - 3} & \cell{c}{0.8 $\pm$ 1.7$\%$\\0.0 $\pm$ 0.0$\%$\\3\\2 - 3} \\ 

   \midrule \texttt{haberman} & \ddcell{$N$&306\\$P$&3\\$\Pr(y\text{=+1})$&74\%\\$\Pr(y\text{=-1})$&26\%} & \bfcell{l}{w. test error\\w. train error\\model size\\model range} & \cell{c}{42.5 $\pm$ 11.3$\%$\\40.6 $\pm$ 1.9$\%$\\2\\2 - 2} & \cell{c}{36.9 $\pm$ 15.0$\%$\\41.0 $\pm$ 9.7$\%$\\3\\3 - 3} & \cell{c}{40.9 $\pm$ 14.0$\%$\\45.1 $\pm$ 12.0$\%$\\1\\1 - 1} & \cell{c}{42.7 $\pm$ 9.4$\%$\\40.4 $\pm$ 8.5$\%$\\2\\0 - 3} & \cell{c}{42.7 $\pm$ 9.4$\%$\\40.4 $\pm$ 8.5$\%$\\2\\1 - 3} & \cell{c}{43.1 $\pm$ 8.0$\%$\\34.3 $\pm$ 2.8$\%$\\6\\4 - 9} & \cell{c}{45.3 $\pm$ 14.7$\%$\\46.0 $\pm$ 3.6$\%$\\3\\3 - 3} & \cell{c}{47.5 $\pm$ 6.2$\%$\\5.4 $\pm$ 1.5$\%$\\4\\4 - 4} & \cell{c}{54.7 $\pm$ 24.3$\%$\\54.7 $\pm$ 24.3$\%$\\1\\0 - 1} & \cell{c}{31.8 $\pm$ 13.1$\%$\\29.3 $\pm$ 1.9$\%$\\3\\3 - 3} \\ 

   \midrule \texttt{mammo} & \ddcell{$N$&961\\$P$&14\\$\Pr(y\text{=+1})$&46\%\\$\Pr(y\text{=-1})$&54\%} & \bfcell{l}{test error\\train error\\model size\\model range} & \cell{c}{19.0 $\pm$ 3.1$\%$\\19.3 $\pm$ 0.3$\%$\\13\\12 - 13} & \cell{c}{19.2 $\pm$ 3.0$\%$\\19.2 $\pm$ 0.4$\%$\\14\\14 - 14} & \cell{c}{19.0 $\pm$ 3.1$\%$\\19.2 $\pm$ 0.3$\%$\\14\\13 - 14} & \cell{c}{20.5 $\pm$ 3.3$\%$\\19.8 $\pm$ 0.3$\%$\\5\\3 - 5} & \cell{c}{20.3 $\pm$ 3.5$\%$\\19.9 $\pm$ 0.3$\%$\\5\\4 - 6} & \cell{c}{20.7 $\pm$ 3.9$\%$\\20.0 $\pm$ 0.6$\%$\\4\\3 - 5} & \cell{c}{20.3 $\pm$ 3.0$\%$\\20.3 $\pm$ 0.4$\%$\\14\\14 - 14} & \cell{c}{19.1 $\pm$ 3.1$\%$\\18.2 $\pm$ 0.4$\%$\\14\\14 - 14} & \cell{c}{21.6 $\pm$ 3.5$\%$\\20.8 $\pm$ 0.3$\%$\\9\\9 - 9} & \cell{c}{19.5 $\pm$ 3.0$\%$\\18.3 $\pm$ 0.3$\%$\\9\\9 - 11} \\ 

   \midrule \texttt{heart} & \ddcell{$N$&303\\$P$&32\\$\Pr(y\text{=+1})$&46\%\\$\Pr(y\text{=-1})$&54\%} & \bfcell{l}{test error\\train error\\model size\\model range} & \cell{c}{15.2 $\pm$ 6.3$\%$\\14.0 $\pm$ 1.0$\%$\\11\\10 - 13} & \cell{c}{14.9 $\pm$ 5.9$\%$\\13.1 $\pm$ 0.8$\%$\\32\\30 - 32} & \cell{c}{14.5 $\pm$ 5.9$\%$\\13.2 $\pm$ 0.6$\%$\\24\\22 - 27} & \cell{c}{21.2 $\pm$ 7.5$\%$\\10.0 $\pm$ 1.8$\%$\\10\\9 - 17} & \cell{c}{23.2 $\pm$ 6.8$\%$\\8.5 $\pm$ 2.0$\%$\\19\\12 - 27} & \cell{c}{19.8 $\pm$ 6.5$\%$\\14.3 $\pm$ 0.9$\%$\\6\\6 - 8} & \cell{c}{15.5 $\pm$ 6.5$\%$\\13.6 $\pm$ 0.5$\%$\\31\\28 - 32} & \cell{c}{15.2 $\pm$ 6.0$\%$\\10.4 $\pm$ 0.8$\%$\\32\\32 - 32} & \cell{c}{23.2 $\pm$ 10.4$\%$\\17.8 $\pm$ 0.8$\%$\\15\\10 - 16} & \cell{c}{18.8 $\pm$ 8.1$\%$\\13.3 $\pm$ 0.9$\%$\\4\\3 - 5} \\ 

   \midrule \texttt{mushroom} & \ddcell{$N$&8124\\$P$&113\\$\Pr(y\text{=+1})$&48\%\\$\Pr(y\text{=-1})$&52\%} & \bfcell{l}{test error\\train error\\model size\\model range} & \cell{c}{0.0 $\pm$ 0.0$\%$\\0.0 $\pm$ 0.0$\%$\\25\\23 - 26} & \cell{c}{1.7 $\pm$ 0.3$\%$\\1.7 $\pm$ 0.0$\%$\\113\\113 - 113} & \cell{c}{0.0 $\pm$ 0.0$\%$\\0.0 $\pm$ 0.0$\%$\\108\\106 - 108} & \cell{c}{0.0 $\pm$ 0.0$\%$\\0.0 $\pm$ 0.0$\%$\\7\\7 - 7} & \cell{c}{0.0 $\pm$ 0.0$\%$\\0.0 $\pm$ 0.0$\%$\\9\\9 - 9} & \cell{c}{1.2 $\pm$ 0.6$\%$\\1.1 $\pm$ 0.3$\%$\\7\\6 - 8} & \cell{c}{0.0 $\pm$ 0.0$\%$\\0.0 $\pm$ 0.0$\%$\\104\\99 - 108} & \cell{c}{0.0 $\pm$ 0.0$\%$\\0.0 $\pm$ 0.0$\%$\\113\\113 - 113} & \cell{c}{0.0 $\pm$ 0.0$\%$\\0.0 $\pm$ 0.0$\%$\\21\\21 - 21} & \cell{c}{0.0 $\pm$ 0.0$\%$\\0.0 $\pm$ 0.0$\%$\\7\\7 - 7} \\ 

   \midrule \texttt{spambase} & \ddcell{$N$&4601\\$P$&57\\$\Pr(y\text{=+1})$&39\%\\$\Pr(y\text{=-1})$&61\%} & \bfcell{l}{test error\\train error\\model size\\model range} & \cell{c}{10.0 $\pm$ 1.7$\%$\\9.5 $\pm$ 0.3$\%$\\28\\28 - 29} & \cell{c}{26.3 $\pm$ 1.7$\%$\\26.1 $\pm$ 0.2$\%$\\57\\57 - 57} & \cell{c}{10.0 $\pm$ 1.7$\%$\\9.6 $\pm$ 0.2$\%$\\28\\28 - 29} & \cell{c}{6.6 $\pm$ 1.3$\%$\\4.2 $\pm$ 0.3$\%$\\27\\23 - 31} & \cell{c}{7.3 $\pm$ 1.0$\%$\\3.9 $\pm$ 0.3$\%$\\69\\56 - 78} & \cell{c}{11.1 $\pm$ 1.4$\%$\\9.8 $\pm$ 0.3$\%$\\7\\6 - 10} & \cell{c}{7.8 $\pm$ 1.5$\%$\\8.1 $\pm$ 0.8$\%$\\57\\57 - 57} & \cell{c}{13.7 $\pm$ 1.4$\%$\\1.3 $\pm$ 0.1$\%$\\57\\57 - 57} & \cell{c}{10.2 $\pm$ 1.2$\%$\\9.6 $\pm$ 0.2$\%$\\26\\24 - 40} & \cell{c}{6.3 $\pm$ 1.2$\%$\\5.7 $\pm$ 0.3$\%$\\32\\28 - 40} \\

   \bottomrule \end{tabular}

}
\caption{Accuracy and sparsity of the most accurate model produced by all methods on all UCI datasets. Here: test error denotes the 10-fold CV test error; train error denotes the 10-fold CV training error; model size corresponds to the 10-fold CV median model size; model range is the 10-fold CV minimum and maximum model-size. We set free parameters to minimize the mean 10-fold CV error so as to reflect the most accurate model produced by each method. We report the weighted 10-fold CV testing and training error for \textds{adult} and \textds{haberman} as we train the methods using a weighted loss function.}
\label{Table::ExpResultsTable}
\end{table}

\subsection{Observations}
We wish to make the following observations regarding our results:

\vspace{0.20cm}
\noindent\textbf{On Computation} 
\vspace{0.10cm}

\noindent There is no evidence that computational issues hurt the performance of SLIM and MN Rules on any of the UCI datasets. In all cases, we were able to obtain reasonable feasible solutions for all datasets in 10 minutes using off-the-shelf settings for CPLEX 12.6. Further, the solver provided a proof of optimality (i.e. a relative MIPGAP of 0.00\%) for all of the models that we trained for \textds{mammo}, \textds{mushroom}, \textds{bankruptcy}, \textds{breastcancer}. We note that a discriminating factor for obtaining a proof of optimality is not necessarily the size of the dataset, but the number of binary features \citep[this seems to be supported by previous work, see for instance][]{asparoukhov1997mathematical}. We also note that additional computational time may lead to overfitting, as shown by the results of numerical experiments in Appendix \ref{Appendix::ExtraComputationalTime} where we solve each IP for 60 minutes. 

\vspace{0.20cm}
\noindent\textbf{On the Accuracy and Sparsity of SLIM and MN Rules} 
\vspace{0.10cm}

\noindent As shown in the leftmost plots in Figures \ref{Fig::ExpPlots1}--\ref{Fig::ExpPlots2}, many methods are unable to produce models that attain the same levels of accuracy and sparsity as SLIM. Among linear methods, SLIM always produces a model that is more accurate than Lasso at some level of sparsity, and sometimes models that are more accurate across the entire regularization path (e.g. \textds{spambase}, \textds{haberman}, \textds{mushroom}, \textds{breastcancer}). MN rules typically produce models that are less accurate than SLIM and Lasso due to their highly limited hypothesis space. However, there are datasets where they do well across the full regularization path (e.g. \textds{spambase}, \textds{mushroom}).

\vspace{0.20cm}
\noindent\textbf{On the Regularization Effect of Discrete Coefficients}
\vspace{0.10cm}

\noindent We expect that methods that optimize the correct functions for accuracy and sparsity will achieve the best possible accuracy at every level of sparsity (i.e. the best possible trade-off between accuracy and sparsity). SLIM and MN rules both optimize true measures of accuracy and sparsity, but are restricted to discrete hypothesis spaces. Given that hypothesis space of SLIM is larger than the hypothesis space of MN rules, we expect the performance of SLIM models to be more accurate than MN rules at each level of sparsity. Thus, the relative difference in accuracy in the regularization paths of SLIM and MN Rules highlight the effects of using a small $\Lset$ (MN Rules) a large $\Lset$ (SLIM). By comparing SLIM and MN Rules to Lasso models, we can identify a baseline effect due the $\Lset$ restriction: in particular, we know that when Lasso's performance dominates that of SLIM, it is very arguably due to the use of a small set of discrete coefficients. This tends to happen mainly at large model sizes (see e.g. the regularization path for \textds{breastcancer}, \textds{heart}, \textds{mammo}).

\vspace{0.20cm}
\noindent\textbf{On the Interpretability of Different Models}
\vspace{0.10cm}

\noindent To provide a focused analysis of interpretability, we limit our observations to models for the \textds{mushroom} dataset, which provides a nice basis for comparison as many methods attain perfect predictive accuracy. We include the sparsest Lasso, SLIM and MN Rules models that achieve predictive accuracy on the \textds{mushroom} data in Figures 16--\ref{Fig::MushroomMN}. We also include the most accurate CART model produced in Figure \ref{Fig::MushroomCART}, which attains a 10-fold CV test error of 1.2 $\pm$ 0.6$\%$. We include models from other methods in Appendix \ref{Appendix::MushroomModels}, noting that they are able to attain perfect or near-perfect predictive accuracy but are omitted because they use far more features.

In this case, the SLIM model uses 7 small integer coefficients and can be expressed as a 5 line scoring system because the \textit{odor=none}, \textit{odor=almond}, and \textit{odor=anise} all had the same coefficient. This model has a high level of interpretability compared to the CART and Lasso models as it is highly sparse and highly expository -- compare this to the Lasso and CART model, where it is far more difficult to gauge the influence of different input variables in the predicted outcome due to the real coefficients and hierarchical structure, respectively. We note that this model is also more accurate and more sparse than the M-of-N rule table produced by \citet{chevaleyre2013rounding} via rounding methods (rounding methods are certainly capable of producing good models, one just needs to get lucky in the rounding, which becomes more difficult as the number of features increases; for instance, with one iteration of randomized rounding, the test error was 40\%, and they required 20 iterations of randomized rounding to find a model with 98\% accuracy).

M-of-N rules do not seem to be a natural form of model to attain perfect accuracy on the \textds{mushroom} dataset. We present a M-of-N rules model with perfect accuracy in Figure \ref{Fig::MushroomMN}. This model is larger than that of \citet{chevaleyre2013rounding} and manages to capture complex relationships between variables through an interesting mechanism using rules (e.g. $odor = creosote$) and anti-rules ($odor \neq creosote$) to create a certain number of ``points" for each categorical variable. For instance, the total number of ``points" for $creosote$ is 3, the total number of ``points" for $foul$ is 4, and the number for $spicy$ is 2. This illustrates that fact that is not one single type of model that is uniformly interpretable for all domains and datasets, and the need for a unified framework that accommodates a tailoring process.

\FloatBarrier

\setlength{\fwidth}{0.425\textwidth}
\begin{figure}[htbp]
\centering
\begin{tabular}{c}
\hspace{0.5in} \includegraphics[trim=1in 1in 1in 1in,clip,scale=0.5]{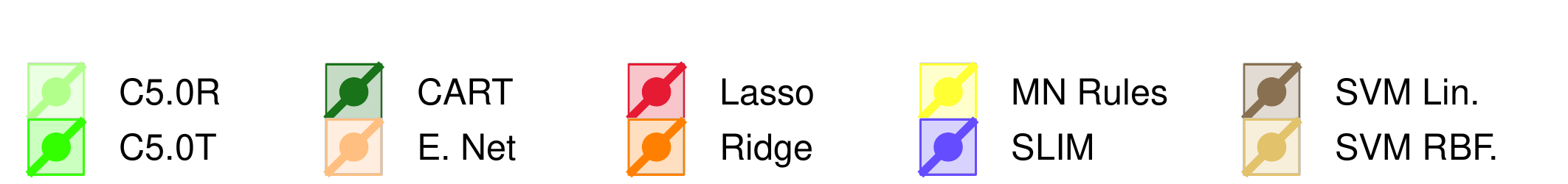} 
\end{tabular}
\begin{tabular}{>{\tiny}m{1cm}c} 
\tabdataname{adult}              & \begin{tabular}{lr}\includegraphics[width=\fwidth]{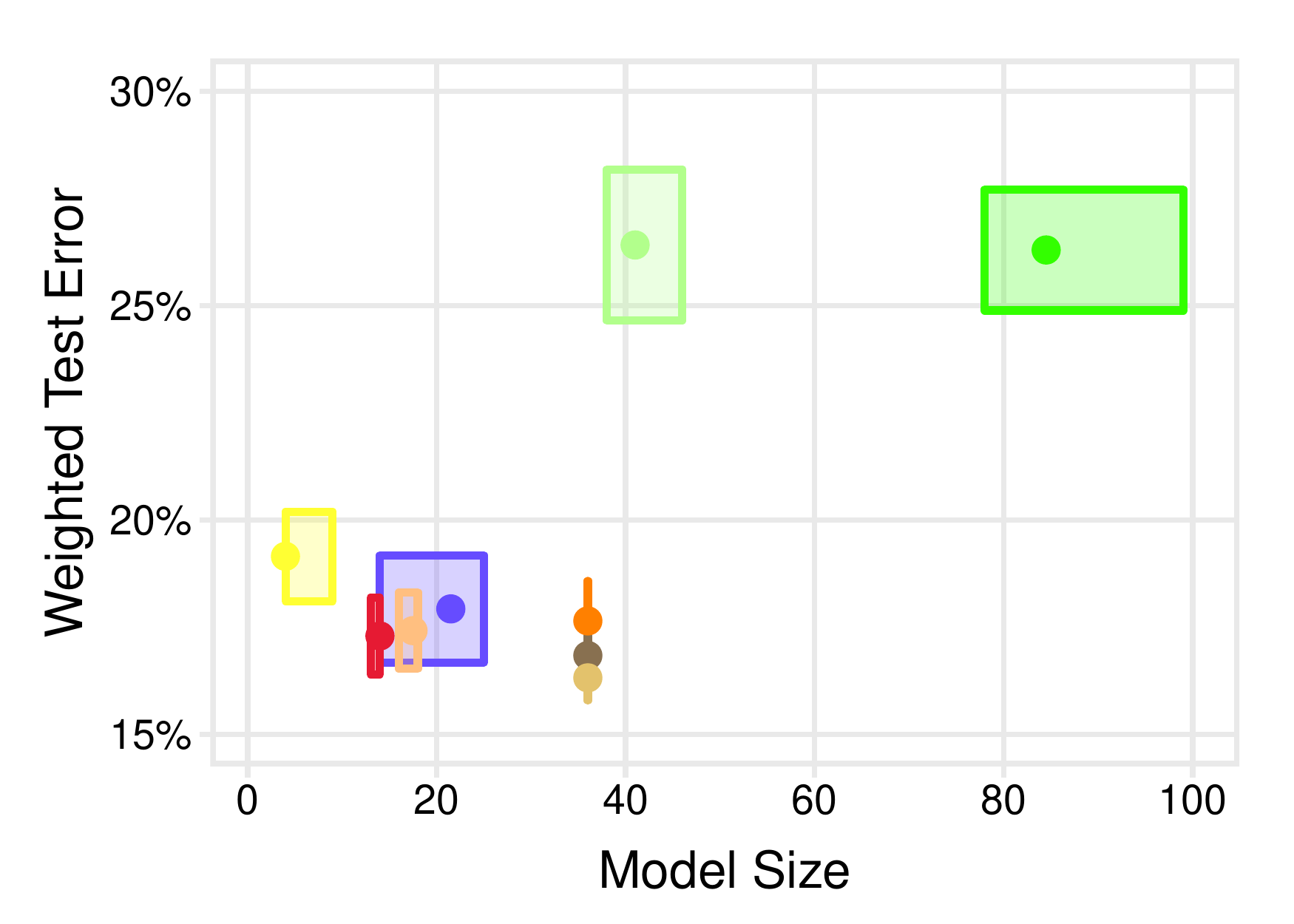}  & \includegraphics[width=\fwidth]{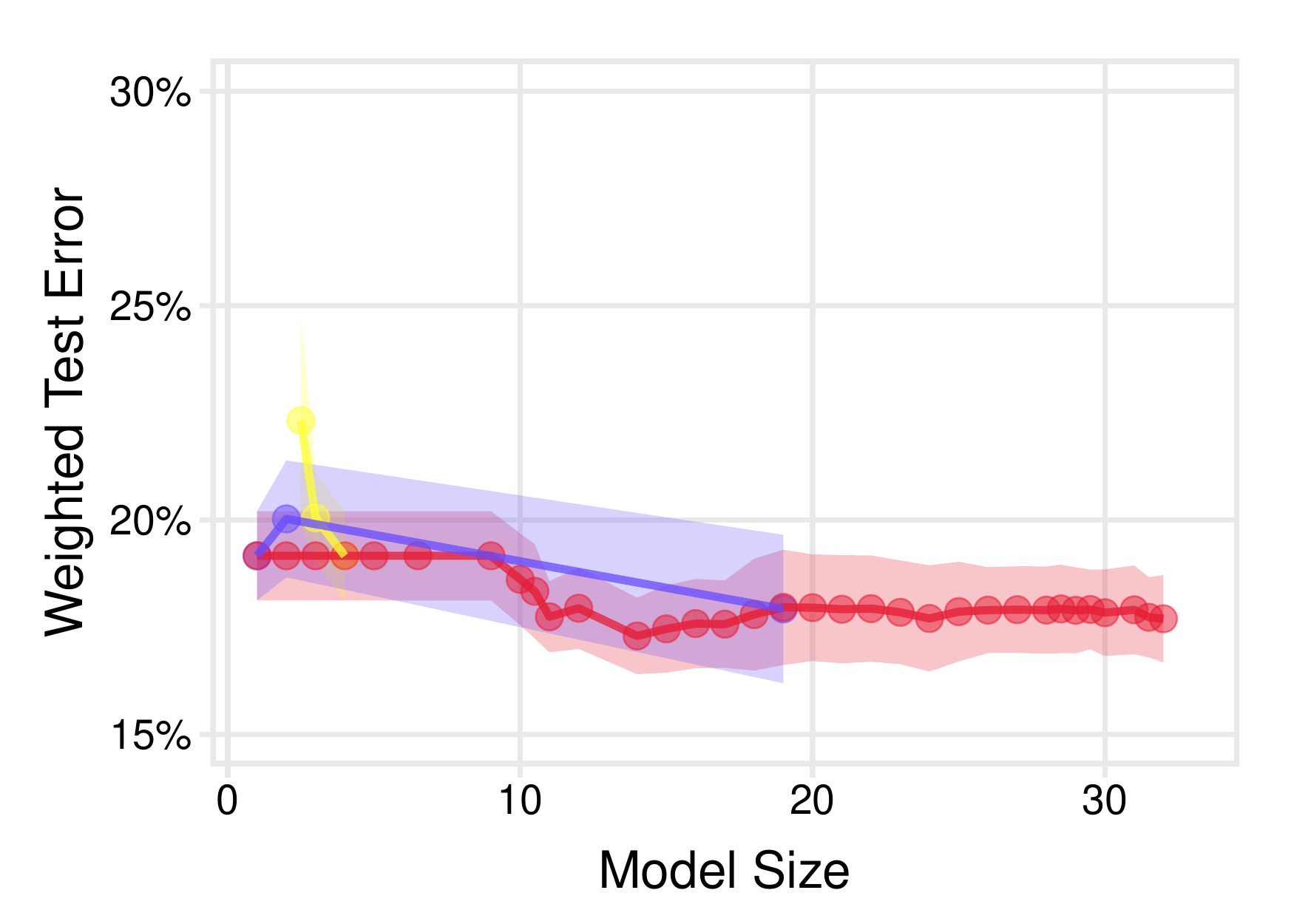} \end{tabular} \\ 
\tabdataname{bcancer}       & \begin{tabular}{lr}\includegraphics[width=\fwidth]{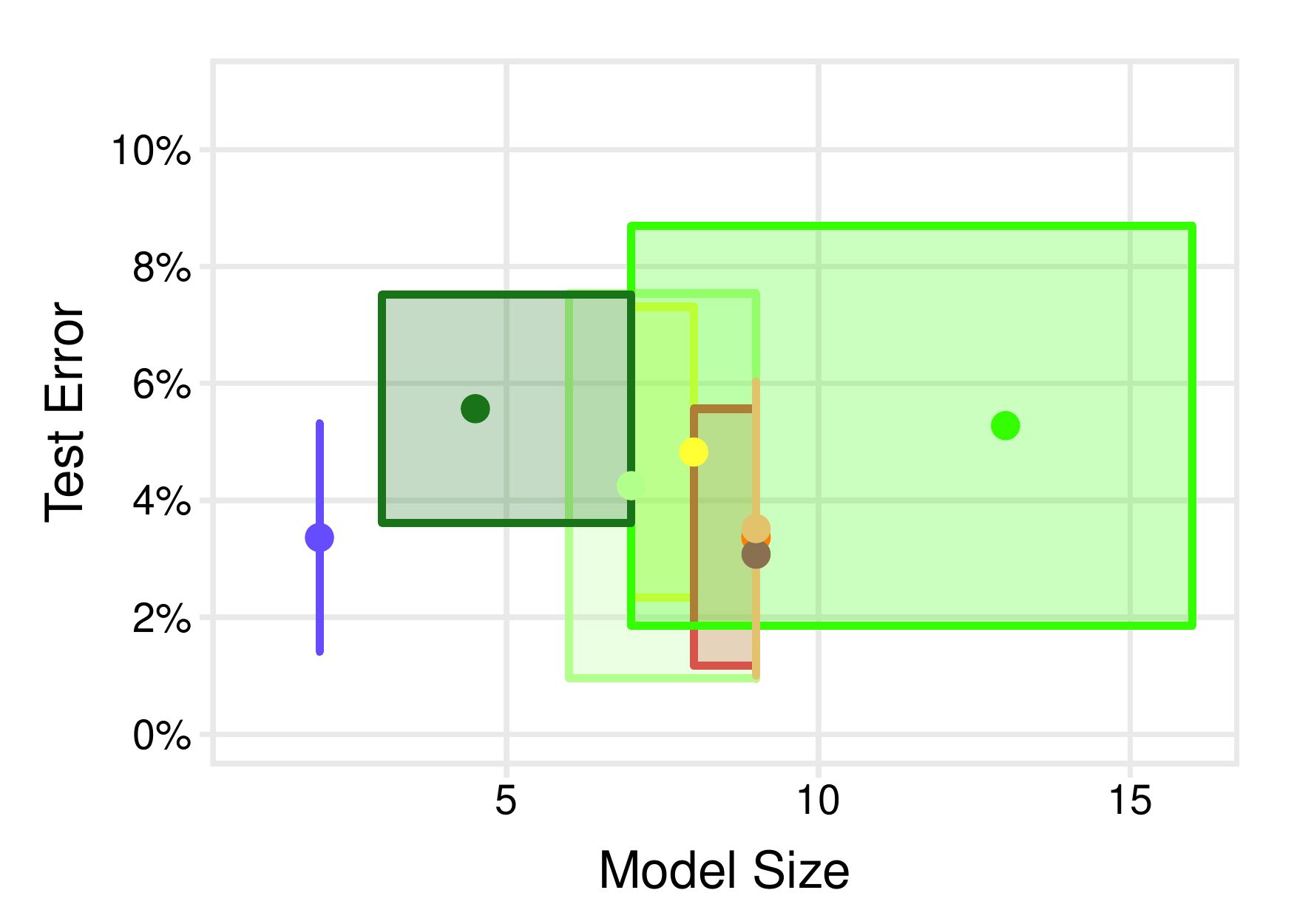} & \includegraphics[width=\fwidth]{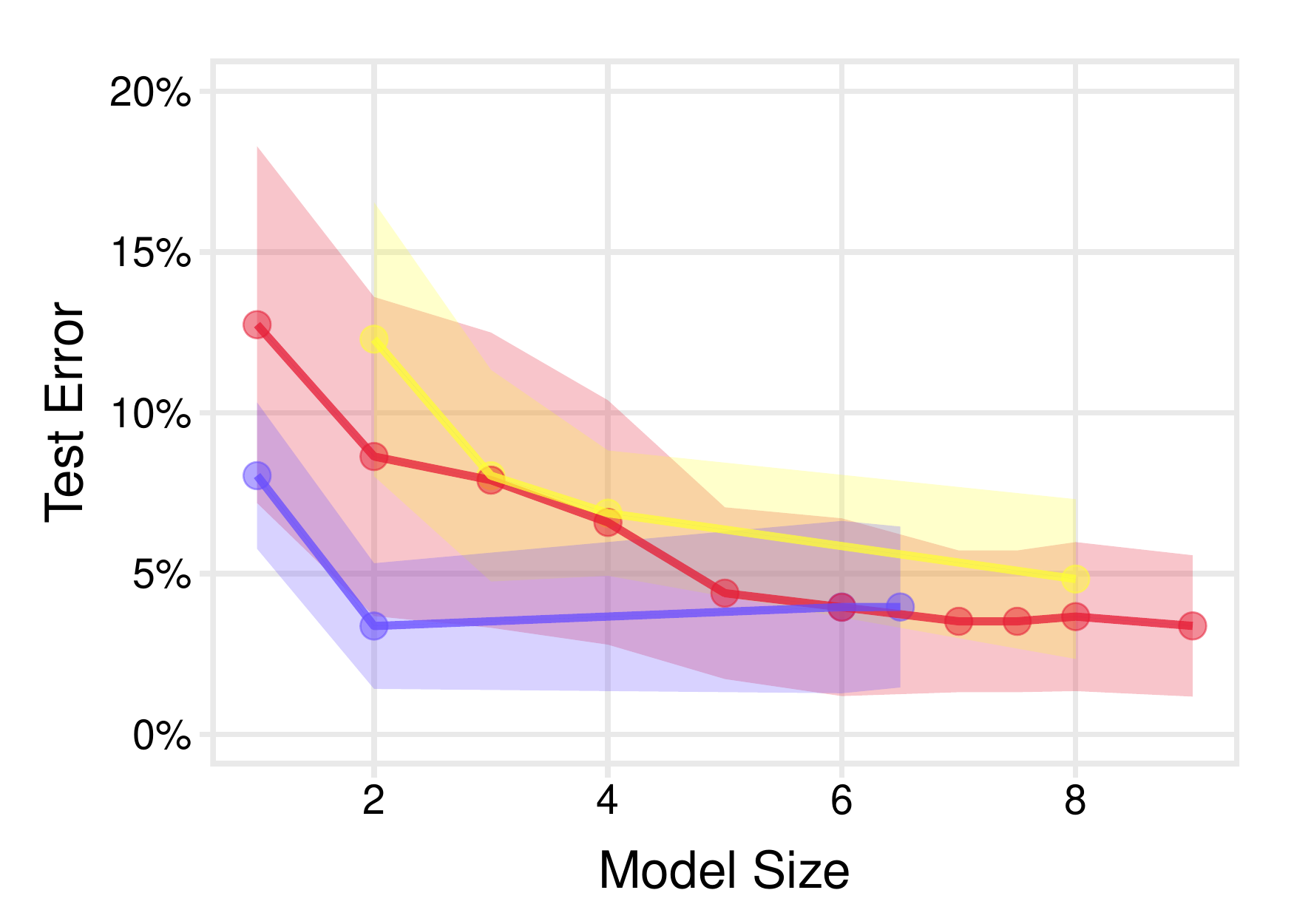} \end{tabular} \\ 
\tabdataname{bankruptcy}         & \begin{tabular}{lr}\includegraphics[width=\fwidth]{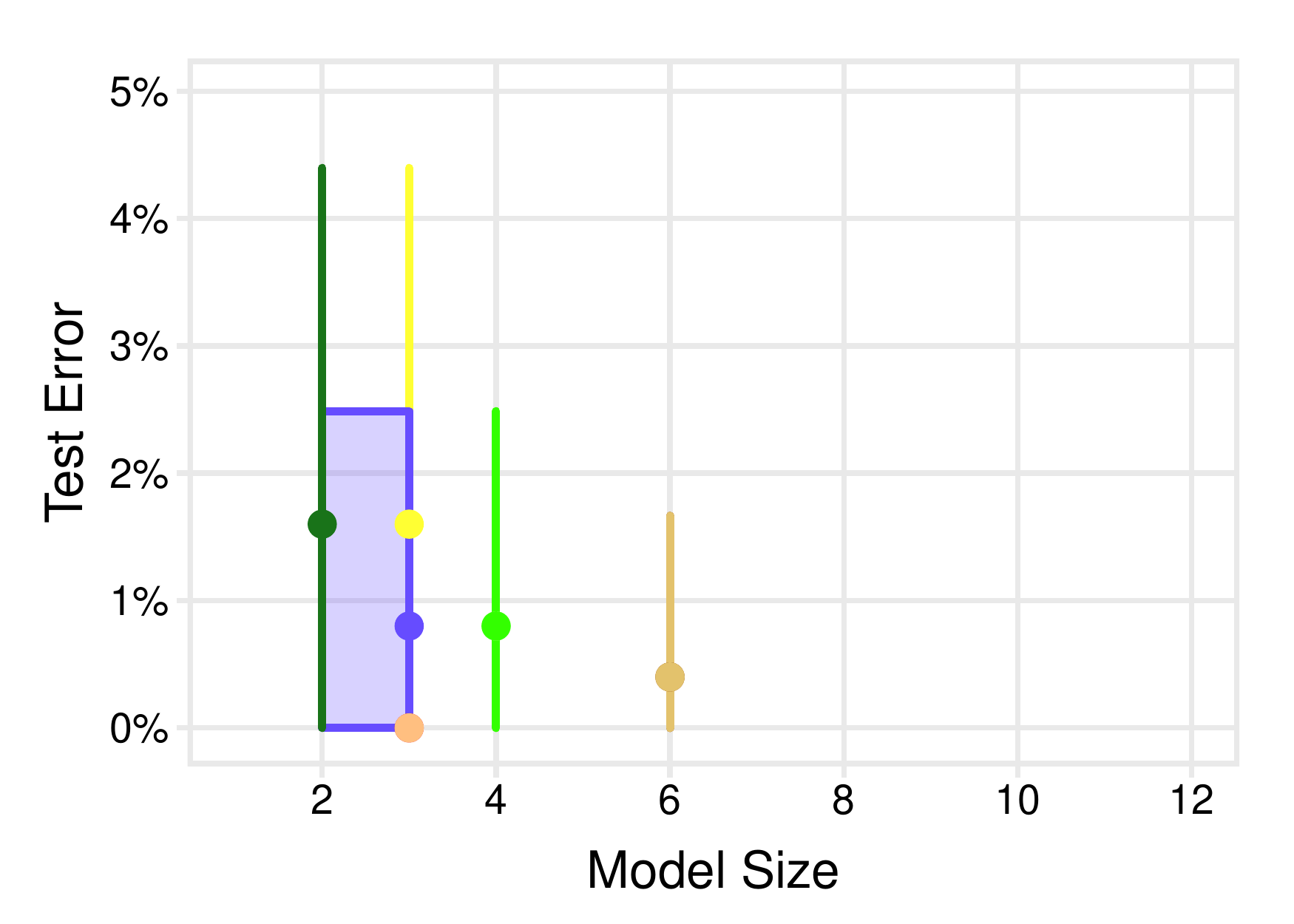}  & \includegraphics[width=\fwidth]{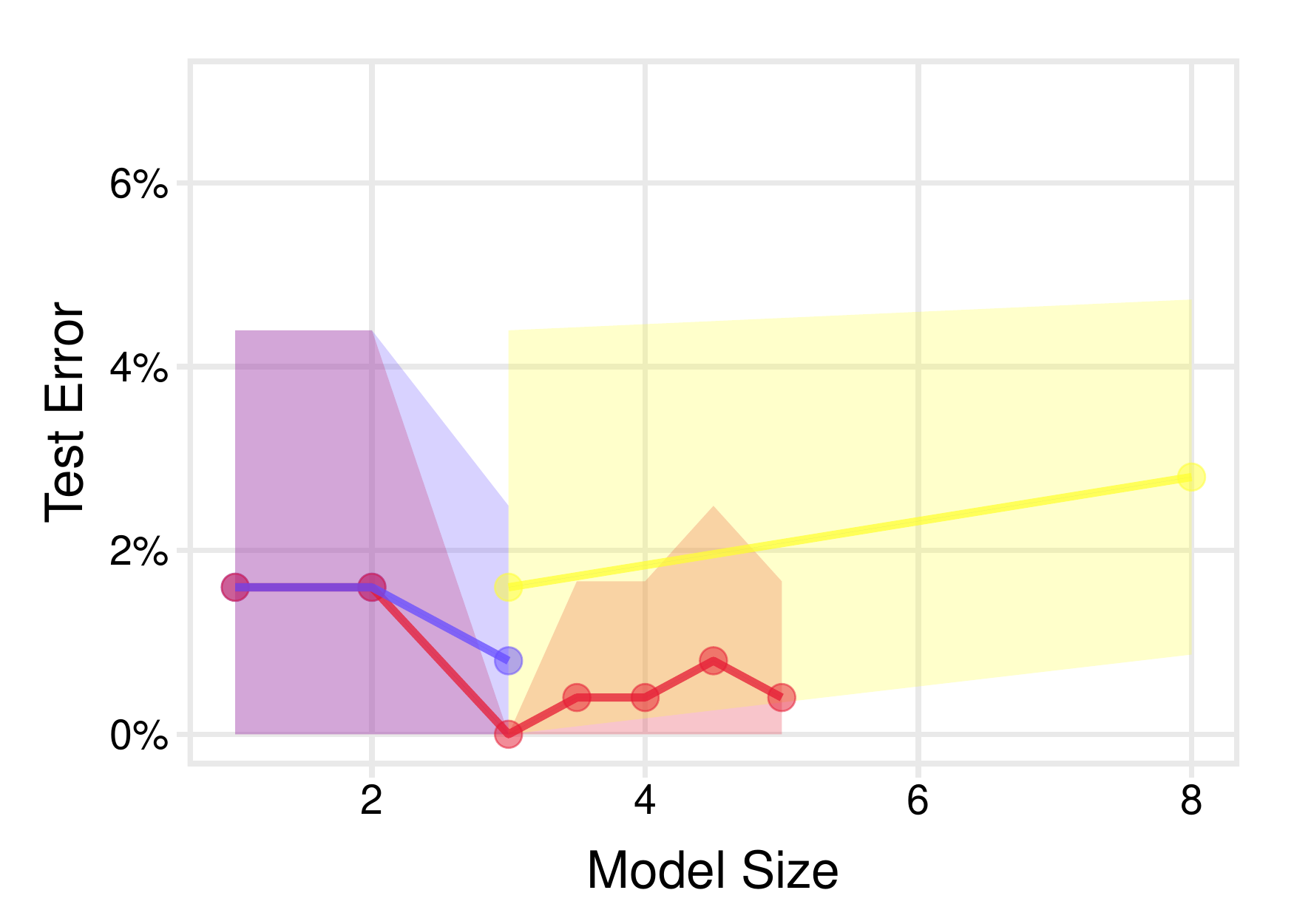} \end{tabular} \\ 
\tabdataname{haberman}           & \begin{tabular}{lr}\includegraphics[width=\fwidth]{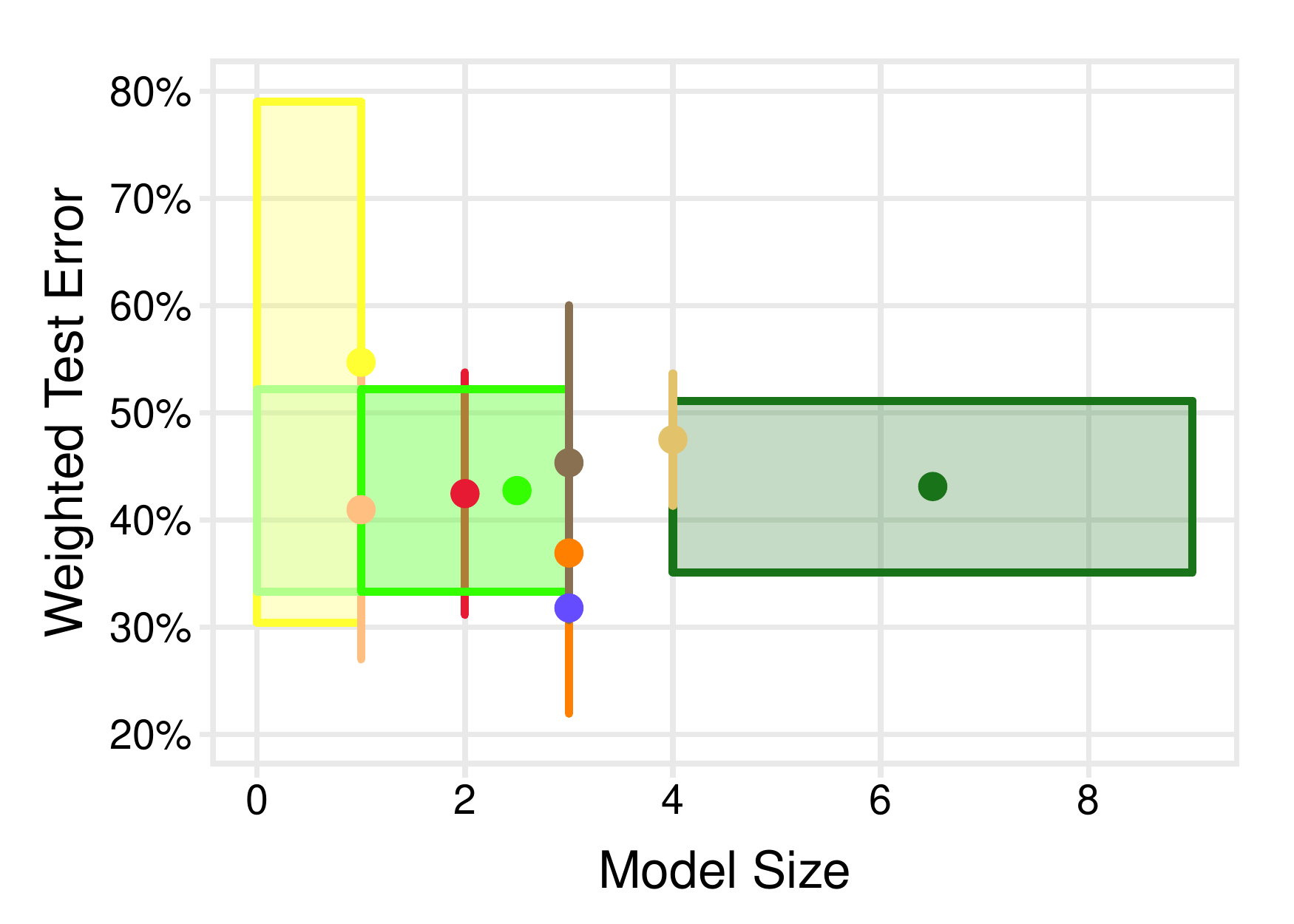}  & \includegraphics[width=\fwidth]{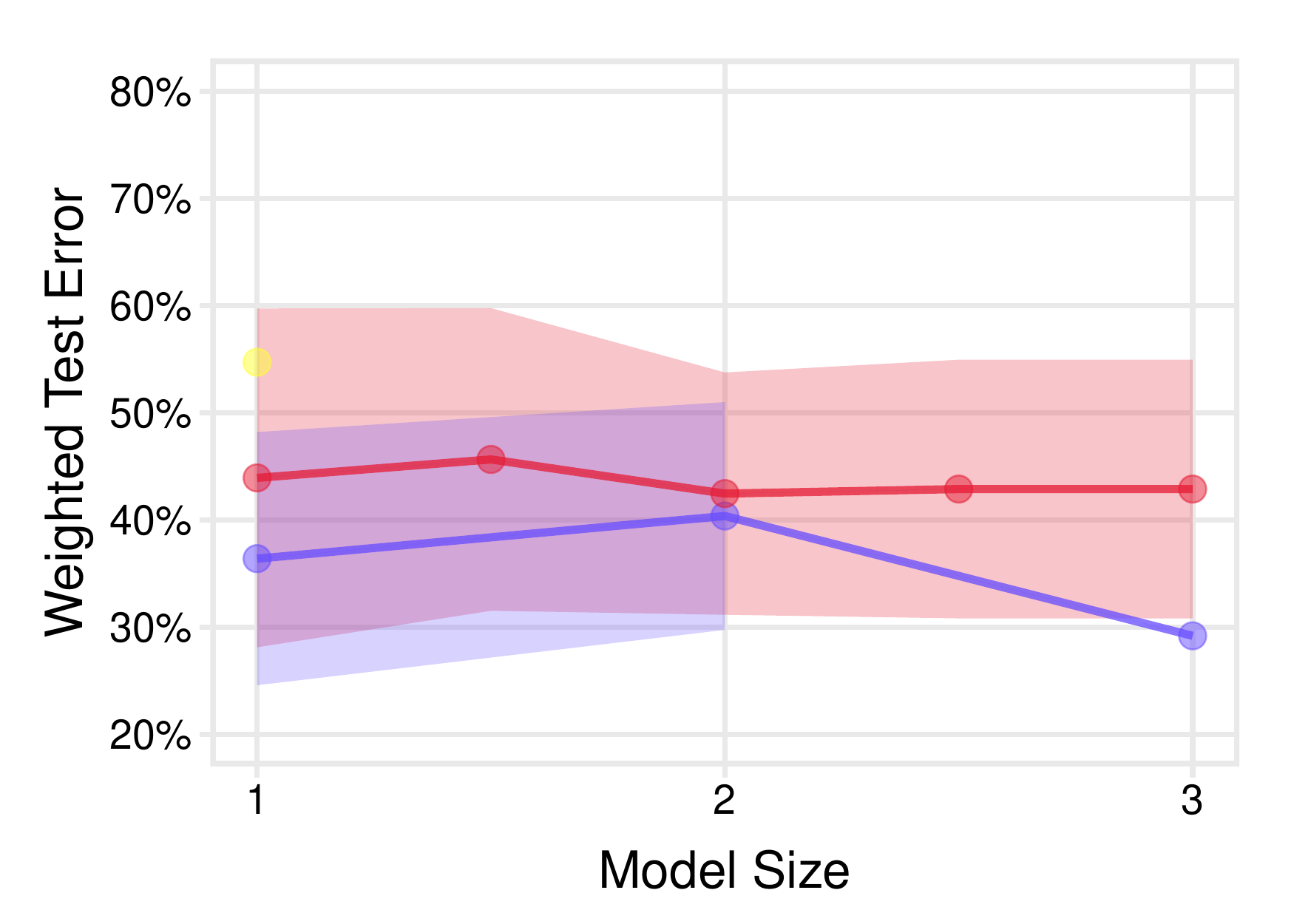} \end{tabular}
\end{tabular}
\caption{Accuracy and sparsity of all classification methods on all datasets. For each dataset, we plot the performance of models when free parameters are set to values that minimize the mean 10-fold CV error (left), and plot the performance of regularized linear models across the full regularization path (right).}
\label{Fig::ExpPlots1}
\end{figure}
\begin{figure}[htbp]
\centering
\begin{tabular}{c}
\hspace{0.5in} \includegraphics[trim=1in 1in 1in 1in,clip,scale=0.5]{plot_legend} 
\end{tabular}
\begin{tabular}{>{\tiny}m{1cm}c} 
\tabdataname{mammo}     & \begin{tabular}{lr}\includegraphics[width=\fwidth]{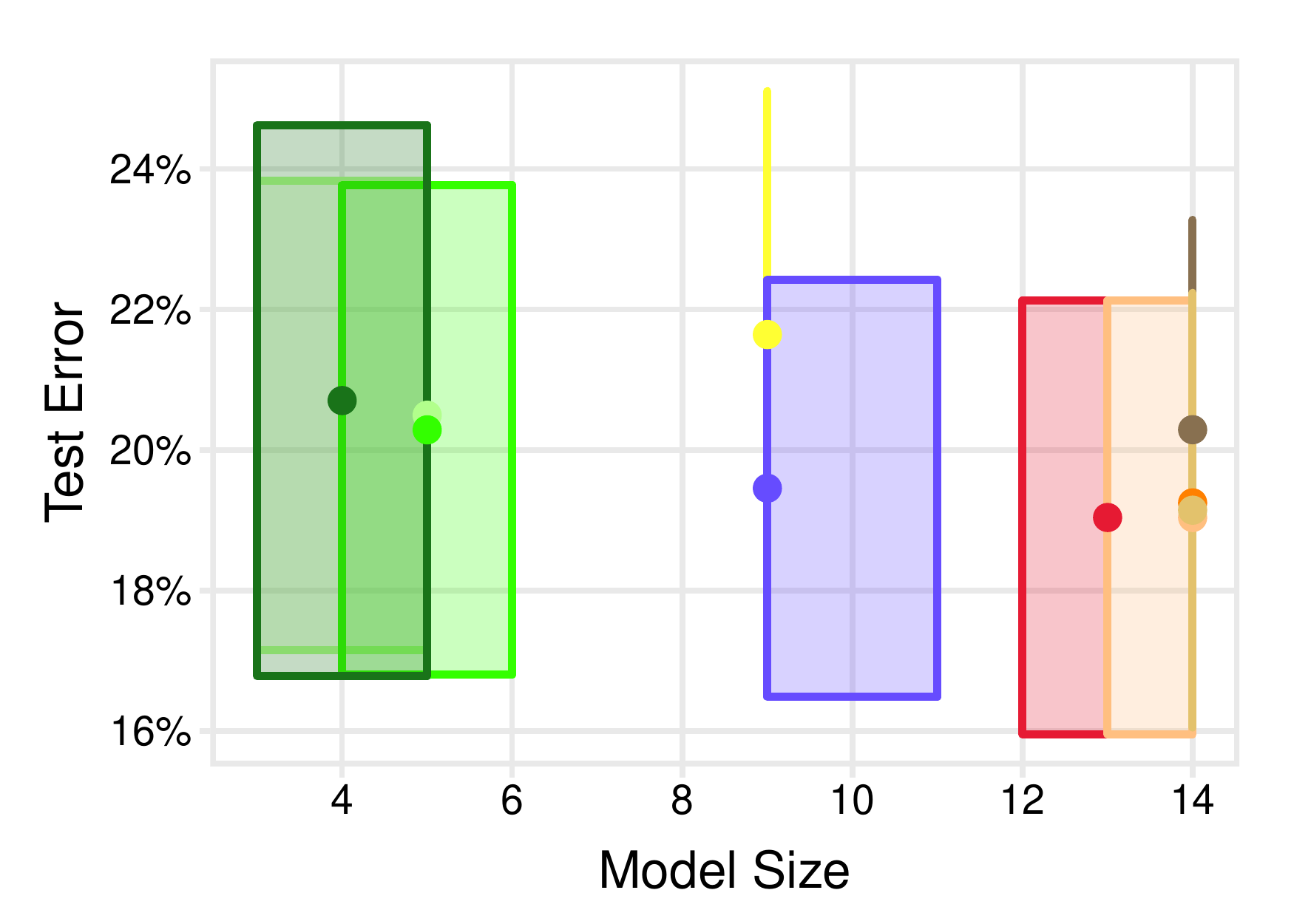}  & \includegraphics[width=\fwidth]{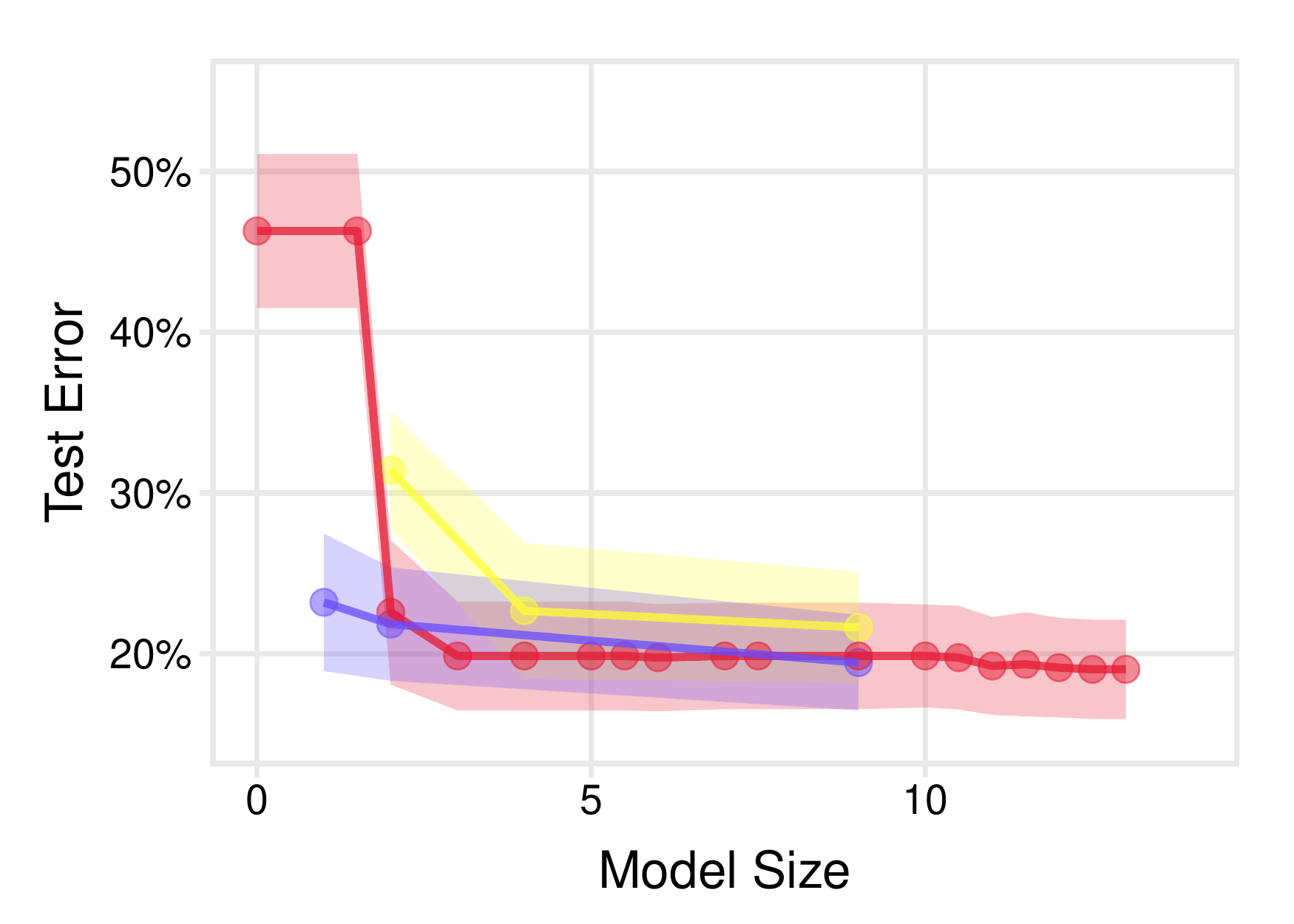} \end{tabular} \\ 
\tabdataname{heart}     & \begin{tabular}{lr}\includegraphics[width=\fwidth]{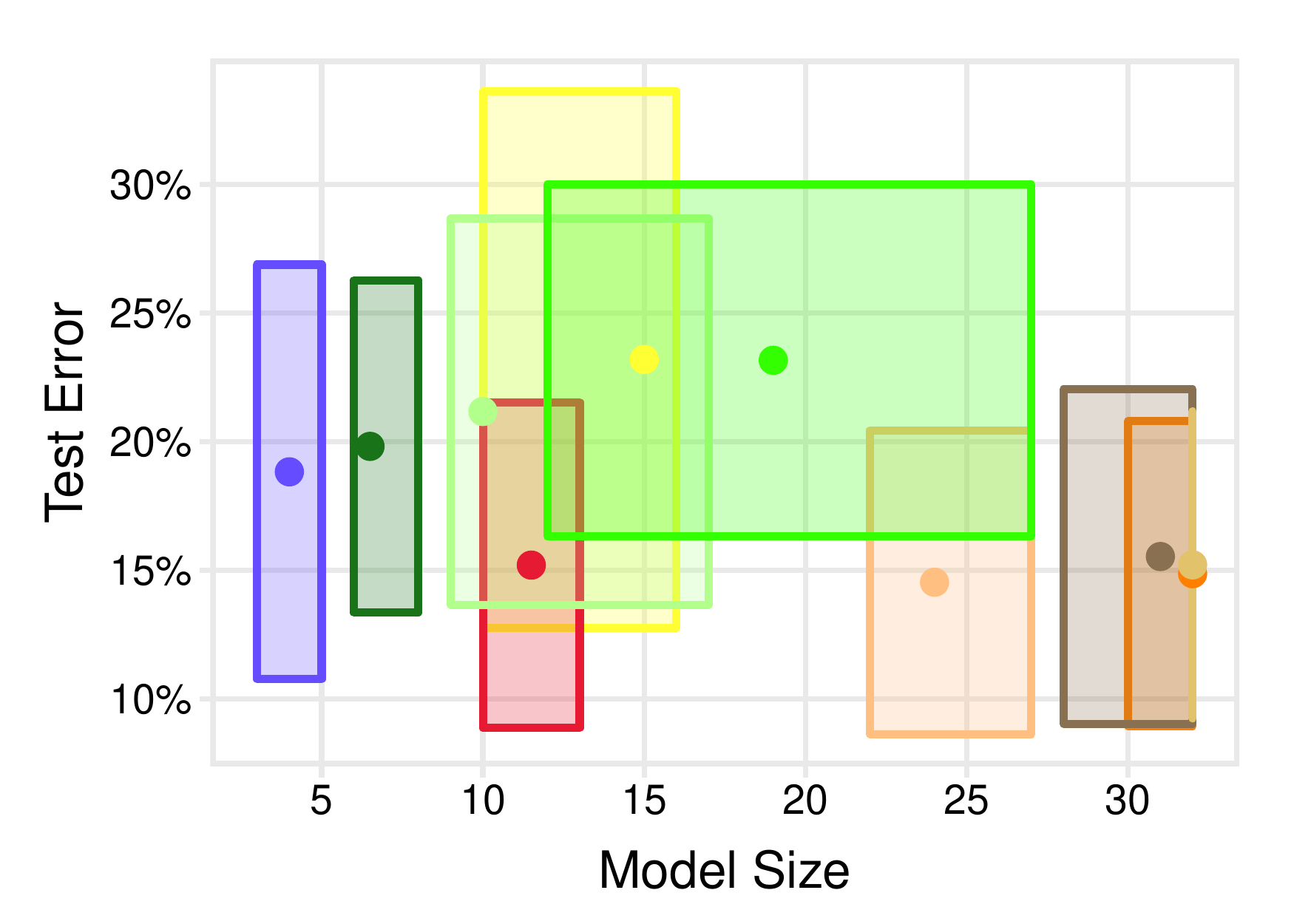}  & \includegraphics[width=\fwidth]{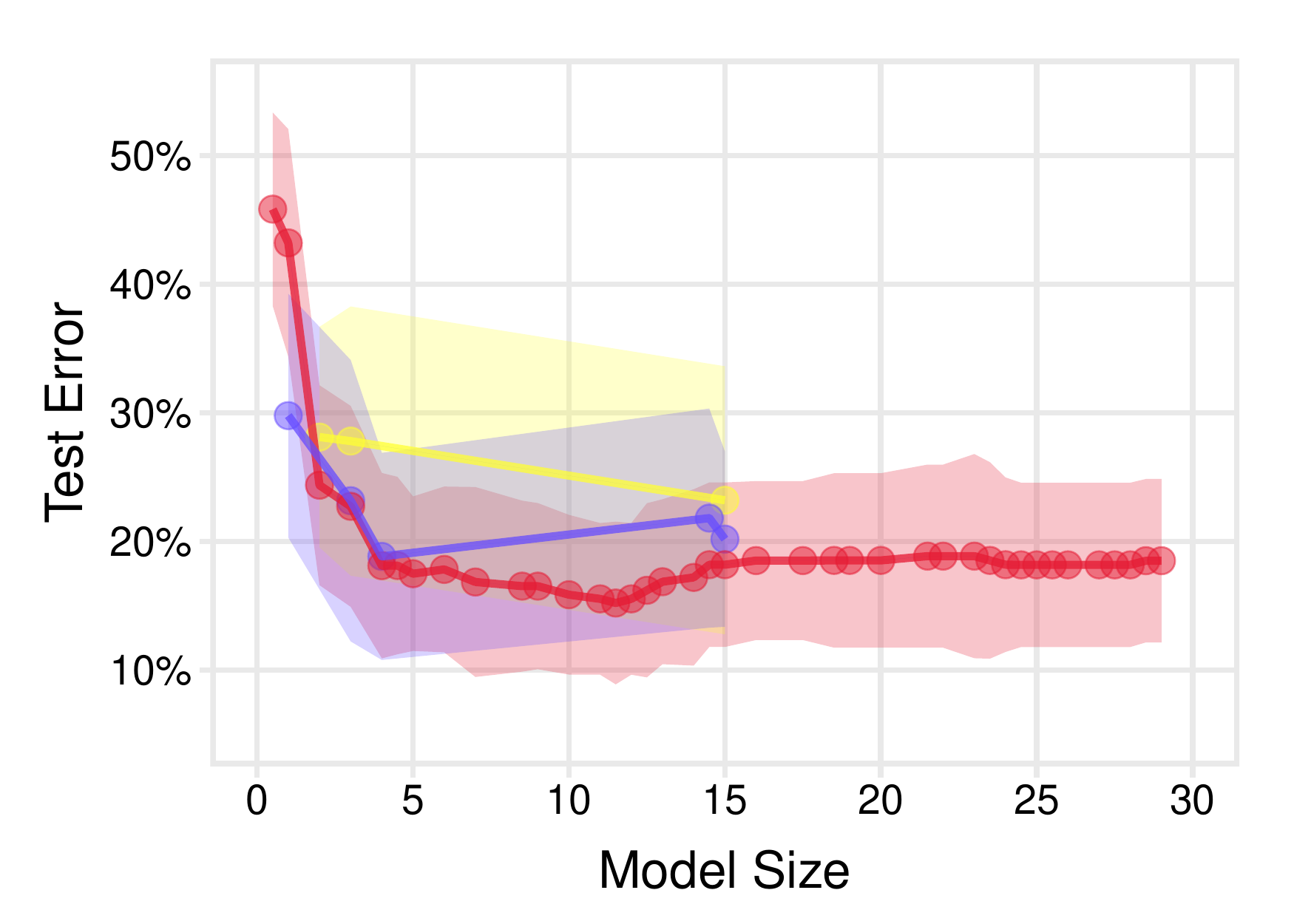} \end{tabular} \\ 
\tabdataname{mushroom}  & \begin{tabular}{lr}\includegraphics[width=\fwidth]{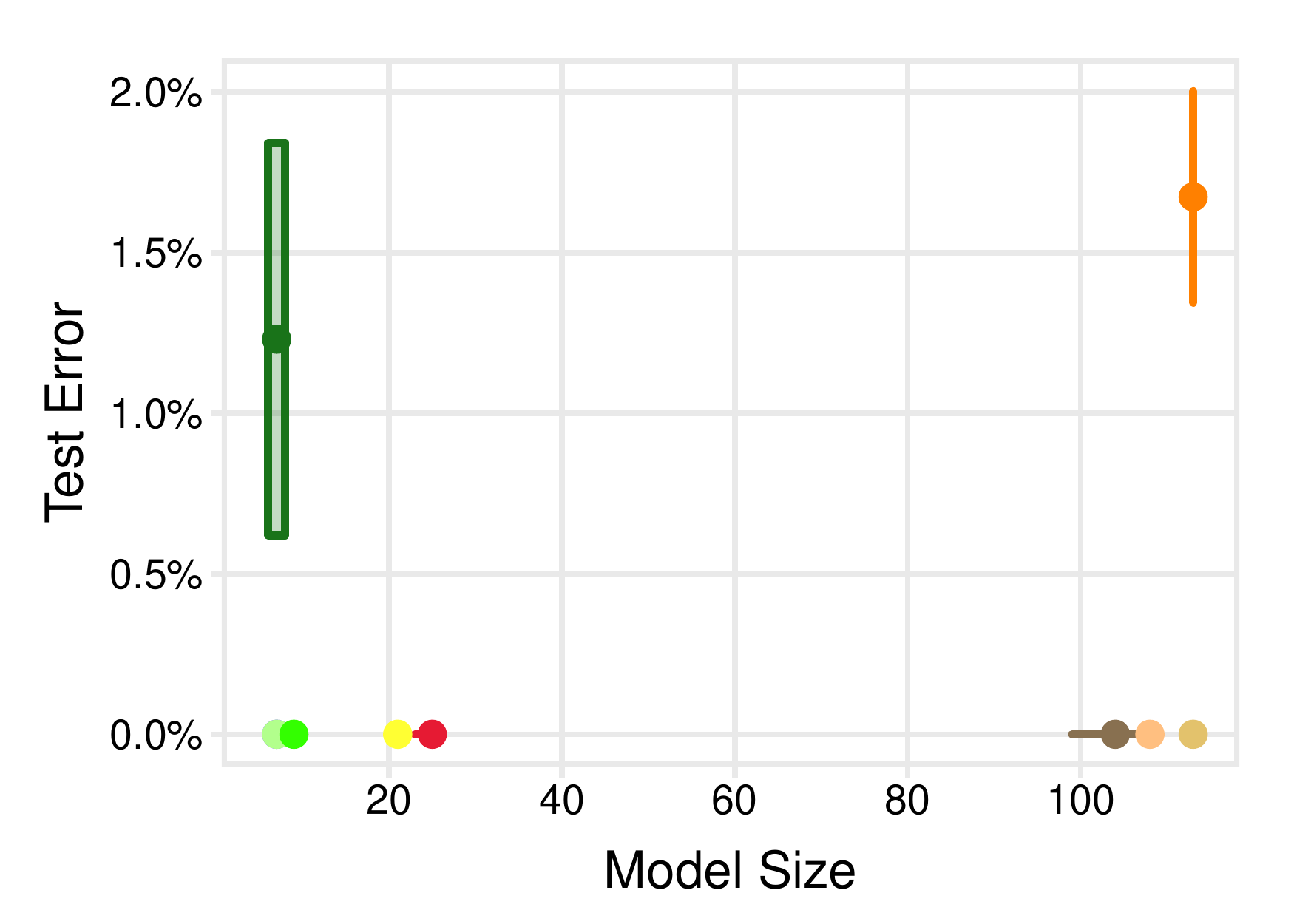}  & \includegraphics[width=\fwidth]{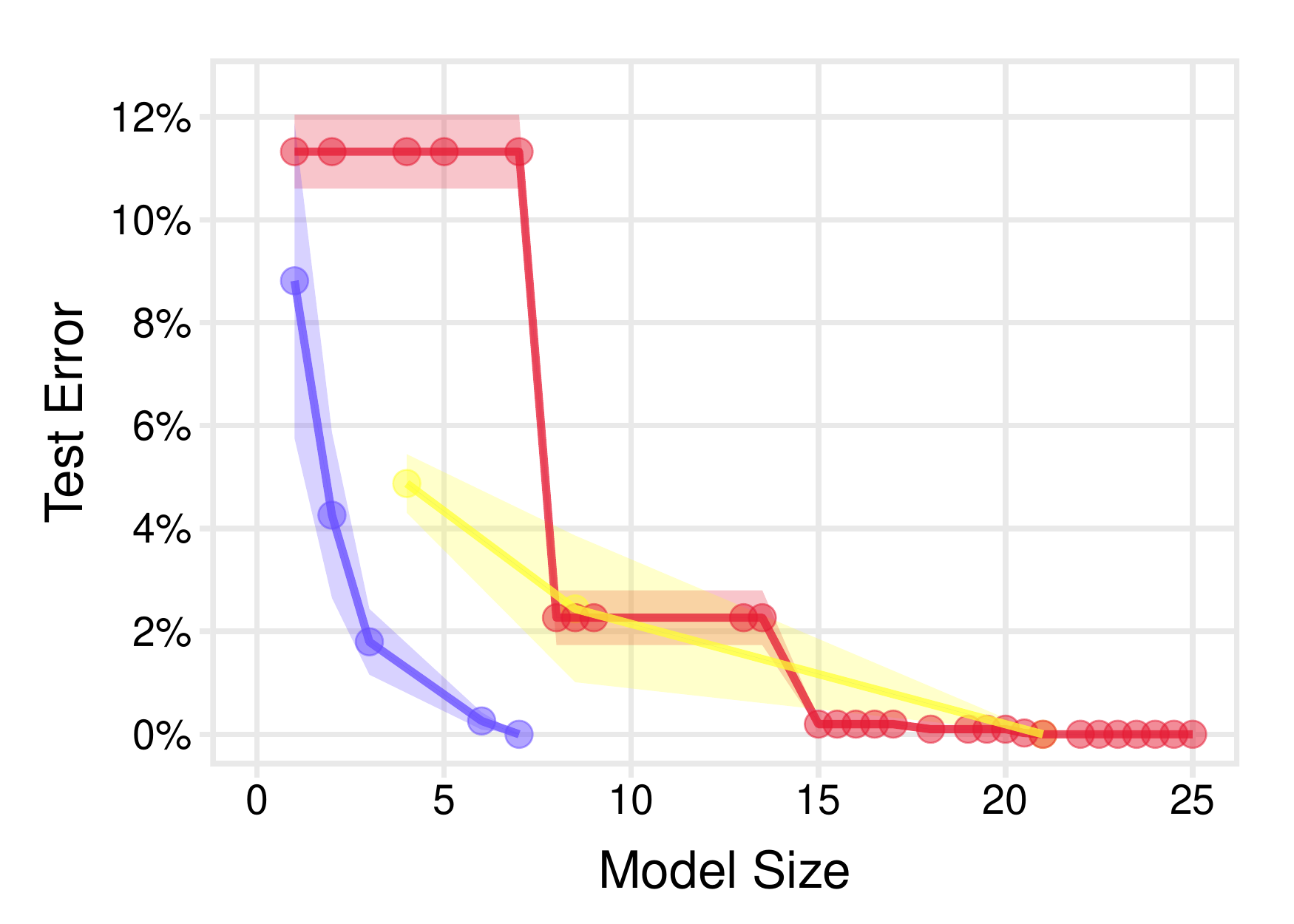} \end{tabular} \\ 
\tabdataname{spambase}  & \begin{tabular}{lr}\includegraphics[width=\fwidth]{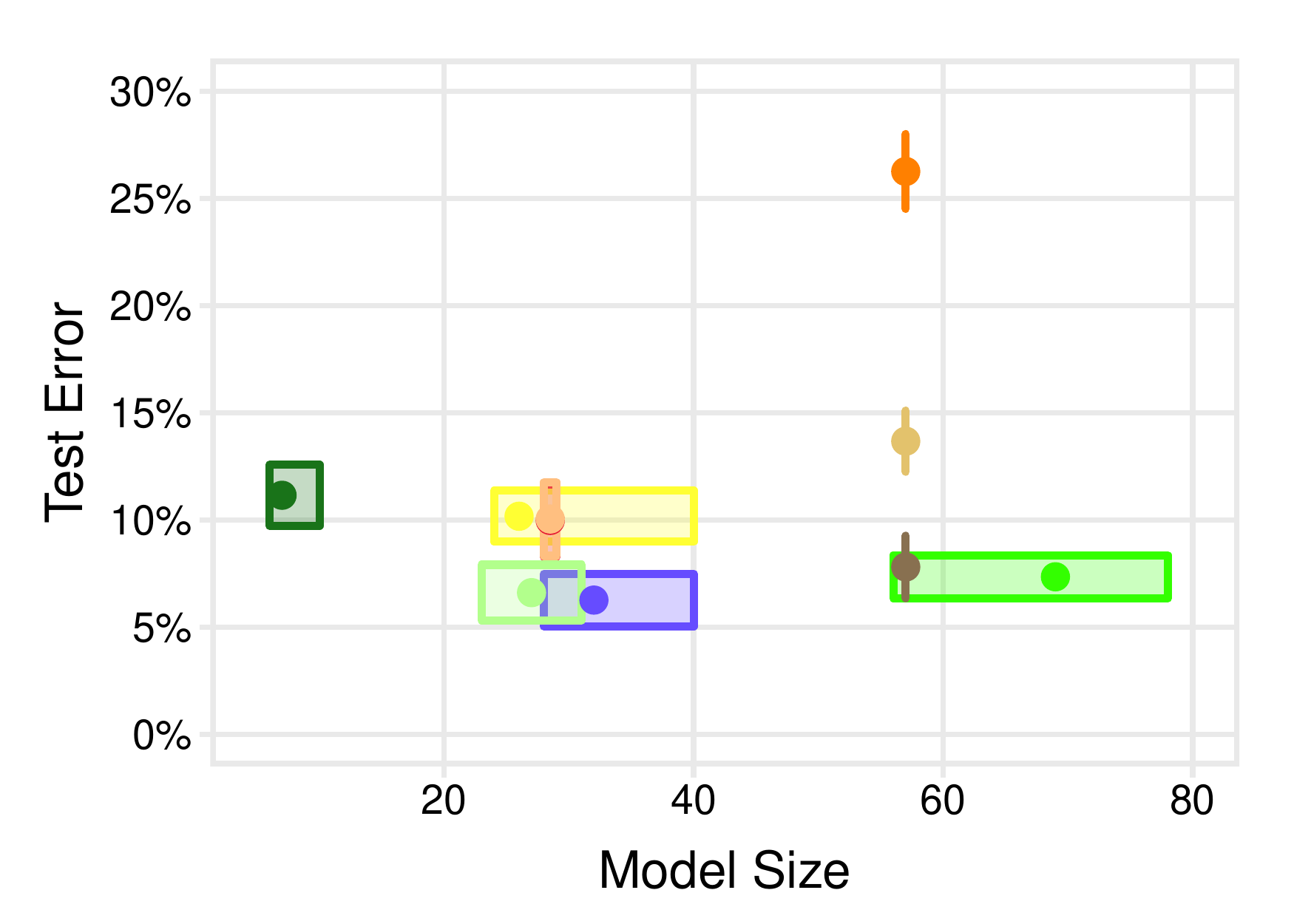}  & \includegraphics[width=\fwidth]{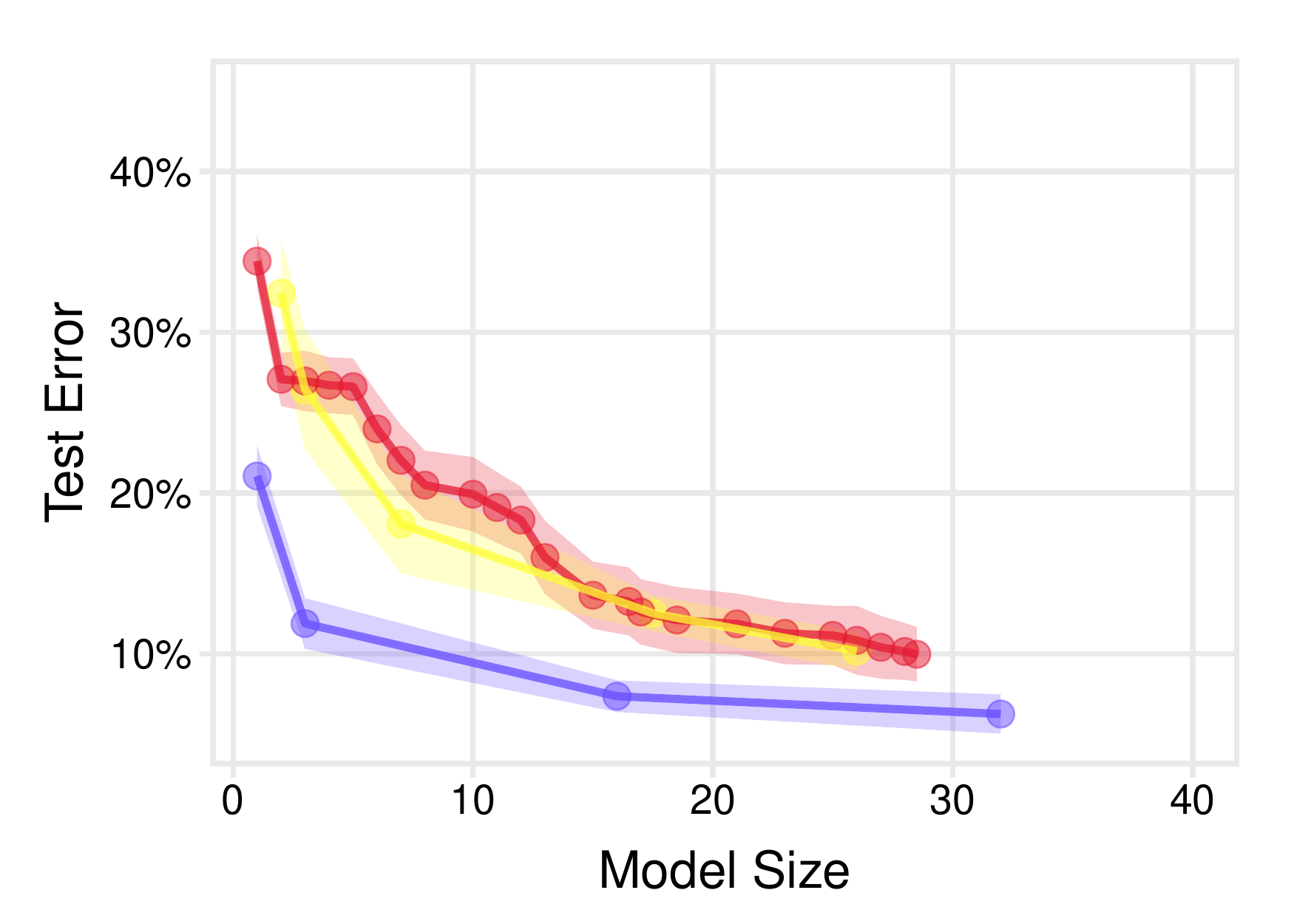} \end{tabular}
\end{tabular}
\caption{Accuracy and sparsity of all classification methods on all datasets. For each dataset, we plot the performance of models when free parameters are set to values that minimize the mean 10-fold CV error (left), and plot the performance of regularized linear models across the full regularization path (right).}
\label{Fig::ExpPlots2}
\end{figure}
\begin{figure}[htbp]
\centering
\includegraphics[width=0.65\textwidth]{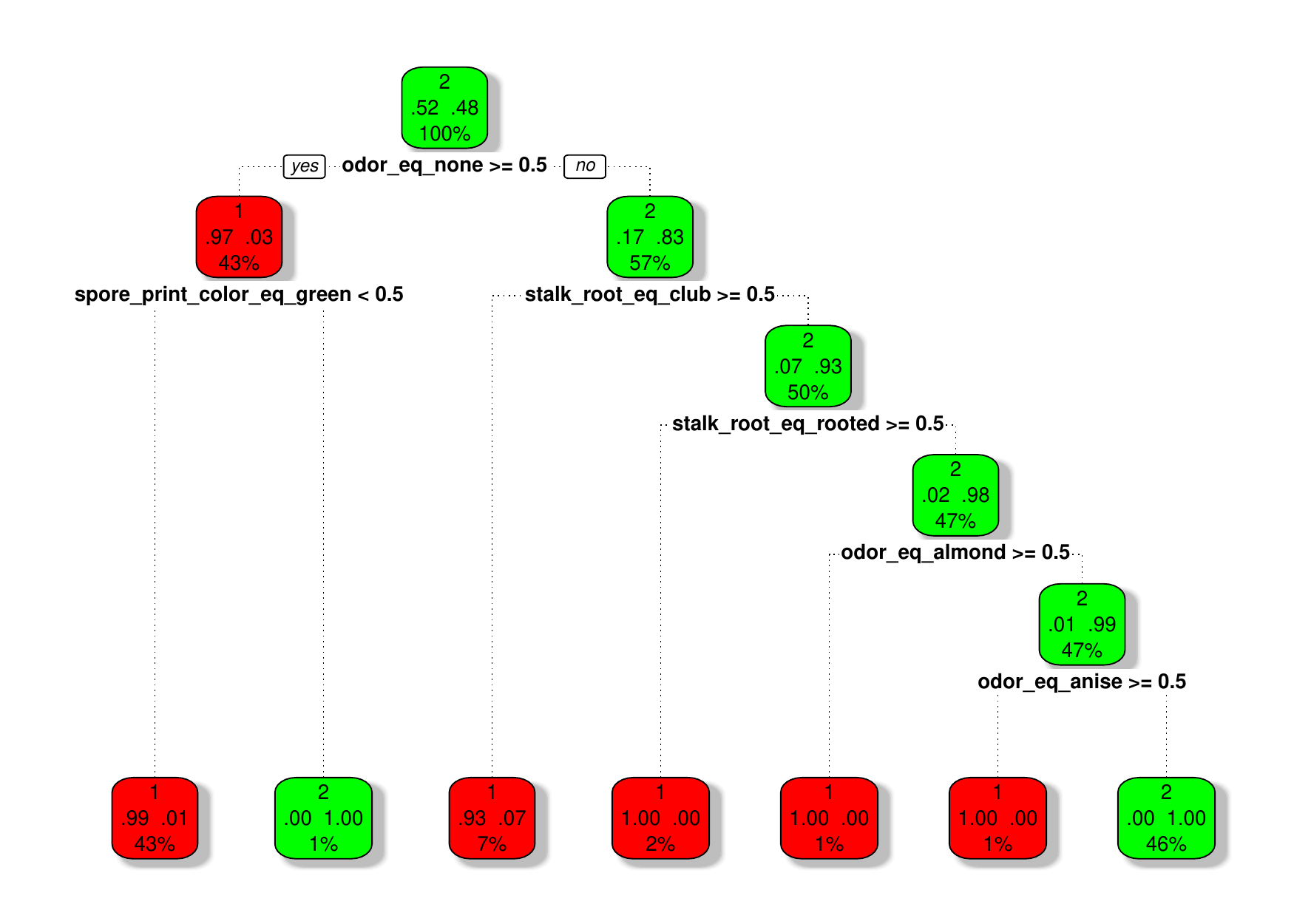} 
\caption{CART decision tree for the \texttt{mushroom} dataset. This model has 7 leaves and a mean 10-fold CV test error of 1.2 $\pm$ 0.6$\%$.}
\label{Fig::MushroomCART}
\end{figure}
\begin{figure}[htbp]
\label{Fig::MushroomLasso}
\centering{
\scriptsize{
\begin{tabularx}{\textwidth}{p{1mm}lp{1mm}lp{1mm}l}
 & $10.86 ~\textrm{spore\_print\_color\_eq\_green}$ & $\scriptsize{+}$ & $4.49 ~\textrm{gill\_size\_eq\_narrow}$ & $\scriptsize{+}$ & $4.29 ~\textrm{odor\_eq\_foul}$ \\ 
 $\scriptsize{+}$ & $2.73 ~\textrm{stalk\_surface\_below\_ring\_eq\_scaly}$ & $\scriptsize{+}$ & $2.60 ~\textrm{stalk\_surface\_above\_ring\_eq\_grooves}$ & $\scriptsize{+}$ & $2.38 ~\textrm{population\_eq\_clustered}$ \\ 
$\scriptsize{+}$ & $0.85 ~\textrm{spore\_print\_color\_eq\_white}$ & $\scriptsize{+}$ & $0.44 ~\textrm{stalk\_root\_eq\_bulbous}$ & $\scriptsize{+}$ & $0.43 ~\textrm{gill\_spacing\_eq\_close}$ \\ 
$\scriptsize{+}$ & $0.38 ~\textrm{cap\_color\_eq\_white}$ & $\scriptsize{+}$ & $0.01 ~\textrm{stalk\_color\_below\_ring\_eq\_yellow}$ & $\scriptsize{-}$ & $8.61 ~\textrm{odor\_eq\_anise}$ \\ 
 $\scriptsize{-}$ & $8.61 ~\textrm{odor\_eq\_almond}$ & $\scriptsize{-}$ & $8.51 ~\textrm{odor\_eq\_none}$ & $\scriptsize{-}$ & $0.53 ~\textrm{cap\_surface\_eq\_fibrous}$ \\ 
 $\scriptsize{-}$ & $0.25 ~\textrm{population\_eq\_solitary}$ & $\scriptsize{-}$ & $0.21 ~\textrm{stalk\_surface\_below\_ring\_eq\_fibrous}$ & $\scriptsize{-}$ & $0.09 ~\textrm{spore\_print\_color\_eq\_brown}$ \\ 
 $\scriptsize{-}$ & $0.00 ~\textrm{cap\_shape\_eq\_convex}$ & $\scriptsize{-}$ & $0.00 ~\textrm{gill\_spacing\_eq\_crowded}$ & $\scriptsize{-}$ & $0.00 ~\textrm{gill\_size\_eq\_broad}$ \\ 
 $\scriptsize{+}$ & $0.25$ &  &  &  &  \\ 
  \end{tabularx}  
  \caption{Lasso score function for the \texttt{mushroom} dataset. This model uses 21 coefficients and has a 10-fold CV test error of 0.0 $\pm$ 0.0$\%$.}
}
}
\end{figure}
\begin{figure}[htbp]
\centering
\renewcommand{\arraystretch}{1.1}
\small{\textbf{PREDICT MUSHROOM IS POISONOUS IF SCORE $> 3$}} \\ 
\vspace{0.5em}
\begin{tabular}{|l l  c | c |}
   \hline
1.   & $spore\_print\_color = green$ & 4 points & $\phantom{+}\quad\cdots\cdots$ \\ 
  2. & $stalk\_surface\_above\_ring = grooves$  & 2 points & $+\quad\cdots\cdots$ \\ 
  3. & $population = clustered$ & 2 points & $+\quad\cdots\cdots$ \\ 
  4. & $gill\_size = broad$ & -2 points & $+\quad\cdots\cdots$ \\ 
  5. & $odor \in \{none, almond, anise\}$ & -4 points & $+\quad\cdots\cdots$ \\ 
   \hline
 & \small{\textbf{ADD POINTS FROM ROWS 1-5}} & \small{\textbf{SCORE}} & $=\quad\cdots\cdots$ \\ 
   \hline
\end{tabular}
\caption{SLIM scoring system for the \texttt{mushroom} dataset. This model has 7 coefficients and a mean 10-fold CV test error of  0.0 $\pm$ 0.0 $\%$. Note that we were able to combine coefficients for categorical variables.} 
\label{Fig::MushroomSLIM}
\end{figure}
\begin{figure}[htbp]
\centering
\small{
\begin{tabular}{c}
  \bfcell{c}{PREDICT MUSHROOM IS POISONOUS \\
 IF AT LEAST 11 OF THE FOLLOWING RULES ARE TRUE}
 \\ 
   \hline
  $odor \in \{pungent,foul,creosote\}$ \\ 
  $habitat \in \{woods,waste,leaves\}$ \\ 
  $gill\_size = narrow$ \\ 
  $gill\_color =  brown$ \\ 
  $stalk\_color\_above\_ring = pink$ \\ 
  $stalk\_surface\_above\_ring = grooves$ \\ 
  $stalk\_surface\_below\_ring = scaly$ \\ 
  $population = scattered$ \\ 
  $odor \neq creosote$ \\ 
  $odor \neq fishy$ \\ 
  $odor \neq spicy$ \\ 
  $gill\_color \neq purple$ \\ 
  $veil\_color \neq yellow$ \\ 
  $spore\_print\_color \neq orange$ \\ 
  $spore\_print\_color \neq buff$ \\ 
  $population \neq scattered$ \\ 
  $population \neq numerous$
  \end{tabular}
  }
\caption{M-of-N rule table for the \texttt{mushroom} dataset. This model has 21 coefficients and a mean 10-fold CV test error of 0.0 $\pm$ 0.0$\%$. Note that we were able to combine coefficients for categorical variables.}
\label{Fig::MushroomMN}
\end{figure}
\FloatBarrier
\section{Conclusions}\label{Sec::Conclusions}
Interpretability is a crucial aspect of applied predictive modeling that has been notoriously difficult to address because it is subjective and multifaceted. 

In this paper, we introduced an integer programming framework to help practitioners to address this problem in a general setting by directly controlling important qualities related to the accuracy and interpretability of their models. We showed how this framework could create many types of interpretable predictive models, and presented specially designed optimization methods to assist with scalability, called loss decomposition and data reduction. We presented theoretical bounds that related the accuracy of our models to the coarseness of the discretization using number-theoretic concepts. Lastly, we presented extensive experimental results to illustrate the flexibility our approach, and its performance relative to state-of-the-art methods.

The major benefit of our framework over existing methods is that we avoid approximations that are designed to achieve faster computation, but can be detrimental to interpretability. Approximations such as surrogate loss functions and $\lone$-regularization hinder the accuracy and interpretability of models as well as the ability of practitioners to control these qualities. Such approximations are no longer needed for many datasets, since using current integer programming software, we can now train classification models for real-world problems, sometimes with hundreds of thousands of examples. Integer programming software also caters to practitioners in other ways, by allowing them to seamlessly benefit from periodic computational improvements without revising their code, and to choose from a pool of interpretable models by mining feasible solutions.

\clearpage
\appendix
\section{Proofs of Theorems}\label{Appendix::Proofs}
In this appendix, we include proofs to the Theorems from Sections \ref{Sec::Methods} and \ref{Sec::Theory}.

\proof{Proof of Theorem \ref{Thm::DataReduction}}
Let us denote the set of classifiers whose objective value is less or equal to $\tilde{Z}(\tilde{f}^*;\data_N)$ as
\begin{align*}
\tilde{\F}^{\varepsilon} = \left \{ f \in \tilde{\F} \; \Big | \; \tilde{Z}(f;\data_N) \leq \tilde{Z}(\tilde{f};\data_N) + \varepsilon \right\}.
\end{align*}
In addition, let us denote the set of points that have been removed by the data reduction algorithm
\begin{align*}
\mathcal{S} = \data_N \setminus \data_M.
\end{align*}
By definition, data reduction only removes an example if its sign is fixed. 
This means that $\sign{f(\xb_i)}=\sign{\tilde{f}(\xb_i)}$ for all $i \in \mathcal{S}$ and $f\in\tilde{\F}^{\varepsilon}.$ Thus, we can see that for all classifiers $f \in \tilde{\F}^{\varepsilon}$, 
\begin{align}
\label{Eq:ReductionConstant}
Z(f;\data_N) = Z(f;\data_M)+\sum_{i\in\mathcal{S}} \indic{y_if(\xb_i)\leq 0}= Z(f;\data_M)+\sum_{i\in\mathcal{S}} \indic{y_i\tilde{f}(\xb_i)\leq 0} =Z(f;\data_M)+C.
\end{align}

We now proceed to prove the statement in \eqref{Eq::ReductionWTS}. When $\mathcal{S}=\emptyset$, then $\data_N = \data_M$, and \eqref{Eq::ReductionWTS} follows trivially. When, $\mathcal{S}\neq\emptyset$, we note that
\begin{align}
\F^*=\argmin_{f\in\F} Z(f;\data_N) &= \argmin_{f\in\F} Z(f;\data_M \cup \mathcal{S}), \nonumber \\
&= \argmin_{f\in\F} Z(f;\data_M) + Z(f; \mathcal{S}), \nonumber \\
&= \argmin_{f\in\F} Z(f;\data_M) + C, \label{Eq:ReductionFinal} \\ 
&= \argmin_{f\in\F} Z(f;\data_M). \nonumber
\end{align}
Here, the statement in \eqref{Eq:ReductionFinal} follows directly from \eqref{Eq:ReductionConstant}.
\qed 
\endproof

\proof{Proof of Theorem \ref{Thm::ReductionSufficientConditions}}
We assume that we have found a proxy function, $\psi$, that satisfies conditions I--IV and choose $C_{\psi} > 2\varepsilon$. 

Our proof uses the following result: if 
$\| \lmip - \lcvx \| > C_{\lambdab}$
then $\lmip$ cannot be a minimizer of $\Zmip{\lambdab}$ because 
this would lead to a contradiction with the definition of $\lmip$. 
To see that this result holds, we use condition III with $\lambdab=\lmip$ to see that $\|\lmip - \lcvx\|>C_{\lambdab}$ implies $\Zcvx{\lmip} - \Zcvx{\lcvx} >C_{\psi}$. Thus,
\begin{align} 
\Zcvx{\lcvx} + C_{\psi} &< \Zcvx{\lmip} \notag \\ 
\Zcvx{\lcvx} + C_{\psi} &< \Zmip{\lmip} + \varepsilon \label{lemmaeq1} \\ 
\Zcvx{\lcvx} + C_{\psi} - \varepsilon &< \Zmip{\lmip} \notag \\ 
\Zcvx{\lcvx} + C_{\psi} - \varepsilon &< \Zcvx{\lmip} \label{lemmaeq3} \\
\Zcvx{\lcvx} + \varepsilon &< \Zcvx{\lmip}.\label{lemmaeq2}
\end{align}
Here the inequality in \eqref{lemmaeq1} follows from condition IV, the inequality in \eqref{lemmaeq3} follows from condition I, and the inequality in \eqref{lemmaeq2} follows from our choice that $C_{\psi} > 2\varepsilon$.

We now proceed by looking at the LHS and RHS of the inequality in \eqref{lemmaeq2} separately. Using condition I on the LHS of \eqref{lemmaeq2} we get that:
\begin{align}
\Zmip{\lcvx} + \varepsilon \leq \Zcvx{\lcvx} + \varepsilon. \label{lemmaeq3}
\end{align}
Using condition IV on the RHS of \eqref{lemmaeq2} we get that:
\begin{align}
\Zcvx{\lmip} \leq \Zmip{\lmip} + \varepsilon. \label{lemmaeq4}
\end{align}
Combining the inequalities in \eqref{lemmaeq3}, \eqref{lemmaeq2} and  \eqref{lemmaeq4}, we get that:
\begin{align}
\Zmip{\lcvx} < \Zmip{\lmip}. \label{lemmaeq5}
\end{align}
The statement in \eqref{lemmaeq5} is a contradiction of the definition of $\lmip$. Thus, we know that our assumption was incorrect and thus $\| \lmip - \lcvx \| \leq C_{\lambdab}$. 
We plug this into the Lipschitz condition II as follows:
\begin{align} 
\Zcvx{\lmip} - \Zcvx{\lcvx} & \leq L \| \lmip - \lcvx \| < L C_{\lambdab},  \notag\\ 
\Zcvx{\lmip} & < L C_{\lambdab} + \Zcvx{\lcvx}. \notag
\end{align}
Thus, we have satisfied the level set condition with $\varepsilon = LC_{\lambdab}$. 
\qed

\endproof

\proof{Proof of Theorem \ref{Thm::MinMarginBound}}
%
%
%

We use normalized versions of the vectors, $\rhob/\vnorm{\rhob}_2$ and $\lambdab/\Lambda$ because the 0--1 loss is scale invariant:
\begin{align*}
 \sum_{i=1}^N \indic{y_i\lambdab^T\xb_i\leq 0} &= \sum_{i=1}^N \indic{y_i\frac{\lambdab^T\xb_i}{\Lambda}\leq 0}, \\
 \sum_{i=1}^N \indic{y_i \rhob^T \xb_i \leq 0} &=  \sum_{i=1}^N \indic{y_i\frac{\rhob^T\xb_i}{\vnorm{\rhob}_2}\leq 0}.
\end{align*}

We set $\Lambda > \frac{X_{\max}\sqrt{P}}{2 \gamma_{\min}}$ as in \eqref{Eq::MinMarginLambda}. Given $\Lambda$, we then define $\lambdab/\Lambda$ element-wise, so that $\lambda_j/\Lambda$ is equal to $\rho_j/\vnorm{\rhob}_2$ rounded to the nearest $1/\Lambda$ for each $j= 1,\ldots,P$.

We first show that our choice of $\Lambda$ and $\lambdab$ ensures that the difference between the margin of $\rhob/\vnorm{\rhob}_2$ and the margin of $\lambdab/\Lambda$ on all training examples is always less than the minimum margin of $\rhob/\vnorm{\rhob}_2$, defined as $\gamma_{\min} = \min_i \frac{|\rhob^T\xb_i|}{\vnorm{\rhob}_2}$. This statement follows from the fact that, for all $i$:
\begin{align}
\left| \frac{\lambdab^T \xb_i}{\Lambda}-\frac{\rhob^T\xb_i}{\vnorm{\rhob}_2} \right| ~
\leq ~ & \left\|\frac{\lambdab}{\Lambda} - \frac{\rhob}{\vnorm{\rhob}_2}\right\|_2\|\xb_i\|_2 \label{ThmPf::A} \\
= ~         & \left( \sum_{j=1}^P  \left| \frac{\lambda_j}{\Lambda} - \frac{\rho_j}{\vnorm{\rhob}_2} \right|^2 \right)^{1/2}\|\xb_i\|_2 \nonumber \\ 
\leq ~      & \left(\sum_{j=1}^P \frac{1}{(2\Lambda)^2} \right)^{1/2}\|\xb_i\|_2 \label{ThmPf::B} \\
= ~         & \frac{\sqrt{P}}{2\Lambda}X_{\max} \nonumber \\ 
< ~         & \frac{\sqrt{P}X_{\max}}{2\left( \frac{X_{\max}\sqrt{P}}{2\min_i  \frac{|\rhob^T\xb_i|}{\vnorm{\rhob}_2}}         \right) }  \label{ThmPf::C} \\ 
= ~ & \min_i \frac{|\rhob^T\xb_i|}{\vnorm{\rhob}_2}. \label{ThmPf::D}
\end{align}
Here: the inequality in \eqref{ThmPf::A} uses the Cauchy-Schwarz inequality; the inequality in \eqref{ThmPf::B} is due to the fact that the distance between $\rho_j/\vnorm{\rhob}_2$ and $\lambda_j/\Lambda$ is at most $1/2\Lambda$; and the inequality in \eqref{ThmPf::C} is due to our choice of $\Lambda$.

Next, we show that our choice of $\Lambda$ and $\lambdab$ ensures that $\rhob/\vnorm{\rhob}_2$ and $\lambdab/\Lambda$ classify each point in the same way. We consider three cases: first, the case where $\xb_i$ lies on the margin; second, the case where $\rhob$ has a positive margin on $\xb_i$; and third, the case where $\rhob$ has a negative margin on $\xb_i$. 
For the case when $\xb_i$ lies on the margin, $\min_i|\rhob^T\xb_i|=0$ and the theorem holds trivially.
For the case where $\rhob$ has positive margin, $\rhob^T\xb_i>0$, the following calculation using (\ref{ThmPf::D}) is relevant:
\begin{eqnarray*}
\frac{\rhob^T\xb_i}{\vnorm{\rhob}_2} - \frac{\lambdab^T\xb_i}{\Lambda}  \leq \left| \frac{\lambdab^T\xb_i}{\Lambda}-\frac{\rhob^T\xb_i}{\vnorm{\rhob}_2}   \right| 
< \min_i \frac{|\rhob^T\xb_i|}{\vnorm{\rhob}_2}.
\end{eqnarray*}
We will use the fact that for any $i^{'}$, by definition of the minimum: 
\begin{align*}
0 \leq \frac{|\rhob^T\xb_{i^{'}}|}{\vnorm{\rhob}_2}- \min_i \frac{|\rhob^T\xb_i |}{\vnorm{\rhob}_2},
\end{align*}
and combine this with a rearrangement of the previous expression to obtain:
\begin{eqnarray*}
0\leq \frac{|\rhob^T\xb_i|}{\vnorm{\rhob}_2}  -  \min_i \frac{|\rhob^T\xb_i |}{\vnorm{\rhob}_2} = \frac{\rhob^T\xb_i}{\vnorm{\rhob}_2}  -  \min_i \frac{|\rhob^T\xb_i |}{\vnorm{\rhob}_2} < \frac{\lambdab^T\xb_i}{\Lambda}.
\end{eqnarray*}
Thus, we have shown that $\lambdab^T\xb_i>0$ whenever $\rhob^T\xb_i>0$.

For the case where $\rhob$ has a negative margin on $\xb_i$, $\rhob^T\xb_i<0$, we perform an analogous calculation:
\begin{eqnarray*}
\frac{\lambdab^T\xb_i}{\|\lambdab\|_2} - \frac{\rhob^T\xb_i}{\vnorm{\rhob}_2}  \leq \left| \frac{\lambdab^T\xb_i}{\Lambda}-\frac{\rhob^T\xb_i}{\vnorm{\rhob}_2}   \right| 
< \min_i \frac{|\rhob^T\xb_i|}{\vnorm{\rhob}_2}.
\end{eqnarray*}
and then using that $\rhob^T\xb_i<0$,
\begin{eqnarray*}
0\leq \frac{|\rhob^T\xb_i|}{\vnorm{\rhob}_2}  -  \min_i \frac{|\rhob^T\xb_i |}{\vnorm{\rhob}_2} = \frac{-\rhob^T\xb_i}{\vnorm{\rhob}_2}  -  \min_i \frac{|\rhob^T\xb_i |}{\vnorm{\rhob}_2} < -\frac{\lambdab^T\xb_i}{\Lambda}.
\end{eqnarray*}
Thus, we have shown $\lambdab^T\xb_i<0$ whenever $\rhob^T\xb_i<0$.

Putting both the positive margin and negative margin cases together, we find that for all $i$,
\[\indic{y_i\rhob^T\xb_i\leq 0} = \indic{y_i\lambdab^T\xb_i\leq 0}.\] 
Summing over $i$ yields the statement of the theorem. \qed 
\endproof

\proof{Proof of Theorem \ref{Thm::L0Bound}}
We note that $\lambdab = 0$ is a feasible solution to the optimization problem used in the training process as $0 \in \Lset$. Since $\lambdab = 0$ achieves an objective value of $Z(0;\data_N) = 1$, any optimal solution, $\lambdab \in \argmin_{\lambda \in \Lset} Z(\lambdab;\data_N)$, must attain an objective value $Z(\lambdab;\data_N) \leq 1$. This implies
\begin{eqnarray*}
Z(\lambdab;\data_N) \leq 1,\\
 C_0\vnorm{\lambdab}_0 \leq \frac{1}{N}\sum_{i=1}^N \indic{y_i \lambdab^T \xb_i \leq 0}  + C_0\vnorm{\lambdab}_0  \leq 1,\\ 
\vnorm{\lambdab}_0 \leq \frac{1}{C_0},\\
\vnorm{\lambdab}_0 \leq \left\lfloor \frac{1}{C_0} \right\rfloor. 
\end{eqnarray*}
The last line uses that $\|\lambdab\|_0$ is an integer. 

Thus, $\mathcal{H}_{P,C_0}$ is large enough to contain all minimizers of $Z(\cdot;\data_N)$ for any $\data_N$. The result then follows from applying Theorem \ref{Thm::NormalBound}.
\endproof

\clearpage
\section{Loss Constraints and Loss Cuts}\label{Appendix::LossConstraints}

\begin{table}[htbp]
\centering
\renewcommand*{\arraystretch}{1.2}
\begin{tabular}{llc}
\toprule
\bf{Loss Function} &  \bf{Loss Variable} & \bf{Loss Constraints} \\ 
\toprule
0--1 & $\loss_i  = \indic{y_i \lambdab^T\xb_i \leq 0}$ & $M_i \loss_i \geq \gamma - y_i \lambdab^T\xb_i $ \\ 
\midrule
Hinge &  $\loss_i = \max (0, 1- y_i \lambdab^T\xb_i)$ &  $\loss_i \geq 1 - y_i \lambdab^T\xb_i$ \\
\midrule
Quadratic & $\loss_i = (1 - y_i\xb_i\cdot \lambdab)^2$ & -  \\
\midrule
Logistic & $\loss_i = \log (1 + \exp(- y_i \lambdab^T\xb_i))$ & -   \\
\midrule
Exponential & $\loss_i = \exp(1- y_i \lambdab^T\xb_i)$ & - \\
\bottomrule
\end{tabular}
\caption{Loss constraints for popular loss functions.}
\label{Table::LossFunctions}
\end{table}

\begin{table}[htbp]
\renewcommand{\arraystretch}{1.5}
\centering
\begin{tabular}{lcc}
\toprule
\bf{Loss Function} &  $\Loss{\lambdab;\data_N}$ & $\dLoss{\lambdab}$ \\ 
\toprule
Hinge Loss &  $\frac{1}{N}\sum_{i=1}^N \max (0, 1- y_i \lambdab^T\xb_i)$ &  $\frac{1}{N}\sum_{i=1}^N -y_i\xb_i^T \indic{y_i\lambdab^T\xb_i\geq 1}$ \\
Quadratic Loss & $\frac{1}{N} \sum_{i=1}^N (1 - y_i\xb_i\cdot \lambdab)^2$ & $\frac{1}{N}\sum_{i=1}^N -2y_i\xb_i^T  (1 - y_i\xb_i\cdot \lambdab)^2$  \\ 
Logistic Loss & $\frac{1}{N} \sum_{i=1}^N \log (1 + \exp(- y_i \lambdab^T\xb_i))$ & $\frac{1}{N}\sum_{i=1}^N \exp(-y_i\lambdab^T\xb_i)$ \\ 
Exponential Loss & $\frac{1}{N} \sum_{i=1}^N \exp(1- y_i \lambdab^T\xb_i)$ & $\frac{1}{N} \sum_{i=1}^N -y_i\xb_i \exp(1- y_i \lambdab^T\xb_i)$ \\ 
\bottomrule
\end{tabular}
\caption{The value and subgradient of the aggregate loss for popular loss functions. These quantities are computed by the oracle function for $\lambdab$, and used to produce cutting planes to the aggregate loss function in the proxy problem at each iteration of a decomposition algorithm.}
\label{Table::LossCuts}
\end{table}
\vspace{3in}

.

\clearpage
\section{Progress of Decomposition Algorithm}\label{Appendix::ProgressOfDecomposition}
In this appendix, we provide an iteration-by-iteration overview of the Benders' decomposition algorithm when we applied it to train the simulated dataset from Section \ref{Sec::Decomposition} with $N= 10000000$ examples. In this case, the algorithm converged after 43 iterations, taking 310.3 seconds. Here, the IP solver only used 0.3 seconds to provide the oracle function with feasible values of $\lambdab$, and the oracle function used 310.0 seconds to compute cutting planes to the aggregate loss function at $\lambdab$. 
\begin{figure}[htbp]
\centering
\includegraphics[scale=0.4,trim=0mm 0mm 0mm 0mm,clip]{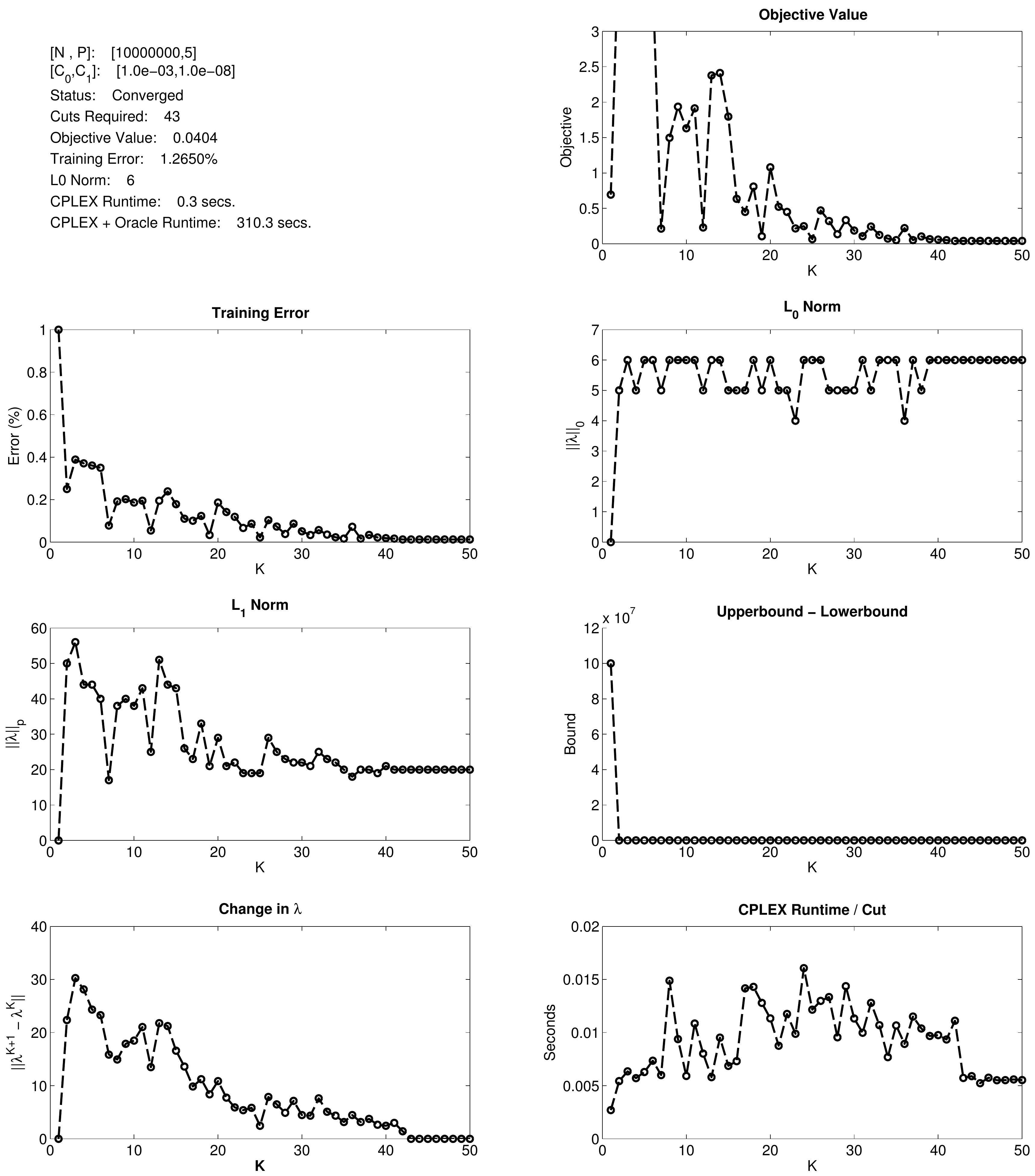} 
\end{figure}
\clearpage
\section{Additional Models for Sleep Apnea Dataset}\label{Appendix::SleepApneaModels}

In this appendix, we include the best models that we were able to produce for the sleep apnea demonstration in \ref{Sec::Demonstrations} using methods that were unable to fulfill the full set of model requirements. Summary statistics associated with each model can be found in Table \ref{Table::DemoTableMaxFPR}. 
\begin{figure}[htbp]
{\scriptsize
\begin{tabularx}{\textwidth}{p{1mm}lp{1mm}lp{1mm}l}
  & $0.24 ~\textrm{htn}$ & $\scriptsize{+}$ & $0.19 ~\textrm{male}$ & $\scriptsize{+}$ & $0.15 ~\textrm{snoring\_2}$ \\ 
 $\scriptsize{+}$ & $0.13 ~\textrm{snoring}$ & $\scriptsize{+}$ & $0.08 ~\textrm{bmi\_geq\_30}$ & $\scriptsize{+}$ & $0.07 ~\textrm{age\_geq\_60}$ \\ 
 $\scriptsize{+}$ & $0.05 ~\textrm{bmi\_geq\_25}$ & $\scriptsize{-}$ & $0.19 ~\textrm{female}$ & $\scriptsize{-}$ & $0.05 ~\textrm{bmi\_lt\_25}$ \\ 
 $\scriptsize{-}$ & $0.49$ &  &  &  &  \\ 
  \end{tabularx}
}
\caption{Elastic Net score function for the \textds{apnea} dataset.}
\end{figure}
\begin{figure}[htbp]
{\centering
{\scriptsize
\begin{tabularx}{\textwidth}{p{1mm}lp{1mm}lp{1mm}l}
  & $0.30 ~\textrm{age\_geq\_60}$ & $\scriptsize{+}$ & $0.29 ~\textrm{htn}$ & $\scriptsize{+}$ & $0.28 ~\textrm{male}$ \\ 
 $\scriptsize{+}$ & $0.23 ~\textrm{snoring\_2}$ & $\scriptsize{+}$ & $0.22 ~\textrm{snoring}$ & $\scriptsize{+}$ & $0.20 ~\textrm{bmi\_geq\_25}$ \\ 
 $\scriptsize{+}$ & $0.19 ~\textrm{retired}$ & $\scriptsize{+}$ & $0.18 ~\textrm{bmi\_geq\_30}$ & $\scriptsize{+}$ & $0.18 ~\textrm{bmi\_geq\_35}$ \\ 
 $\scriptsize{+}$ & $0.18 ~\textrm{stopbreathing}$ & $\scriptsize{+}$ & $0.17 ~\textrm{age\_geq\_30}$ & $\scriptsize{+}$ & $0.14 ~\textrm{bmi\_geq\_40}$ \\ 
 $\scriptsize{+}$ & $0.13 ~\textrm{diabetes}$ & $\scriptsize{+}$ & $0.12 ~\textrm{stopbreathing\_2}$ & $\scriptsize{+}$ & $0.09 ~\textrm{ESS5\_ge\_2}$ \\ 
$\scriptsize{+}$ & $0.09 ~\textrm{nocturnal}$ & $\scriptsize{+}$ & $0.08 ~\textrm{drymouth\_am}$ & $\scriptsize{+}$ & $0.08 ~\textrm{bronchitis}$ \\ 
 $\scriptsize{+}$ & $0.08 ~\textrm{fallbacks\_easily}$ & $\scriptsize{+}$ & $0.07 ~\textrm{finished\_high\_school}$ & $\scriptsize{+}$ & $0.07 ~\textrm{ESS2\_ge\_2}$ \\ 
 $\scriptsize{+}$ & $0.07 ~\textrm{latency\_le\_10}$ & $\scriptsize{+}$ & $0.06 ~\textrm{arthritis}$ & $\scriptsize{+}$ & $0.06 ~\textrm{gasping\_arousals}$ \\ 
 $\scriptsize{+}$ & $0.06 ~\textrm{ESS1\_ge\_2}$ & $\scriptsize{+}$ & $0.06 ~\textrm{nap\_refreshes}$ & $\scriptsize{+}$ & $0.06 ~\textrm{some\_college}$ \\ 
 $\scriptsize{+}$ & $0.06 ~\textrm{cancer}$ & $\scriptsize{+}$ & $0.05 ~\textrm{some\_high\_school}$ & $\scriptsize{+}$ & $0.05 ~\textrm{coronary\_dz}$ \\ 
 $\scriptsize{+}$ & $0.04 ~\textrm{not\_enough\_time}$ & $\scriptsize{+}$ & $0.04 ~\textrm{any\_other\_disorder}$ & $\scriptsize{+}$ & $0.04 ~\textrm{chf}$ \\ 
 $\scriptsize{+}$ & $0.04 ~\textrm{no\_problem}$ & $\scriptsize{+}$ & $0.04 ~\textrm{ESS3\_ge\_2}$ & $\scriptsize{+}$ & $0.03 ~\textrm{kidney\_disorder}$ \\ 
 $\scriptsize{+}$ & $0.03 ~\textrm{caffeine}$ & $\scriptsize{+}$ & $0.02 ~\textrm{bedtime\_stable}$ & $\scriptsize{+}$ & $0.02 ~\textrm{memory\_problems}$ \\ 
 $\scriptsize{+}$ & $0.02 ~\textrm{alcohol\_for\_sleep}$ & $\scriptsize{+}$ & $0.02 ~\textrm{sleeps\_always\_with\_partner}$ & $\scriptsize{+}$ & $0.02 ~\textrm{sleeps\_with\_noone}$ \\ 
 $\scriptsize{+}$ & $0.01 ~\textrm{wakes\_up\_more\_than\_3}$ & $\scriptsize{+}$ & $0.01 ~\textrm{sleep\_paralysis}$ & $\scriptsize{+}$ & $0.01 ~\textrm{ESS6\_ge\_2}$ \\ 
 $\scriptsize{+}$ & $0.01 ~\textrm{bad\_sleep\_habits}$ & $\scriptsize{+}$ & $0.01 ~\textrm{timezone\_travel}$ & $\scriptsize{+}$ & $0.01 ~\textrm{ESS7\_ge\_2}$ \\ 
 $\scriptsize{+}$ & $0.01 ~\textrm{multiple\_other\_disorders}$ & $\scriptsize{+}$ & $0.00 ~\textrm{hypothyroid}$ & $\scriptsize{+}$ & $0.00 ~\textrm{irregular\_heart}$ \\ 
 $\scriptsize{-}$ & $0.28 ~\textrm{female}$ & $\scriptsize{-}$ & $0.20 ~\textrm{bmi\_lt\_25}$ & $\scriptsize{-}$ & $0.17 ~\textrm{age\_lt\_30}$ \\ 
 $\scriptsize{-}$ & $0.14 ~\textrm{insomia}$ & $\scriptsize{-}$ & $0.13 ~\textrm{headaches}$ & $\scriptsize{-}$ & $0.09 ~\textrm{anxiety}$ \\ 
 $\scriptsize{-}$ & $0.08 ~\textrm{insomnia\_onset}$ & $\scriptsize{-}$ & $0.07 ~\textrm{fibromyalgia}$ & $\scriptsize{-}$ & $0.07 ~\textrm{asthma}$ \\ 
 $\scriptsize{-}$ & $0.07 ~\textrm{finished\_college}$ & $\scriptsize{-}$ & $0.06 ~\textrm{parasomnia}$ & $\scriptsize{-}$ & $0.06 ~\textrm{better\_if\_move}$ \\ 
 $\scriptsize{-}$ & $0.06 ~\textrm{ESS4\_ge\_2}$ & $\scriptsize{-}$ & $0.06 ~\textrm{worse\_night}$ & $\scriptsize{-}$ & $0.05 ~\textrm{abnormal\_heart\_rhythm}$ \\ 
 $\scriptsize{-}$ & $0.05 ~\textrm{no\_timing}$ & $\scriptsize{-}$ & $0.05 ~\textrm{latency\_ge\_60}$ & $\scriptsize{-}$ & $0.05 ~\textrm{sleep\_hallucinations}$ \\ 
 $\scriptsize{-}$ & $0.04 ~\textrm{disabled}$ & $\scriptsize{-}$ & $0.04 ~\textrm{insomnia\_earlywake}$ & $\scriptsize{-}$ & $0.04 ~\textrm{leg\_jerks}$ \\ 
 $\scriptsize{-}$ & $0.04 ~\textrm{sleep\_twitches}$ & $\scriptsize{-}$ & $0.04 ~\textrm{shift\_work}$ & $\scriptsize{-}$ & $0.04 ~\textrm{bipolar}$ \\ 
 $\scriptsize{-}$ & $0.04 ~\textrm{depression}$ & $\scriptsize{-}$ & $0.03 ~\textrm{ged}$ & $\scriptsize{-}$ & $0.03 ~\textrm{head\_trauma}$ \\ 
 $\scriptsize{-}$ & $0.03 ~\textrm{wake\_twitches}$ & $\scriptsize{-}$ & $0.03 ~\textrm{bed\_partner\_disturbs\_me}$ & $\scriptsize{-}$ & $0.03 ~\textrm{wakes\_up\_1\_to\_3}$ \\ 
$\scriptsize{-}$ & $0.03 ~\textrm{bedtime\_variable}$ & $\scriptsize{-}$ & $0.03 ~\textrm{latency\_30\_to\_60}$ & $\scriptsize{-}$ & $0.03 ~\textrm{tired\_regardless\_sleep\_duration}$ \\ 
 $\scriptsize{-}$ & $0.03 ~\textrm{insomnia\_maint}$ & $\scriptsize{-}$ & $0.03 ~\textrm{substance\_abuse}$ & $\scriptsize{-}$ & $0.03 ~\textrm{sleeps\_sometimes\_with\_partner}$ \\ 
$\scriptsize{-}$ & $0.03 ~\textrm{graduate\_degree}$ & $\scriptsize{-}$ & $0.02 ~\textrm{latency\_10\_to\_30}$ & $\scriptsize{-}$ & $0.02 ~\textrm{sleepiness}$ \\ 
 $\scriptsize{-}$ & $0.02 ~\textrm{technical\_school}$ & $\scriptsize{-}$ & $0.02 ~\textrm{no\_diff}$ & $\scriptsize{-}$ & $0.02 ~\textrm{cataplexy}$ \\ 
$\scriptsize{-}$ & $0.02 ~\textrm{worse\_AM}$ & $\scriptsize{-}$ & $0.02 ~\textrm{stroke}$ & $\scriptsize{-}$ & $0.02 ~\textrm{COPD}$ \\ 
 $\scriptsize{-}$ & $0.02 ~\textrm{worse\_if\_move}$ & $\scriptsize{-}$ & $0.01 ~\textrm{smoking}$ & $\scriptsize{-}$ & $0.01 ~\textrm{ESS\_sum\_geq\_9}$ \\ 
 $\scriptsize{-}$ & $0.01 ~\textrm{associates\_degree}$ & $\scriptsize{-}$ & $0.01 ~\textrm{pacemaker}$ & $\scriptsize{-}$ & $0.01 ~\textrm{PTSD}$ \\ 
 $\scriptsize{-}$ & $0.01 ~\textrm{hyperthyroid}$ & $\scriptsize{-}$ & $0.00 ~\textrm{prefer\_other\_language}$ & $\scriptsize{-}$ & $0.00 ~\textrm{unemployed}$ \\ 
 $\scriptsize{-}$ & $0.00 ~\textrm{seizures}$ & $\scriptsize{-}$ & $0.00 ~\textrm{fallsback\_slowly}$ & $\scriptsize{-}$ & $0.00 ~\textrm{ESS8\_ge\_2}$ \\ 
 $\scriptsize{-}$ & $0.00 ~\textrm{wakes\_up\_never}$ & $\scriptsize{-}$ & $0.00 ~\textrm{meningitis}$ & $\scriptsize{-}$ & $1.13$ \\ 
  \end{tabularx}
  \caption{Ridge score function for the \textds{apnea} dataset.}
}
}
\end{figure}
\begin{figure}[htbp]
{\centering
\scriptsize
\begin{tabularx}{\textwidth}{p{1mm}lp{1mm}lp{1mm}l}
  & $0.82 ~\textrm{age\_geq\_60}$ & $\scriptsize{+}$ & $0.77 ~\textrm{bmi\_geq\_40}$ & $\scriptsize{+}$ & $0.62 ~\textrm{retired}$ \\ 
 $\scriptsize{+}$ & $0.59 ~\textrm{kidney\_disorder}$ & $\scriptsize{+}$ & $0.56 ~\textrm{stopbreathing}$ & $\scriptsize{+}$ & $0.55 ~\textrm{fallbacks\_easily}$ \\ 
$\scriptsize{+}$ & $0.55 ~\textrm{male}$ & $\scriptsize{+}$ & $0.49 ~\textrm{htn}$ & $\scriptsize{+}$ & $0.49 ~\textrm{fallsback\_slowly}$ \\ 
 $\scriptsize{+}$ & $0.48 ~\textrm{some\_high\_school}$ & $\scriptsize{+}$ & $0.46 ~\textrm{sleeps\_with\_noone}$ & $\scriptsize{+}$ & $0.44 ~\textrm{age\_geq\_30}$ \\ 
 $\scriptsize{+}$ & $0.43 ~\textrm{bronchitis}$ & $\scriptsize{+}$ & $0.42 ~\textrm{sleep\_paralysis}$ & $\scriptsize{+}$ & $0.42 ~\textrm{cancer}$ \\ 
 $\scriptsize{+}$ & $0.39 ~\textrm{snoring}$ & $\scriptsize{+}$ & $0.38 ~\textrm{ESS5\_ge\_2}$ & $\scriptsize{+}$ & $0.38 ~\textrm{diabetes}$ \\ 
 $\scriptsize{+}$ & $0.36 ~\textrm{bmi\_geq\_35}$ & $\scriptsize{+}$ & $0.31 ~\textrm{sleeps\_always\_with\_partner}$ & $\scriptsize{+}$ & $0.28 ~\textrm{wake\_twitches}$ \\ 
 $\scriptsize{+}$ & $0.28 ~\textrm{bmi\_geq\_25}$ & $\scriptsize{+}$ & $0.27 ~\textrm{sleep\_twitches}$ & $\scriptsize{+}$ & $0.27 ~\textrm{alcohol\_for\_sleep}$ \\ 
 $\scriptsize{+}$ & $0.26 ~\textrm{finished\_high\_school}$ & $\scriptsize{+}$ & $0.26 ~\textrm{some\_college}$ & $\scriptsize{+}$ & $0.24 ~\textrm{ESS2\_ge\_2}$ \\ 
 $\scriptsize{+}$ & $0.23 ~\textrm{not\_enough\_time}$ & $\scriptsize{+}$ & $0.21 ~\textrm{snoring\_2}$ & $\scriptsize{+}$ & $0.21 ~\textrm{ESS1\_ge\_2}$ \\ 
 $\scriptsize{+}$ & $0.20 ~\textrm{depression}$ & $\scriptsize{+}$ & $0.19 ~\textrm{no\_diff}$ & $\scriptsize{+}$ & $0.19 ~\textrm{seizures}$ \\ 
 $\scriptsize{+}$ & $0.19 ~\textrm{latency\_le\_10}$ & $\scriptsize{+}$ & $0.17 ~\textrm{gasping\_arousals}$ & $\scriptsize{+}$ & $0.17 ~\textrm{ESS3\_ge\_2}$ \\ 
 $\scriptsize{+}$ & $0.15 ~\textrm{timezone\_travel}$ & $\scriptsize{+}$ & $0.15 ~\textrm{bed\_partner\_disturbs\_me}$ & $\scriptsize{+}$ & $0.14 ~\textrm{memory\_problems}$ \\ 
 $\scriptsize{+}$ & $0.14 ~\textrm{sleeps\_sometimes\_with\_partner}$ & $\scriptsize{+}$ & $0.12 ~\textrm{ESS7\_ge\_2}$ & $\scriptsize{+}$ & $0.12 ~\textrm{chf}$ \\ 
 $\scriptsize{+}$ & $0.12 ~\textrm{latency\_30\_to\_60}$ & $\scriptsize{+}$ & $0.10 ~\textrm{irregular\_heart}$ & $\scriptsize{+}$ & $0.10 ~\textrm{PTSD}$ \\ 
 $\scriptsize{+}$ & $0.09 ~\textrm{tired\_regardless\_sleep\_duration}$ & $\scriptsize{+}$ & $0.08 ~\textrm{drymouth\_am}$ & $\scriptsize{+}$ & $0.07 ~\textrm{no\_problem}$ \\ 
 $\scriptsize{+}$ & $0.06 ~\textrm{nap\_refreshes}$ & $\scriptsize{+}$ & $0.06 ~\textrm{bmi\_geq\_30}$ & $\scriptsize{+}$ & $0.04 ~\textrm{ESS8\_ge\_2}$ \\ 
 $\scriptsize{+}$ & $0.04 ~\textrm{coronary\_dz}$ & $\scriptsize{+}$ & $0.04 ~\textrm{bad\_sleep\_habits}$ & $\scriptsize{+}$ & $0.04 ~\textrm{nocturnal}$ \\ 
 $\scriptsize{+}$ & $0.04 ~\textrm{arthritis}$ & $\scriptsize{+}$ & $0.03 ~\textrm{latency\_10\_to\_30}$ & $\scriptsize{+}$ & $0.02 ~\textrm{caffeine}$ \\ 
 $\scriptsize{+}$ & $0.02 ~\textrm{smoking}$ & $\scriptsize{+}$ & $0.02 ~\textrm{associates\_degree}$ & $\scriptsize{+}$ & $0.02 ~\textrm{hypothyroid}$ \\ 
 $\scriptsize{+}$ & $0.01 ~\textrm{graduate\_degree}$ & $\scriptsize{+}$ & $0.01 ~\textrm{insomnia\_onset}$ & $\scriptsize{+}$ & $0.00 ~\textrm{latency\_ge\_60}$ \\ 
 $\scriptsize{+}$ & $0.00 ~\textrm{age\_geq\_45}$ & $\scriptsize{-}$ & $0.69 ~\textrm{COPD}$ & $\scriptsize{-}$ & $0.61 ~\textrm{fibromyalgia}$ \\ 
 $\scriptsize{-}$ & $0.58 ~\textrm{worse\_night}$ & $\scriptsize{-}$ & $0.55 ~\textrm{female}$ & $\scriptsize{-}$ & $0.49 ~\textrm{worse\_if\_move}$ \\ 
 $\scriptsize{-}$ & $0.44 ~\textrm{age\_lt\_30}$ & $\scriptsize{-}$ & $0.44 ~\textrm{worse\_AM}$ & $\scriptsize{-}$ & $0.42 ~\textrm{bipolar}$ \\ 
 $\scriptsize{-}$ & $0.38 ~\textrm{wakes\_up\_1\_to\_3}$ & $\scriptsize{-}$ & $0.36 ~\textrm{technical\_school}$ & $\scriptsize{-}$ & $0.35 ~\textrm{head\_trauma}$ \\ 
 $\scriptsize{-}$ & $0.34 ~\textrm{abnormal\_heart\_rhythm}$ & $\scriptsize{-}$ & $0.33 ~\textrm{asthma}$ & $\scriptsize{-}$ & $0.32 ~\textrm{insomia}$ \\ 
 $\scriptsize{-}$ & $0.31 ~\textrm{shift\_work}$ & $\scriptsize{-}$ & $0.31 ~\textrm{bedtime\_stable}$ & $\scriptsize{-}$ & $0.30 ~\textrm{multiple\_other\_disorders}$ \\ 
 $\scriptsize{-}$ & $0.29 ~\textrm{no\_timing}$ & $\scriptsize{-}$ & $0.28 ~\textrm{prefer\_other\_language}$ & $\scriptsize{-}$ & $0.28 ~\textrm{bmi\_lt\_25}$ \\ 
 $\scriptsize{-}$ & $0.26 ~\textrm{ged}$ & $\scriptsize{-}$ & $0.26 ~\textrm{insomnia\_earlywake}$ & $\scriptsize{-}$ & $0.25 ~\textrm{parasomnia}$ \\ 
 $\scriptsize{-}$ & $0.23 ~\textrm{substance\_abuse}$ & $\scriptsize{-}$ & $0.21 ~\textrm{stopbreathing\_2}$ & $\scriptsize{-}$ & $0.21 ~\textrm{stroke}$ \\ 
 $\scriptsize{-}$ & $0.20 ~\textrm{ESS\_sum\_geq\_9}$ & $\scriptsize{-}$ & $0.19 ~\textrm{ESS4\_ge\_2}$ & $\scriptsize{-}$ & $0.17 ~\textrm{hyperthyroid}$ \\ 
 $\scriptsize{-}$ & $0.16 ~\textrm{bedtime\_variable}$ & $\scriptsize{-}$ & $0.16 ~\textrm{cataplexy}$ & $\scriptsize{-}$ & $0.15 ~\textrm{sleep\_hallucinations}$ \\ 
 $\scriptsize{-}$ & $0.13 ~\textrm{better\_if\_move}$ & $\scriptsize{-}$ & $0.12 ~\textrm{wakes\_up\_more\_than\_3}$ & $\scriptsize{-}$ & $0.12 ~\textrm{pacemaker}$ \\ 
 $\scriptsize{-}$ & $0.10 ~\textrm{leg\_jerks}$ & $\scriptsize{-}$ & $0.10 ~\textrm{sleepiness}$ & $\scriptsize{-}$ & $0.08 ~\textrm{finished\_college}$ \\ 
 $\scriptsize{-}$ & $0.08 ~\textrm{anxiety}$ & $\scriptsize{-}$ & $0.06 ~\textrm{headaches}$ & $\scriptsize{-}$ & $0.05 ~\textrm{meningitis}$ \\ 
 $\scriptsize{-}$ & $0.05 ~\textrm{disabled}$ & $\scriptsize{-}$ & $0.05 ~\textrm{insomnia\_maint}$ & $\scriptsize{-}$ & $0.03 ~\textrm{ESS6\_ge\_2}$ \\ 
 $\scriptsize{-}$ & $0.03 ~\textrm{any\_other\_disorder}$ & $\scriptsize{-}$ & $0.03 ~\textrm{unemployed}$ & $\scriptsize{-}$ & $0.01 ~\textrm{wakes\_up\_never}$ \\ 
 $\scriptsize{-}$ & $2.19$ &  &  &  &  \\ 
  \end{tabularx}
}
\caption{SVM Linear score function for the \textds{apnea} dataset. }
\end{figure}
\begin{figure}[htbp]
\includegraphics[width=\textwidth]{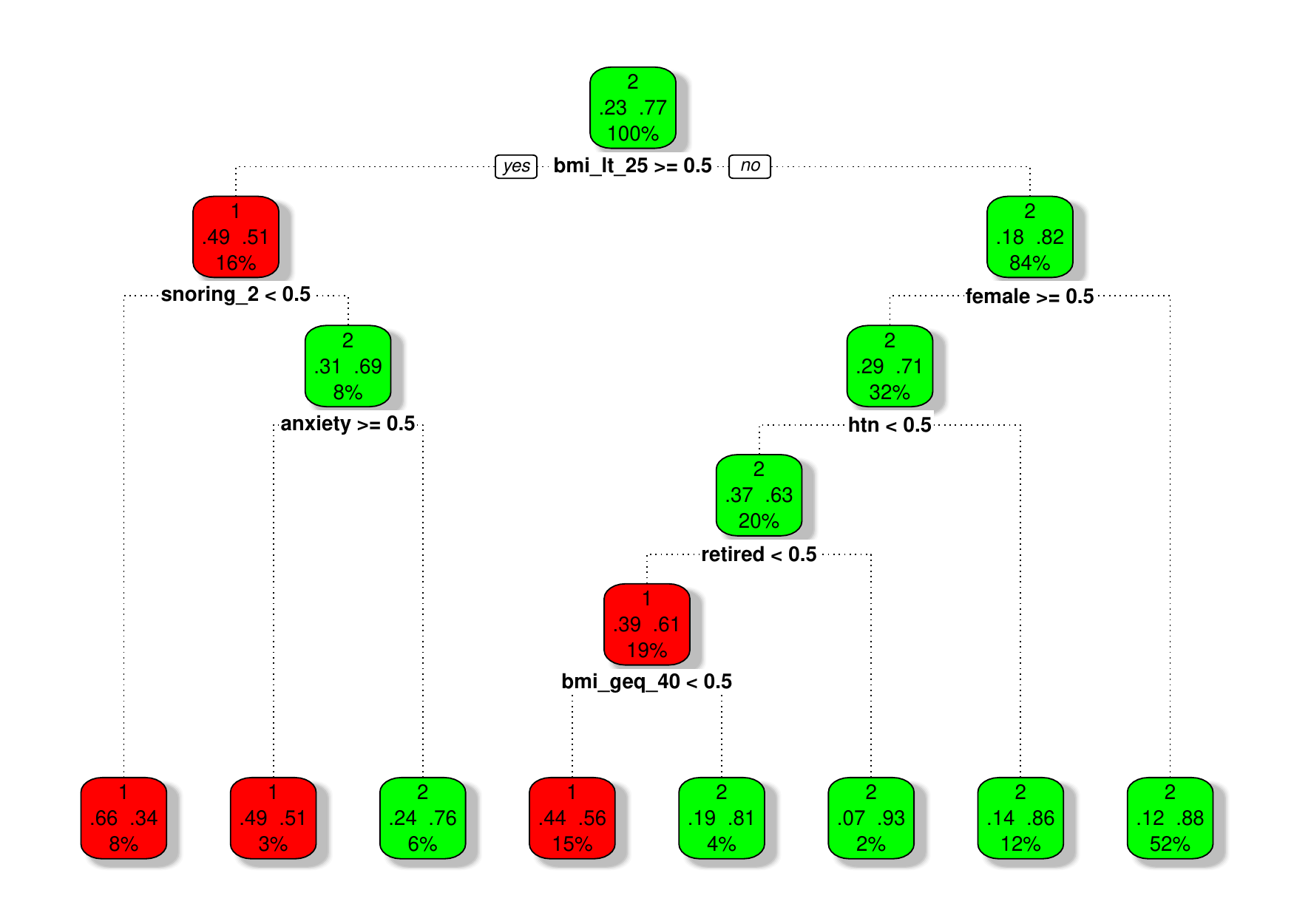} 
\caption{CART model for the \textds{apnea} dataset.}
\end{figure}
\begin{figure}[htbp]
\tiny
\begin{verbatim}
Rule 1: (7, lift 3.8)
	age_lt_30 > 0
	bmi_lt_25 > 0
	snoring > 0
	headaches > 0
	->  class 0  [0.889]

Rule 2: (5, lift 3.7)
	bmi_lt_25 > 0
	snoring <= 0
	insomnia_onset > 0
	htn > 0
	graduate_degree <= 0
	->  class 0  [0.857]

Rule 3: (122/29, lift 3.3)
	bmi_lt_25 > 0
	snoring <= 0
	stopbreathing <= 0
	htn <= 0
	retired <= 0
	->  class 0  [0.758]

Rule 4: (67/16, lift 3.3)
	male <= 0
	age_lt_30 > 0
	bmi_geq_40 <= 0
	htn <= 0
	->  class 0  [0.754]

Rule 5: (57/19, lift 2.9)
	bmi_lt_25 > 0
	latency_30_to_60 > 0
	retired <= 0
	->  class 0  [0.661]

Rule 6: (47/17, lift 2.7)
	bmi_geq_35 <= 0
	leg_jerks > 0
	snoring_2 <= 0
	retired <= 0
	->  class 0  [0.633]

Rule 7: (276/111, lift 2.6)
	age_geq_60 <= 0
	bmi_geq_35 <= 0
	snoring_2 <= 0
	retired <= 0
	->  class 0  [0.597]

Rule 8: (40/17, lift 2.5)
	male <= 0
	irregular_heart > 0
	htn <= 0
	->  class 0  [0.571]

Rule 9: (297/137, lift 2.3)
	male <= 0
	bmi_geq_40 <= 0
	hypothyroid <= 0
	htn <= 0
	some_college <= 0
	retired <= 0
	->  class 0  [0.538]

Rule 10: (792/510, lift 1.5)
	male <= 0
	->  class 0  [0.356]
	}
\end{verbatim}
\caption{Rules 1 through 10 for the C5.0T model for the \textds{apnea} dataset. }
\end{figure}

\begin{figure}[htbp]
\tiny
\begin{verbatim}
Rule 11: (30, lift 1.3)
	male > 0
	age_geq_60 > 0
	bmi_lt_25 <= 0
	leg_jerks <= 0
	snoring_2 <= 0
	retired <= 0
	->  class 1  [0.969]

Rule 12: (27, lift 1.3)
	male <= 0
	age_lt_30 <= 0
	bmi_lt_25 <= 0
	wakes_up_more_than_3 <= 0
	hypothyroid > 0
	headaches <= 0
	retired <= 0
	->  class 1  [0.966]

Rule 13: (47/1, lift 1.2)
	male <= 0
	bmi_geq_40 > 0
	sleeps_sometimes_with_partner <= 0
	irregular_heart <= 0
	htn <= 0
	substance_abuse <= 0
	associates_degree <= 0
	->  class 1  [0.959]

Rule 14: (222/10, lift 1.2)
	bmi_lt_25 <= 0
	retired > 0
	->  class 1  [0.951]

Rule 15: (58/2, lift 1.2)
	male > 0
	bmi_geq_35 > 0
	snoring_2 <= 0
	->  class 1  [0.950]

Rule 16: (277/20, lift 1.2)
	retired > 0
	->  class 1  [0.925]

Rule 17: (539/43, lift 1.2)
	bmi_lt_25 <= 0
	shift_work <= 0
	htn > 0
	->  class 1  [0.919]

Rule 18: (789/69, lift 1.2)
	male > 0
	bmi_lt_25 <= 0
	snoring_2 > 0
	->  class 1  [0.912]

Rule 19: (579/65, lift 1.2)
	stopbreathing > 0
	snoring_2 > 0
	->  class 1  [0.886]

Rule 20: (807/101, lift 1.1)
	snoring > 0
	headaches <= 0
	->  class 1  [0.874]

Default class: 1
\end{verbatim}
\caption{Rules 11 through 20 for the C5.0T model for the \textds{apnea} dataset. }
\end{figure}

\begin{figure}[htbp]
\tiny
\begin{verbatim}
bmi_lt_25 > 0:
:...retired > 0:
:   :...asthma <= 0: 1 (48/7)
:   :   asthma > 0: 0 (7/4)
:   retired <= 0:
:   :...snoring <= 0:
:       :...htn <= 0:
:       :   :...snoring_2 <= 0: 0 (117/24)
:       :   :   snoring_2 > 0:
:       :   :   :...stopbreathing <= 0: 0 (19/8)
:       :   :       stopbreathing > 0: 1 (4)
:       :   htn > 0:
:       :   :...insomnia_onset <= 0: 1 (8)
:       :       insomnia_onset > 0:
:       :       :...graduate_degree <= 0: 0 (5)
:       :           graduate_degree > 0: 1 (3)
:       snoring > 0:
:       :...age_lt_30 <= 0:
:           :...latency_30_to_60 <= 0: 1 (71/15)
:           :   latency_30_to_60 > 0:
:           :   :...ESS4_ge_2 > 0: 0 (5)
:           :       ESS4_ge_2 <= 0:
:           :       :...sleep_twitches <= 0: 1 (8/1)
:           :           sleep_twitches > 0: 0 (2)
:           age_lt_30 > 0:
:           :...headaches > 0: 0 (7)
:               headaches <= 0:
:               :...drymouth_am <= 0: 1 (4)
:                   drymouth_am > 0:
:                   :...sleeps_sometimes_with_partner <= 0: 0 (4)
:                       sleeps_sometimes_with_partner > 0: 1 (2)
bmi_lt_25 <= 0:
:...retired > 0: 1 (222/10)
    retired <= 0:
    :...male > 0:
        :...snoring_2 > 0: 1 (703/66)
        :   snoring_2 <= 0:
        :   :...bmi_geq_35 > 0: 1 (43/2)
        :       bmi_geq_35 <= 0:
        :       :...age_geq_60 <= 0: 0 (96/56)
        :           age_geq_60 > 0:
        :           :...leg_jerks <= 0: 1 (16)
        :               leg_jerks > 0: 0 (3/1)
        male <= 0:
        :...htn > 0:
            :...shift_work <= 0: 1 (137/20)
            :   shift_work > 0:
            :   :...stopbreathing <= 0: 0 (18/9)
            :       stopbreathing > 0: 1 (6)
            htn <= 0:
            :...bmi_geq_40 <= 0:
                :...sleep_hallucinations > 0: 0 (19/5)
                :   sleep_hallucinations <= 0:
                :   :...age_lt_30 > 0: 0 (25/8)
                :       age_lt_30 <= 0:
                :       :...hypothyroid <= 0:
                :           :...some_college <= 0: 0 (175/97)
                :           :   some_college > 0: 1 (38/9)
                :           hypothyroid > 0:
                :           :...stopbreathing > 0: 1 (7)
                :               stopbreathing <= 0:
                :               :...headaches > 0: 0 (12/4)
                :                   headaches <= 0:
                :                   :...wakes_up_more_than_3 <= 0: 1 (14)
                :                       wakes_up_more_than_3 > 0: 0 (2)
                bmi_geq_40 > 0:
                :...not_enough_time > 0: 1 (17)
                    not_enough_time <= 0:
                    :...sleeps_sometimes_with_partner > 0:
                        :...latency_le_10 <= 0: 0 (8/1)
                        :   latency_le_10 > 0: 1 (2)
                        sleeps_sometimes_with_partner <= 0:
                        :...irregular_heart > 0: 0 (2)
                            irregular_heart <= 0:
                            :...substance_abuse > 0: 0 (2)
                                substance_abuse <= 0:
                                :...associates_degree <= 0: 1 (36/1)
                                    associates_degree > 0:
                                    :...insomia <= 0: 1 (3)
                                        insomia > 0: 0 (2)
\end{verbatim}
\caption{C5.0T model for the \textds{apnea} dataset}
\end{figure}

\clearpage
\section{Additional Models for the \texttt{mushroom} Dataset}\label{Appendix::MushroomModels}
In this appendix, we include the remaining models that were trained on the $\texttt{mushroom}$ dataset.
\begin{figure}[htbp]
{\scriptsize
\begin{tabularx}{0.8\textwidth}{p{1mm}lp{1mm}lp{1mm}l}
  & $0.79 ~\textrm{odor\_eq\_foul}$ & $\scriptsize{+}$ & $0.75 ~\textrm{gill\_size\_eq\_narrow}$ & $\scriptsize{+}$ & $0.53 ~\textrm{spore\_print\_color\_eq\_chocolate}$ \\ 
   $\scriptsize{+}$ & $0.47 ~\textrm{population\_eq\_several}$ & $\scriptsize{+}$ & $0.47 ~\textrm{gill\_color\_eq\_buff}$ & $\scriptsize{+}$ & $0.46 ~\textrm{stalk\_surface\_above\_ring\_eq\_grooves}$ \\ 
   $\scriptsize{+}$ & $0.45 ~\textrm{gill\_spacing\_eq\_close}$ & $\scriptsize{+}$ & $0.45 ~\textrm{stalk\_root\_eq\_bulbous}$ & $\scriptsize{+}$ & $0.44 ~\textrm{odor\_eq\_pungent}$ \\ 
   $\scriptsize{+}$ & $0.37 ~\textrm{odor\_eq\_creosote}$ & $\scriptsize{+}$ & $0.36 ~\textrm{stalk\_surface\_below\_ring\_eq\_grooves}$ & $\scriptsize{+}$ & $0.33 ~\textrm{spore\_print\_color\_eq\_green}$ \\ 
   $\scriptsize{+}$ & $0.31 ~\textrm{ring\_type\_eq\_large}$ & $\scriptsize{+}$ & $0.27 ~\textrm{cap\_surface\_eq\_smooth}$ & $\scriptsize{+}$ & $0.21 ~\textrm{habitat\_eq\_urban}$ \\ 
   $\scriptsize{+}$ & $0.18 ~\textrm{odor\_eq\_spicy}$ & $\scriptsize{+}$ & $0.18 ~\textrm{odor\_eq\_fishy}$ & $\scriptsize{+}$ & $0.18 ~\textrm{cap\_color\_eq\_buff}$ \\ 
   $\scriptsize{+}$ & $0.17 ~\textrm{cap\_color\_eq\_pink}$ & $\scriptsize{+}$ & $0.16 ~\textrm{stalk\_shape\_eq\_elarging}$ & $\scriptsize{+}$ & $0.15 ~\textrm{spore\_print\_color\_eq\_white}$ \\ 
   $\scriptsize{+}$ & $0.12 ~\textrm{population\_eq\_scattered}$ & $\scriptsize{+}$ & $0.11 ~\textrm{stalk\_color\_below\_ring\_eq\_pink}$ & $\scriptsize{+}$ & $0.11 ~\textrm{gill\_color\_eq\_green}$ \\ 
   $\scriptsize{+}$ & $0.10 ~\textrm{stalk\_color\_above\_ring\_eq\_buff}$ & $\scriptsize{+}$ & $0.10 ~\textrm{habitat\_eq\_paths}$ & $\scriptsize{+}$ & $0.10 ~\textrm{stalk\_color\_above\_ring\_eq\_pink}$ \\ 
   $\scriptsize{+}$ & $0.10 ~\textrm{stalk\_color\_below\_ring\_eq\_buff}$ & $\scriptsize{+}$ & $0.10 ~\textrm{gill\_color\_eq\_chocolate}$ & $\scriptsize{+}$ & $0.08 ~\textrm{stalk\_color\_above\_ring\_eq\_brown}$ \\ 
   $\scriptsize{+}$ & $0.08 ~\textrm{ring\_type\_eq\_none}$ & $\scriptsize{+}$ & $0.08 ~\textrm{stalk\_color\_below\_ring\_eq\_cinnamon}$ & $\scriptsize{+}$ & $0.08 ~\textrm{stalk\_color\_above\_ring\_eq\_cinnamon}$ \\ 
   $\scriptsize{+}$ & $0.08 ~\textrm{odor\_eq\_musty}$ & $\scriptsize{+}$ & $0.08 ~\textrm{veil\_color\_eq\_white}$ & $\scriptsize{+}$ & $0.07 ~\textrm{gill\_attachment\_eq\_free}$ \\ 
   $\scriptsize{+}$ & $0.07 ~\textrm{gill\_color\_eq\_gray}$ & $\scriptsize{+}$ & $0.06 ~\textrm{stalk\_color\_below\_ring\_eq\_yellow}$ & $\scriptsize{+}$ & $0.06 ~\textrm{cap\_color\_eq\_white}$ \\ 
   $\scriptsize{+}$ & $0.06 ~\textrm{ring\_type\_eq\_evanescent}$ & $\scriptsize{+}$ & $0.06 ~\textrm{cap\_shape\_eq\_flat}$ & $\scriptsize{+}$ & $0.05 ~\textrm{stalk\_root\_eq\_equal}$ \\ 
   $\scriptsize{+}$ & $0.03 ~\textrm{veil\_color\_eq\_yellow}$ & $\scriptsize{+}$ & $0.03 ~\textrm{stalk\_color\_above\_ring\_eq\_yellow}$ & $\scriptsize{+}$ & $0.02 ~\textrm{cap\_shape\_eq\_convex}$ \\ 
   $\scriptsize{+}$ & $0.02 ~\textrm{cap\_shape\_eq\_knobbed}$ & $\scriptsize{+}$ & $0.02 ~\textrm{cap\_surface\_eq\_grooves}$ & $\scriptsize{+}$ & $0.02 ~\textrm{cap\_shape\_eq\_conical}$ \\ 
   $\scriptsize{+}$ & $0.01 ~\textrm{gill\_color\_eq\_yellow}$ & $\scriptsize{+}$ & $0.01 ~\textrm{habitat\_eq\_grasses}$ & $\scriptsize{+}$ & $0.01 ~\textrm{stalk\_color\_below\_ring\_eq\_white}$ \\ 
   $\scriptsize{+}$ & $0.01 ~\textrm{stalk\_surface\_above\_ring\_eq\_scaly}$ & $\scriptsize{+}$ & $0.01 ~\textrm{cap\_color\_eq\_red}$ & $\scriptsize{+}$ & $0.00 ~\textrm{cap\_surface\_eq\_scaly}$ \\ 
   $\scriptsize{-}$ & $1.33 ~\textrm{odor\_eq\_none}$ & $\scriptsize{-}$ & $0.75 ~\textrm{gill\_size\_eq\_broad}$ & $\scriptsize{-}$ & $0.45 ~\textrm{gill\_spacing\_eq\_crowded}$ \\ 
   $\scriptsize{-}$ & $0.43 ~\textrm{spore\_print\_color\_eq\_brown}$ & $\scriptsize{-}$ & $0.42 ~\textrm{stalk\_surface\_above\_ring\_eq\_smooth}$ & $\scriptsize{-}$ & $0.37 ~\textrm{spore\_print\_color\_eq\_black}$ \\ 
   $\scriptsize{-}$ & $0.36 ~\textrm{odor\_eq\_anise}$ & $\scriptsize{-}$ & $0.36 ~\textrm{odor\_eq\_almond}$ & $\scriptsize{-}$ & $0.30 ~\textrm{bruises\_eq\_TRUE}$ \\ 
   $\scriptsize{-}$ & $0.30 ~\textrm{ring\_type\_eq\_pendant}$ & $\scriptsize{-}$ & $0.29 ~\textrm{cap\_surface\_eq\_fibrous}$ & $\scriptsize{-}$ & $0.25 ~\textrm{stalk\_root\_eq\_club}$ \\ 
   $\scriptsize{-}$ & $0.24 ~\textrm{population\_eq\_numerous}$ & $\scriptsize{-}$ & $0.23 ~\textrm{gill\_color\_eq\_brown}$ & $\scriptsize{-}$ & $0.22 ~\textrm{population\_eq\_solitary}$ \\ 
   $\scriptsize{-}$ & $0.20 ~\textrm{stalk\_surface\_below\_ring\_eq\_smooth}$ & $\scriptsize{-}$ & $0.18 ~\textrm{gill\_color\_eq\_white}$ & $\scriptsize{-}$ & $0.17 ~\textrm{stalk\_color\_above\_ring\_eq\_gray}$ \\ 
   $\scriptsize{-}$ & $0.17 ~\textrm{stalk\_color\_below\_ring\_eq\_gray}$ & $\scriptsize{-}$ & $0.16 ~\textrm{stalk\_shape\_eq\_tapering}$ & $\scriptsize{-}$ & $0.16 ~\textrm{stalk\_surface\_below\_ring\_eq\_fibrous}$ \\ 
   $\scriptsize{-}$ & $0.16 ~\textrm{ring\_type\_eq\_flaring}$ & $\scriptsize{-}$ & $0.15 ~\textrm{stalk\_root\_eq\_rooted}$ & $\scriptsize{-}$ & $0.15 ~\textrm{habitat\_eq\_woods}$ \\ 
   $\scriptsize{-}$ & $0.14 ~\textrm{habitat\_eq\_waste}$ & $\scriptsize{-}$ & $0.11 ~\textrm{cap\_color\_eq\_yellow}$ & $\scriptsize{-}$ & $0.11 ~\textrm{cap\_color\_eq\_brown}$ \\ 
   $\scriptsize{-}$ & $0.11 ~\textrm{stalk\_color\_above\_ring\_eq\_orange}$ & $\scriptsize{-}$ & $0.11 ~\textrm{stalk\_color\_below\_ring\_eq\_orange}$ & $\scriptsize{-}$ & $0.11 ~\textrm{spore\_print\_color\_eq\_purple}$ \\ 
   $\scriptsize{-}$ & $0.10 ~\textrm{population\_eq\_abundant}$ & $\scriptsize{-}$ & $0.10 ~\textrm{gill\_color\_eq\_purple}$ & $\scriptsize{-}$ & $0.08 ~\textrm{gill\_color\_eq\_black}$ \\ 
   $\scriptsize{-}$ & $0.08 ~\textrm{ring\_number\_eq\_2}$ & $\scriptsize{-}$ & $0.08 ~\textrm{gill\_color\_eq\_red}$ & $\scriptsize{-}$ & $0.07 ~\textrm{stalk\_color\_above\_ring\_eq\_red}$ \\ 
   $\scriptsize{-}$ & $0.07 ~\textrm{stalk\_color\_below\_ring\_eq\_red}$ & $\scriptsize{-}$ & $0.07 ~\textrm{gill\_attachment\_eq\_attached}$ & $\scriptsize{-}$ & $0.07 ~\textrm{cap\_shape\_eq\_bell}$ \\ 
   $\scriptsize{-}$ & $0.06 ~\textrm{gill\_color\_eq\_pink}$ & $\scriptsize{-}$ & $0.06 ~\textrm{veil\_color\_eq\_orange}$ & $\scriptsize{-}$ & $0.06 ~\textrm{veil\_color\_eq\_brown}$ \\ 
   $\scriptsize{-}$ & $0.05 ~\textrm{cap\_shape\_eq\_sunken}$ & $\scriptsize{-}$ & $0.05 ~\textrm{cap\_color\_eq\_green}$ & $\scriptsize{-}$ & $0.05 ~\textrm{cap\_color\_eq\_purple}$ \\ 
   $\scriptsize{-}$ & $0.05 ~\textrm{cap\_color\_eq\_cinnamon}$ & $\scriptsize{-}$ & $0.05 ~\textrm{stalk\_color\_above\_ring\_eq\_white}$ & $\scriptsize{-}$ & $0.05 ~\textrm{stalk\_surface\_above\_ring\_eq\_fibrous}$ \\ 
   $\scriptsize{-}$ & $0.04 ~\textrm{cap\_color\_eq\_gray}$ & $\scriptsize{-}$ & $0.04 ~\textrm{gill\_color\_eq\_orange}$ & $\scriptsize{-}$ & $0.03 ~\textrm{population\_eq\_clustered}$ \\ 
   $\scriptsize{-}$ & $0.03 ~\textrm{spore\_print\_color\_eq\_buff}$ & $\scriptsize{-}$ & $0.03 ~\textrm{spore\_print\_color\_eq\_yellow}$ & $\scriptsize{-}$ & $0.03 ~\textrm{spore\_print\_color\_eq\_orange}$ \\ 
   $\scriptsize{-}$ & $0.02 ~\textrm{stalk\_color\_below\_ring\_eq\_brown}$ & $\scriptsize{-}$ & $0.02 ~\textrm{habitat\_eq\_leaves}$ & $\scriptsize{-}$ & $0.02 ~\textrm{habitat\_eq\_meadows}$ \\ 
   $\scriptsize{-}$ & $0.01 ~\textrm{stalk\_surface\_below\_ring\_eq\_scaly}$ & $\scriptsize{-}$ & $0.01 ~\textrm{ring\_number\_eq\_1}$ & $\scriptsize{+}$ & $0.30$ \\ 
  \end{tabularx}
}
\caption{Ridge score function trained on the full \texttt{mushroom} dataset. This model uses 113 coefficients and has a 10-fold CV test error of 1.7 $\pm$ 0.3$\%$.}
\end{figure}

\begin{figure}[htbp]
{\scriptsize
\begin{tabularx}{\textwidth}{p{1mm}lp{1mm}lp{1mm}l}
& $3.44 ~\textrm{spore\_print\_color\_eq\_green}$ & $\scriptsize{+}$ & $2.65 ~\textrm{odor\_eq\_foul}$ & $\scriptsize{+}$ & $2.53 ~\textrm{odor\_eq\_creosote}$ \\ 
 $\scriptsize{+}$ & $2.46 ~\textrm{stalk\_root\_eq\_bulbous}$ & $\scriptsize{+}$ & $2.16 ~\textrm{odor\_eq\_pungent}$ & $\scriptsize{+}$ & $2.09 ~\textrm{gill\_size\_eq\_narrow}$ \\ 
 $\scriptsize{+}$ & $1.76 ~\textrm{gill\_color\_eq\_buff}$ & $\scriptsize{+}$ & $1.63 ~\textrm{stalk\_surface\_above\_ring\_eq\_grooves}$ & $\scriptsize{+}$ & $1.43 ~\textrm{gill\_spacing\_eq\_close}$ \\ 
 $\scriptsize{+}$ & $1.24 ~\textrm{population\_eq\_clustered}$ & $\scriptsize{+}$ & $1.23 ~\textrm{stalk\_surface\_below\_ring\_eq\_scaly}$ & $\scriptsize{+}$ & $0.90 ~\textrm{spore\_print\_color\_eq\_chocolate}$ \\ 
 $\scriptsize{+}$ & $0.84 ~\textrm{cap\_color\_eq\_buff}$ & $\scriptsize{+}$ & $0.84 ~\textrm{stalk\_color\_below\_ring\_eq\_yellow}$ & $\scriptsize{+}$ & $0.76 ~\textrm{cap\_color\_eq\_pink}$ \\ 
 $\scriptsize{+}$ & $0.76 ~\textrm{odor\_eq\_fishy}$ & $\scriptsize{+}$ & $0.76 ~\textrm{odor\_eq\_spicy}$ & $\scriptsize{+}$ & $0.65 ~\textrm{habitat\_eq\_meadows}$ \\ 
 $\scriptsize{+}$ & $0.62 ~\textrm{ring\_type\_eq\_large}$ & $\scriptsize{+}$ & $0.62 ~\textrm{habitat\_eq\_grasses}$ & $\scriptsize{+}$ & $0.61 ~\textrm{cap\_surface\_eq\_grooves}$ \\ 
 $\scriptsize{+}$ & $0.59 ~\textrm{stalk\_shape\_eq\_elarging}$ & $\scriptsize{+}$ & $0.59 ~\textrm{gill\_color\_eq\_green}$ & $\scriptsize{+}$ & $0.56 ~\textrm{veil\_color\_eq\_yellow}$ \\ 
 $\scriptsize{+}$ & $0.56 ~\textrm{stalk\_color\_above\_ring\_eq\_yellow}$ & $\scriptsize{+}$ & $0.51 ~\textrm{ring\_type\_eq\_evanescent}$ & $\scriptsize{+}$ & $0.46 ~\textrm{cap\_color\_eq\_white}$ \\ 
 $\scriptsize{+}$ & $0.40 ~\textrm{cap\_shape\_eq\_conical}$ & $\scriptsize{+}$ & $0.33 ~\textrm{stalk\_surface\_below\_ring\_eq\_grooves}$ & $\scriptsize{+}$ & $0.32 ~\textrm{population\_eq\_several}$ \\ 
 $\scriptsize{+}$ & $0.32 ~\textrm{odor\_eq\_musty}$ & $\scriptsize{+}$ & $0.32 ~\textrm{ring\_type\_eq\_none}$ & $\scriptsize{+}$ & $0.32 ~\textrm{stalk\_color\_above\_ring\_eq\_cinnamon}$ \\ 
 $\scriptsize{+}$ & $0.32 ~\textrm{stalk\_color\_below\_ring\_eq\_cinnamon}$ & $\scriptsize{+}$ & $0.31 ~\textrm{habitat\_eq\_urban}$ & $\scriptsize{+}$ & $0.29 ~\textrm{cap\_shape\_eq\_bell}$ \\ 
 $\scriptsize{+}$ & $0.28 ~\textrm{stalk\_color\_below\_ring\_eq\_pink}$ & $\scriptsize{+}$ & $0.26 ~\textrm{population\_eq\_scattered}$ & $\scriptsize{+}$ & $0.24 ~\textrm{stalk\_color\_above\_ring\_eq\_pink}$ \\ 
 $\scriptsize{+}$ & $0.21 ~\textrm{stalk\_color\_below\_ring\_eq\_white}$ & $\scriptsize{+}$ & $0.18 ~\textrm{cap\_surface\_eq\_smooth}$ & $\scriptsize{+}$ & $0.14 ~\textrm{gill\_color\_eq\_yellow}$ \\ 
 $\scriptsize{+}$ & $0.14 ~\textrm{stalk\_root\_eq\_equal}$ & $\scriptsize{+}$ & $0.14 ~\textrm{spore\_print\_color\_eq\_white}$ & $\scriptsize{+}$ & $0.12 ~\textrm{stalk\_color\_above\_ring\_eq\_buff}$ \\ 
 $\scriptsize{+}$ & $0.09 ~\textrm{gill\_attachment\_eq\_free}$ & $\scriptsize{+}$ & $0.07 ~\textrm{habitat\_eq\_leaves}$ & $\scriptsize{+}$ & $0.07 ~\textrm{bruises\_eq\_TRUE}$ \\ 
 $\scriptsize{+}$ & $0.06 ~\textrm{stalk\_color\_below\_ring\_eq\_buff}$ & $\scriptsize{+}$ & $0.05 ~\textrm{stalk\_surface\_above\_ring\_eq\_scaly}$ & $\scriptsize{+}$ & $0.05 ~\textrm{gill\_color\_eq\_chocolate}$ \\ 
 $\scriptsize{-}$ & $4.04 ~\textrm{odor\_eq\_none}$ & $\scriptsize{-}$ & $2.73 ~\textrm{odor\_eq\_anise}$ & $\scriptsize{-}$ & $2.73 ~\textrm{odor\_eq\_almond}$ \\ 
 $\scriptsize{-}$ & $2.05 ~\textrm{gill\_size\_eq\_broad}$ & $\scriptsize{-}$ & $1.51 ~\textrm{ring\_type\_eq\_flaring}$ & $\scriptsize{-}$ & $1.47 ~\textrm{spore\_print\_color\_eq\_brown}$ \\ 
 $\scriptsize{-}$ & $1.41 ~\textrm{gill\_spacing\_eq\_crowded}$ & $\scriptsize{-}$ & $1.37 ~\textrm{spore\_print\_color\_eq\_purple}$ & $\scriptsize{-}$ & $1.26 ~\textrm{spore\_print\_color\_eq\_black}$ \\ 
 $\scriptsize{-}$ & $1.15 ~\textrm{population\_eq\_solitary}$ & $\scriptsize{-}$ & $1.12 ~\textrm{habitat\_eq\_waste}$ & $\scriptsize{-}$ & $1.10 ~\textrm{stalk\_surface\_below\_ring\_eq\_fibrous}$ \\ 
 $\scriptsize{-}$ & $1.08 ~\textrm{stalk\_surface\_above\_ring\_eq\_smooth}$ & $\scriptsize{-}$ & $0.89 ~\textrm{stalk\_color\_below\_ring\_eq\_brown}$ & $\scriptsize{-}$ & $0.82 ~\textrm{stalk\_root\_eq\_rooted}$ \\ 
 $\scriptsize{-}$ & $0.79 ~\textrm{cap\_surface\_eq\_fibrous}$ & $\scriptsize{-}$ & $0.76 ~\textrm{population\_eq\_numerous}$ & $\scriptsize{-}$ & $0.75 ~\textrm{cap\_color\_eq\_cinnamon}$ \\ 
 $\scriptsize{-}$ & $0.73 ~\textrm{stalk\_root\_eq\_club}$ & $\scriptsize{-}$ & $0.60 ~\textrm{stalk\_surface\_above\_ring\_eq\_fibrous}$ & $\scriptsize{-}$ & $0.58 ~\textrm{gill\_color\_eq\_brown}$ \\ 
 $\scriptsize{-}$ & $0.58 ~\textrm{stalk\_shape\_eq\_tapering}$ & $\scriptsize{-}$ & $0.46 ~\textrm{gill\_color\_eq\_red}$ & $\scriptsize{-}$ & $0.45 ~\textrm{stalk\_surface\_below\_ring\_eq\_smooth}$ \\ 
 $\scriptsize{-}$ & $0.41 ~\textrm{cap\_shape\_eq\_sunken}$ & $\scriptsize{-}$ & $0.39 ~\textrm{gill\_color\_eq\_pink}$ & $\scriptsize{-}$ & $0.39 ~\textrm{stalk\_color\_above\_ring\_eq\_red}$ \\ 
 $\scriptsize{-}$ & $0.38 ~\textrm{cap\_color\_eq\_brown}$ & $\scriptsize{-}$ & $0.37 ~\textrm{habitat\_eq\_woods}$ & $\scriptsize{-}$ & $0.34 ~\textrm{stalk\_color\_below\_ring\_eq\_red}$ \\ 
 $\scriptsize{-}$ & $0.33 ~\textrm{stalk\_color\_below\_ring\_eq\_orange}$ & $\scriptsize{-}$ & $0.33 ~\textrm{stalk\_color\_above\_ring\_eq\_orange}$ & $\scriptsize{-}$ & $0.32 ~\textrm{gill\_color\_eq\_black}$ \\ 
 $\scriptsize{-}$ & $0.30 ~\textrm{gill\_color\_eq\_white}$ & $\scriptsize{-}$ & $0.29 ~\textrm{cap\_color\_eq\_green}$ & $\scriptsize{-}$ & $0.29 ~\textrm{cap\_color\_eq\_purple}$ \\ 
 $\scriptsize{-}$ & $0.29 ~\textrm{stalk\_color\_above\_ring\_eq\_gray}$ & $\scriptsize{-}$ & $0.28 ~\textrm{habitat\_eq\_paths}$ & $\scriptsize{-}$ & $0.26 ~\textrm{stalk\_color\_below\_ring\_eq\_gray}$ \\ 
 $\scriptsize{-}$ & $0.20 ~\textrm{cap\_color\_eq\_yellow}$ & $\scriptsize{-}$ & $0.19 ~\textrm{gill\_color\_eq\_purple}$ & $\scriptsize{-}$ & $0.18 ~\textrm{ring\_number\_eq\_2}$ \\ 
 $\scriptsize{-}$ & $0.13 ~\textrm{cap\_shape\_eq\_convex}$ & $\scriptsize{-}$ & $0.11 ~\textrm{veil\_color\_eq\_white}$ & $\scriptsize{-}$ & $0.11 ~\textrm{veil\_color\_eq\_orange}$ \\ 
 $\scriptsize{-}$ & $0.11 ~\textrm{veil\_color\_eq\_brown}$ & $\scriptsize{-}$ & $0.10 ~\textrm{stalk\_color\_above\_ring\_eq\_brown}$ & $\scriptsize{-}$ & $0.09 ~\textrm{gill\_attachment\_eq\_attached}$ \\ 
 $\scriptsize{-}$ & $0.06 ~\textrm{ring\_type\_eq\_pendant}$ & $\scriptsize{-}$ & $0.05 ~\textrm{gill\_color\_eq\_orange}$ & $\scriptsize{-}$ & $0.03 ~\textrm{ring\_number\_eq\_1}$ \\ 
 $\scriptsize{-}$ & $0.02 ~\textrm{spore\_print\_color\_eq\_buff}$ & $\scriptsize{-}$ & $0.02 ~\textrm{spore\_print\_color\_eq\_yellow}$ & $\scriptsize{-}$ & $0.02 ~\textrm{spore\_print\_color\_eq\_orange}$ \\ 
 $\scriptsize{-}$ & $0.02 ~\textrm{cap\_shape\_eq\_flat}$ & $\scriptsize{-}$ & $0.01 ~\textrm{stalk\_color\_above\_ring\_eq\_white}$ & $\scriptsize{-}$ & $0.00 ~\textrm{cap\_surface\_eq\_scaly}$ \\ 
 $\scriptsize{+}$ & $0.42$ &  &  &  &  \\ 
  \end{tabularx}
}
\caption{Elastic Net score function trained on the full \textds{mushroom} dataset. This model uses 108 coefficients and has a 10-fold CV test error of 0.0 $\pm$ 0.0$\%$.}
\end{figure}

\begin{figure}[htbp]
\centering
\tiny
\begin{verbatim}
Rule 1: (3216, lift 1.9)
	odor_eq_none > 0
	gill_size_eq_narrow <= 0
	spore_print_color_eq_green <= 0
	->  class 0  [1.000]

Rule 2: (1440, lift 1.9)
	bruises_eq_TRUE <= 0
	odor_eq_none > 0
	stalk_surface_below_ring_eq_scaly <= 0
	->  class 0  [0.999]

Rule 3: (400, lift 1.9)
	odor_eq_almond > 0
	->  class 0  [0.998]

Rule 4: (400, lift 1.9)
	odor_eq_anise > 0
	->  class 0  [0.998]

Rule 5: (3796, lift 2.1)
	odor_eq_almond <= 0
	odor_eq_anise <= 0
	odor_eq_none <= 0
	->  class 1  [1.000]

Rule 6: (72, lift 2.0)
	spore_print_color_eq_green > 0
	->  class 1  [0.986]

Rule 7: (40, lift 2.0)
	gill_size_eq_narrow > 0
	stalk_surface_below_ring_eq_scaly > 0
	->  class 1  [0.976]
	
Default class: 1
\end{verbatim}
\caption{C5.0R model trained the full \textds{mushroom} dataset. This model has 7 rules and a 10-fold CV test error of $\pm$ 0.0$\%$.}
\end{figure}

\begin{figure}[htbp]
\centering
\tiny
\begin{verbatim}
odor_eq_none <= 0:
:...odor_eq_almond > 0: 0 (400)
:   odor_eq_almond <= 0:
:   :...odor_eq_anise <= 0: 1 (3796)
:       odor_eq_anise > 0: 0 (400)
odor_eq_none > 0:
:...spore_print_color_eq_green > 0: 1 (72)
    spore_print_color_eq_green <= 0:
    :...stalk_surface_below_ring_eq_scaly > 0:
        :...gill_size_eq_narrow <= 0: 0 (16)
        :   gill_size_eq_narrow > 0: 1 (40)
        stalk_surface_below_ring_eq_scaly <= 0:
        :...gill_size_eq_narrow <= 0: 0 (3200)
            gill_size_eq_narrow > 0:
            :...bruises_eq_TRUE <= 0: 0 (192)
                bruises_eq_TRUE > 0: 1 (8)
\end{verbatim}
\caption{C5.0T model trained on the full \textds{mushroom} dataset. This model has 9 leaves and a 10-fold CV test error of $\pm$ 0.0$\%$.}
\end{figure}

\begin{figure}[htbp]
{\scriptsize
\begin{tabularx}{\textwidth}{p{1mm}lp{1mm}lp{1mm}l}
& $1.45 ~\textrm{spore\_print\_color\_eq\_green}$ & $\scriptsize{+}$ & $1.04 ~\textrm{odor\_eq\_creosote}$ & $\scriptsize{+}$ & $0.83 ~\textrm{stalk\_surface\_above\_ring\_eq\_grooves}$ \\ 
 $\scriptsize{+}$ & $0.69 ~\textrm{odor\_eq\_pungent}$ & $\scriptsize{+}$ & $0.64 ~\textrm{population\_eq\_clustered}$ & $\scriptsize{+}$ & $0.61 ~\textrm{gill\_color\_eq\_buff}$ \\ 
 $\scriptsize{+}$ & $0.55 ~\textrm{odor\_eq\_foul}$ & $\scriptsize{+}$ & $0.55 ~\textrm{stalk\_root\_eq\_bulbous}$ & $\scriptsize{+}$ & $0.47 ~\textrm{gill\_size\_eq\_narrow}$ \\ 
 $\scriptsize{+}$ & $0.41 ~\textrm{stalk\_surface\_below\_ring\_eq\_scaly}$ & $\scriptsize{+}$ & $0.39 ~\textrm{bruises\_eq\_TRUE}$ & $\scriptsize{+}$ & $0.39 ~\textrm{gill\_spacing\_eq\_close}$ \\ 
 $\scriptsize{+}$ & $0.31 ~\textrm{odor\_eq\_fishy}$ & $\scriptsize{+}$ & $0.31 ~\textrm{odor\_eq\_spicy}$ & $\scriptsize{+}$ & $0.30 ~\textrm{habitat\_eq\_meadows}$ \\ 
 $\scriptsize{+}$ & $0.30 ~\textrm{habitat\_eq\_grasses}$ & $\scriptsize{+}$ & $0.27 ~\textrm{cap\_surface\_eq\_grooves}$ & $\scriptsize{+}$ & $0.25 ~\textrm{ring\_number\_eq\_1}$ \\ 
 $\scriptsize{+}$ & $0.25 ~\textrm{ring\_type\_eq\_pendant}$ & $\scriptsize{+}$ & $0.22 ~\textrm{ring\_type\_eq\_evanescent}$ & $\scriptsize{+}$ & $0.18 ~\textrm{cap\_shape\_eq\_conical}$ \\ 
 $\scriptsize{+}$ & $0.18 ~\textrm{stalk\_shape\_eq\_elarging}$ & $\scriptsize{+}$ & $0.17 ~\textrm{stalk\_color\_above\_ring\_eq\_yellow}$ & $\scriptsize{+}$ & $0.17 ~\textrm{stalk\_color\_below\_ring\_eq\_yellow}$ \\ 
 $\scriptsize{+}$ & $0.17 ~\textrm{veil\_color\_eq\_yellow}$ & $\scriptsize{+}$ & $0.12 ~\textrm{stalk\_color\_below\_ring\_eq\_pink}$ & $\scriptsize{+}$ & $0.12 ~\textrm{stalk\_color\_below\_ring\_eq\_white}$ \\ 
 $\scriptsize{+}$ & $0.11 ~\textrm{habitat\_eq\_urban}$ & $\scriptsize{+}$ & $0.10 ~\textrm{gill\_attachment\_eq\_free}$ & $\scriptsize{+}$ & $0.05 ~\textrm{stalk\_color\_above\_ring\_eq\_pink}$ \\ 
 $\scriptsize{+}$ & $0.05 ~\textrm{stalk\_color\_above\_ring\_eq\_white}$ & $\scriptsize{+}$ & $0.04 ~\textrm{ring\_type\_eq\_large}$ & $\scriptsize{+}$ & $0.04 ~\textrm{spore\_print\_color\_eq\_chocolate}$ \\ 
 $\scriptsize{+}$ & $0.02 ~\textrm{stalk\_root\_eq\_club}$ & $\scriptsize{+}$ & $0.00 ~\textrm{cap\_color\_eq\_pink}$ & $\scriptsize{+}$ & $0.00 ~\textrm{cap\_color\_eq\_buff}$ \\ 
 $\scriptsize{+}$ & $0.00 ~\textrm{cap\_color\_eq\_white}$ & $\scriptsize{+}$ & $0.00 ~\textrm{cap\_color\_eq\_yellow}$ & $\scriptsize{+}$ & $0.00 ~\textrm{cap\_color\_eq\_gray}$ \\ 
 $\scriptsize{-}$ & $0.99 ~\textrm{odor\_eq\_anise}$ & $\scriptsize{-}$ & $0.99 ~\textrm{odor\_eq\_almond}$ & $\scriptsize{-}$ & $0.92 ~\textrm{odor\_eq\_none}$ \\ 
 $\scriptsize{-}$ & $0.52 ~\textrm{stalk\_root\_eq\_rooted}$ & $\scriptsize{-}$ & $0.51 ~\textrm{ring\_type\_eq\_flaring}$ & $\scriptsize{-}$ & $0.47 ~\textrm{gill\_size\_eq\_broad}$ \\ 
 $\scriptsize{-}$ & $0.45 ~\textrm{spore\_print\_color\_eq\_purple}$ & $\scriptsize{-}$ & $0.45 ~\textrm{spore\_print\_color\_eq\_brown}$ & $\scriptsize{-}$ & $0.45 ~\textrm{spore\_print\_color\_eq\_black}$ \\ 
 $\scriptsize{-}$ & $0.43 ~\textrm{habitat\_eq\_waste}$ & $\scriptsize{-}$ & $0.41 ~\textrm{stalk\_root\_eq\_equal}$ & $\scriptsize{-}$ & $0.39 ~\textrm{stalk\_surface\_above\_ring\_eq\_fibrous}$ \\ 
 $\scriptsize{-}$ & $0.39 ~\textrm{stalk\_surface\_above\_ring\_eq\_smooth}$ & $\scriptsize{-}$ & $0.39 ~\textrm{gill\_spacing\_eq\_crowded}$ & $\scriptsize{-}$ & $0.31 ~\textrm{stalk\_color\_below\_ring\_eq\_brown}$ \\ 
 $\scriptsize{-}$ & $0.25 ~\textrm{ring\_number\_eq\_2}$ & $\scriptsize{-}$ & $0.21 ~\textrm{habitat\_eq\_paths}$ & $\scriptsize{-}$ & $0.18 ~\textrm{stalk\_shape\_eq\_tapering}$ \\ 
 $\scriptsize{-}$ & $0.18 ~\textrm{stalk\_color\_above\_ring\_eq\_brown}$ & $\scriptsize{-}$ & $0.18 ~\textrm{population\_eq\_solitary}$ & $\scriptsize{-}$ & $0.16 ~\textrm{population\_eq\_several}$ \\ 
 $\scriptsize{-}$ & $0.16 ~\textrm{population\_eq\_scattered}$ & $\scriptsize{-}$ & $0.16 ~\textrm{population\_eq\_numerous}$ & $\scriptsize{-}$ & $0.14 ~\textrm{stalk\_surface\_below\_ring\_eq\_fibrous}$ \\ 
 $\scriptsize{-}$ & $0.14 ~\textrm{stalk\_surface\_below\_ring\_eq\_smooth}$ & $\scriptsize{-}$ & $0.14 ~\textrm{stalk\_surface\_below\_ring\_eq\_grooves}$ & $\scriptsize{-}$ & $0.10 ~\textrm{gill\_attachment\_eq\_attached}$ \\ 
 $\scriptsize{-}$ & $0.10 ~\textrm{stalk\_color\_above\_ring\_eq\_orange}$ & $\scriptsize{-}$ & $0.10 ~\textrm{stalk\_color\_below\_ring\_eq\_orange}$ & $\scriptsize{-}$ & $0.09 ~\textrm{cap\_surface\_eq\_fibrous}$ \\ 
 $\scriptsize{-}$ & $0.09 ~\textrm{cap\_surface\_eq\_scaly}$ & $\scriptsize{-}$ & $0.09 ~\textrm{cap\_surface\_eq\_smooth}$ & $\scriptsize{-}$ & $0.07 ~\textrm{veil\_color\_eq\_white}$ \\ 
 $\scriptsize{-}$ & $0.06 ~\textrm{gill\_color\_eq\_red}$ & $\scriptsize{-}$ & $0.06 ~\textrm{gill\_color\_eq\_pink}$ & $\scriptsize{-}$ & $0.06 ~\textrm{gill\_color\_eq\_brown}$ \\ 
 $\scriptsize{-}$ & $0.06 ~\textrm{gill\_color\_eq\_black}$ & $\scriptsize{-}$ & $0.06 ~\textrm{gill\_color\_eq\_white}$ & $\scriptsize{-}$ & $0.06 ~\textrm{gill\_color\_eq\_gray}$ \\ 
 $\scriptsize{-}$ & $0.06 ~\textrm{gill\_color\_eq\_purple}$ & $\scriptsize{-}$ & $0.06 ~\textrm{gill\_color\_eq\_chocolate}$ & $\scriptsize{-}$ & $0.05 ~\textrm{gill\_color\_eq\_yellow}$ \\ 
 $\scriptsize{-}$ & $0.05 ~\textrm{veil\_color\_eq\_orange}$ & $\scriptsize{-}$ & $0.05 ~\textrm{gill\_color\_eq\_orange}$ & $\scriptsize{-}$ & $0.05 ~\textrm{veil\_color\_eq\_brown}$ \\ 
 $\scriptsize{-}$ & $0.05 ~\textrm{stalk\_surface\_above\_ring\_eq\_scaly}$ & $\scriptsize{-}$ & $0.04 ~\textrm{spore\_print\_color\_eq\_white}$ & $\scriptsize{-}$ & $0.04 ~\textrm{cap\_shape\_eq\_flat}$ \\ 
 $\scriptsize{-}$ & $0.04 ~\textrm{cap\_shape\_eq\_convex}$ & $\scriptsize{-}$ & $0.04 ~\textrm{cap\_shape\_eq\_sunken}$ & $\scriptsize{-}$ & $0.04 ~\textrm{cap\_shape\_eq\_knobbed}$ \\ 
 $\scriptsize{-}$ & $0.04 ~\textrm{cap\_shape\_eq\_bell}$ & $\scriptsize{-}$ & $0.03 ~\textrm{spore\_print\_color\_eq\_yellow}$ & $\scriptsize{-}$ & $0.03 ~\textrm{spore\_print\_color\_eq\_orange}$ \\ 
 $\scriptsize{-}$ & $0.03 ~\textrm{spore\_print\_color\_eq\_buff}$ & $\scriptsize{-}$ & $0.03 ~\textrm{habitat\_eq\_leaves}$ & $\scriptsize{-}$ & $0.03 ~\textrm{habitat\_eq\_woods}$ \\ 
    $\scriptsize{-}$ & $0.00 ~\textrm{cap\_color\_eq\_cinnamon}$ & $\scriptsize{-}$ & $0.00 ~\textrm{cap\_color\_eq\_brown}$ & $\scriptsize{-}$ & $0.21$ \\ 
  \end{tabularx}
}
\caption{SVM score function trained on the full \texttt{mushroom} dataset.  This model uses 99 coefficients and has a 10-fold CV test error of 0.0 $\pm$ 0.0$\%$.}
\end{figure}

\clearpage
\section{Numerical Experiments with Longer Training Time}\label{Appendix::ExtraComputationalTime}
In this section, we show the results of the numerical experiments from Section \ref{Sec::NumericalExperiments} when train the IP-based methods from our framework for a longer period of time. Here, we use exactly the same setup as in Section \ref{Sec::NumericalExperiments} but allocate 60 minutes to solve each IP associated with SLIM and MN Rules (as opposed to 10 minutes/IP). Thus, the training process for SLIM and MN Rules involves at most 6 hours of total computing time for each dataset.

When we compare these results to those in Section \ref{Sec::NumericalExperiments}, we find that the IP-based classifiers that were trained for longer have a lower objective values as well as a better guarantee on optimality (i.e. mipgap). However, these classifiers do not necessarily show an improvement in terms of test error or sparsity. As shown in the regularization path plots, we find that MN Rules and the smaller SLIM models (trained at small values of $C_0$) have the same predictive accuracy. However, the larger SLIM models (trained at smaller values of $C_0$) typically have worse predictive accuracy. This suggests that using a large $\Lset$ may overfit the data -- a result that we would expect given that a larger $\Lset$ allow for more complex models. In this case, we can counteract the overfitting can be counteracted by using good feasible solutions instead the optimal solution.

\begin{table}[htbp]
\scriptsize
\centering
\resizebox{\textwidth}{!} {
\setlength{\tabcolsep}{1.5pt}
\begin{tabular}{>{\scriptsize}l>{\scriptsize}c>{\scriptsize}lcccccccccc}
   \toprule 
   \bf{Dataset}&\bf{Details}&\bf{Metric}&\bf{Lasso}&\bf{Ridge}&\bf{E. Net}&\bf{C5.0R}&\bf{C5.0T}&\bf{CART}&\bf{SVM Lin.}&\bf{SVM RBF}&\bf{MN Rules}&\bf{SLIM} \\ \toprule 
   
   \texttt{adult} & \ddcell{$N$&32561\\$P$&36\\$\Pr(y\text{=+1})$&24\%\\$\Pr(y\text{=-1})$&76\%} & \bfcell{l}{test error\\train error\\model size \\model range} & \cell{c}{17.3 $\pm$ 0.9$\%$\\17.2 $\pm$ 0.1$\%$\\14\\13 - 14} & \cell{c}{17.6 $\pm$ 0.9$\%$\\17.6 $\pm$ 0.1$\%$\\36\\36 - 36} & \cell{c}{17.4 $\pm$ 0.9$\%$\\17.4 $\pm$ 0.1$\%$\\17\\16 - 18} & \cell{c}{26.4 $\pm$ 1.8$\%$\\25.3 $\pm$ 0.4$\%$\\41\\38 - 46} & \cell{c}{26.3 $\pm$ 1.4$\%$\\24.9 $\pm$ 0.4$\%$\\84\\78 - 99} & \cell{c}{75.9 $\pm$ 0.0$\%$\\75.9 $\pm$ 0.0$\%$\\4\\4 - 4} & \cell{c}{16.8 $\pm$ 0.8$\%$\\16.7 $\pm$ 0.1$\%$\\36\\36 - 36} & \cell{c}{16.3 $\pm$ 0.5$\%$\\16.3 $\pm$ 0.1$\%$\\36\\36 - 36} & \cell{c}{19.2 $\pm$ 1.0$\%$\\19.2 $\pm$ 0.2$\%$\\9\\4 - 19} & \cell{c}{17.7 $\pm$ 1.0$\%$\\17.6 $\pm$ 1.0$\%$\\22\\21 - 28} \\ \midrule 
   
   \texttt{breastcancer} & \ddcell{$N$&683\\$P$&9\\$\Pr(y\text{=+1})$&35\%\\$\Pr(y\text{=-1})$&65\%} & \bfcell{l}{test error\\train error\\model size \\model range} & \cell{c}{3.4 $\pm$ 2.2$\%$\\2.9 $\pm$ 0.3$\%$\\9\\8 - 9} & \cell{c}{3.4 $\pm$ 1.7$\%$\\3.0 $\pm$ 0.3$\%$\\9\\9 - 9} & \cell{c}{3.1 $\pm$ 2.1$\%$\\2.8 $\pm$ 0.3$\%$\\9\\9 - 9} & \cell{c}{4.3 $\pm$ 3.3$\%$\\2.1 $\pm$ 0.3$\%$\\7\\6 - 9} & \cell{c}{5.3 $\pm$ 3.4$\%$\\1.6 $\pm$ 0.4$\%$\\13\\7 - 16} & \cell{c}{5.6 $\pm$ 1.9$\%$\\3.6 $\pm$ 0.3$\%$\\4\\3 - 7} & \cell{c}{3.1 $\pm$ 2.0$\%$\\2.7 $\pm$ 0.2$\%$\\9\\9 - 9} & \cell{c}{3.5 $\pm$ 2.5$\%$\\0.3 $\pm$ 0.1$\%$\\9\\9 - 9} & \cell{c}{4.8 $\pm$ 2.5$\%$\\4.1 $\pm$ 0.2$\%$\\8\\7 - 8} & \cell{c}{4.0 $\pm$ 2.5$\%$\\1.6 $\pm$ 0.2$\%$\\6\\5 - 9} \\ \midrule 
   
   \texttt{bankruptcy} & \ddcell{$N$&250\\$P$&6\\$\Pr(y\text{=+1})$&57\%\\$\Pr(y\text{=-1})$&43\%} & \bfcell{l}{test error\\train error\\model size \\model range} & \cell{c}{0.0 $\pm$ 0.0$\%$\\0.0 $\pm$ 0.0$\%$\\3\\3 - 3} & \cell{c}{0.4 $\pm$ 1.3$\%$\\0.4 $\pm$ 0.1$\%$\\6\\6 - 6} & \cell{c}{0.0 $\pm$ 0.0$\%$\\0.4 $\pm$ 0.7$\%$\\3\\3 - 3} & \cell{c}{0.8 $\pm$ 1.7$\%$\\0.4 $\pm$ 0.2$\%$\\4\\4 - 4} & \cell{c}{0.8 $\pm$ 1.7$\%$\\0.4 $\pm$ 0.2$\%$\\4\\4 - 4} & \cell{c}{1.6 $\pm$ 2.8$\%$\\1.6 $\pm$ 0.3$\%$\\2\\2 - 2} & \cell{c}{0.4 $\pm$ 1.3$\%$\\0.4 $\pm$ 0.1$\%$\\6\\6 - 6} & \cell{c}{0.4 $\pm$ 1.3$\%$\\0.4 $\pm$ 0.1$\%$\\6\\6 - 6} & \cell{c}{2.8 $\pm$ 1.9$\%$\\0.7 $\pm$ 0.2$\%$\\8\\7 - 12} & \cell{c}{0.8 $\pm$ 1.7$\%$\\0.0 $\pm$ 0.0$\%$\\3\\2 - 3} \\    \midrule

 \texttt{haberman} & \ddcell{$N$&306\\$P$&3\\$\Pr(y\text{=+1})$&74\%\\$\Pr(y\text{=-1})$&26\%} & \bfcell{l}{test error\\train error\\model size \\model range} & \cell{c}{42.5 $\pm$ 11.3$\%$\\40.6 $\pm$ 1.9$\%$\\2\\2 - 2} & \cell{c}{36.9 $\pm$ 15.0$\%$\\41.0 $\pm$ 9.7$\%$\\3\\3 - 3} & \cell{c}{40.9 $\pm$ 14.0$\%$\\45.1 $\pm$ 12.0$\%$\\1\\1 - 1} & \cell{c}{42.7 $\pm$ 9.4$\%$\\40.4 $\pm$ 8.5$\%$\\2\\0 - 3} & \cell{c}{42.7 $\pm$ 9.4$\%$\\40.4 $\pm$ 8.5$\%$\\2\\1 - 3} & \cell{c}{43.1 $\pm$ 8.0$\%$\\34.3 $\pm$ 2.8$\%$\\6\\4 - 9} & \cell{c}{45.3 $\pm$ 14.7$\%$\\46.0 $\pm$ 3.6$\%$\\3\\3 - 3} & \cell{c}{47.5 $\pm$ 6.2$\%$\\5.4 $\pm$ 1.5$\%$\\4\\4 - 4} & \cell{c}{54.7 $\pm$ 24.3$\%$\\54.7 $\pm$ 24.3$\%$\\1\\0 - 1} & \cell{c}{38.4 $\pm$ 10.2$\%$\\35.8 $\pm$ 1.3$\%$\\3\\3 - 4} \\    \midrule

 \texttt{mammo} & \ddcell{$N$&961\\$P$&14\\$\Pr(y\text{=+1})$&46\%\\$\Pr(y\text{=-1})$&54\%} & \bfcell{l}{test error\\train error\\model size \\model range} & \cell{c}{19.0 $\pm$ 3.1$\%$\\19.3 $\pm$ 0.3$\%$\\13\\12 - 13} & \cell{c}{19.2 $\pm$ 3.0$\%$\\19.2 $\pm$ 0.4$\%$\\14\\14 - 14} & \cell{c}{19.0 $\pm$ 3.1$\%$\\19.2 $\pm$ 0.3$\%$\\14\\13 - 14} & \cell{c}{20.5 $\pm$ 3.3$\%$\\19.8 $\pm$ 0.3$\%$\\5\\3 - 5} & \cell{c}{20.3 $\pm$ 3.5$\%$\\19.9 $\pm$ 0.3$\%$\\5\\4 - 6} & \cell{c}{20.7 $\pm$ 3.9$\%$\\20.0 $\pm$ 0.6$\%$\\4\\3 - 5} & \cell{c}{20.3 $\pm$ 3.0$\%$\\20.3 $\pm$ 0.4$\%$\\14\\14 - 14} & \cell{c}{19.1 $\pm$ 3.1$\%$\\18.2 $\pm$ 0.4$\%$\\14\\14 - 14} & \cell{c}{21.6 $\pm$ 3.5$\%$\\20.8 $\pm$ 0.3$\%$\\9\\9 - 9} & \cell{c}{19.5 $\pm$ 3.0$\%$\\18.3 $\pm$ 0.3$\%$\\9\\9 - 11} \\    \midrule

 \texttt{heart} & \ddcell{$N$&303\\$P$&32\\$\Pr(y\text{=+1})$&46\%\\$\Pr(y\text{=-1})$&54\%} & \bfcell{l}{test error\\train error\\model size \\model range} & \cell{c}{15.2 $\pm$ 6.3$\%$\\14.0 $\pm$ 1.0$\%$\\11\\10 - 13} & \cell{c}{14.9 $\pm$ 5.9$\%$\\13.1 $\pm$ 0.8$\%$\\32\\30 - 32} & \cell{c}{14.5 $\pm$ 5.9$\%$\\13.2 $\pm$ 0.6$\%$\\24\\22 - 27} & \cell{c}{21.2 $\pm$ 7.5$\%$\\10.0 $\pm$ 1.8$\%$\\10\\9 - 17} & \cell{c}{23.2 $\pm$ 6.8$\%$\\8.5 $\pm$ 2.0$\%$\\19\\12 - 27} & \cell{c}{19.8 $\pm$ 6.5$\%$\\14.3 $\pm$ 0.9$\%$\\6\\6 - 8} & \cell{c}{15.5 $\pm$ 6.5$\%$\\13.6 $\pm$ 0.5$\%$\\31\\28 - 32} & \cell{c}{15.2 $\pm$ 6.0$\%$\\10.4 $\pm$ 0.8$\%$\\32\\32 - 32} & \cell{c}{23.2 $\pm$ 10.4$\%$\\17.8 $\pm$ 0.8$\%$\\15\\10 - 16} & \cell{c}{19.2 $\pm$ 6.7$\%$\\7.6 $\pm$ 0.8$\%$\\16\\13 - 18} \\    \midrule

 \texttt{mushroom} & \ddcell{$N$&8124\\$P$&113\\$\Pr(y\text{=+1})$&48\%\\$\Pr(y\text{=-1})$&52\%} & \bfcell{l}{test error\\train error\\model size \\model range} & \cell{c}{0.0 $\pm$ 0.0$\%$\\0.0 $\pm$ 0.0$\%$\\25\\23 - 26} & \cell{c}{1.7 $\pm$ 0.3$\%$\\1.7 $\pm$ 0.0$\%$\\113\\113 - 113} & \cell{c}{0.0 $\pm$ 0.0$\%$\\0.0 $\pm$ 0.0$\%$\\108\\106 - 108} & \cell{c}{0.0 $\pm$ 0.0$\%$\\0.0 $\pm$ 0.0$\%$\\7\\7 - 7} & \cell{c}{0.0 $\pm$ 0.0$\%$\\0.0 $\pm$ 0.0$\%$\\9\\9 - 9} & \cell{c}{1.2 $\pm$ 0.6$\%$\\1.1 $\pm$ 0.3$\%$\\7\\6 - 8} & \cell{c}{0.0 $\pm$ 0.0$\%$\\0.0 $\pm$ 0.0$\%$\\104\\99 - 108} & \cell{c}{0.0 $\pm$ 0.0$\%$\\0.0 $\pm$ 0.0$\%$\\113\\113 - 113} & \cell{c}{0.0 $\pm$ 0.0$\%$\\0.0 $\pm$ 0.0$\%$\\21\\21 - 21} & \cell{c}{0.0 $\pm$ 0.0$\%$\\0.0 $\pm$ 0.0$\%$\\7\\7 - 7} \\    \midrule

 \texttt{spambase} & \ddcell{$N$&4601\\$P$&57\\$\Pr(y\text{=+1})$&39\%\\$\Pr(y\text{=-1})$&61\%} & \bfcell{l}{test error\\train error\\model size \\model range} & \cell{c}{10.0 $\pm$ 1.7$\%$\\9.5 $\pm$ 0.3$\%$\\28\\28 - 29} & \cell{c}{26.3 $\pm$ 1.7$\%$\\26.1 $\pm$ 0.2$\%$\\57\\57 - 57} & \cell{c}{10.0 $\pm$ 1.7$\%$\\9.6 $\pm$ 0.2$\%$\\28\\28 - 29} & \cell{c}{6.6 $\pm$ 1.3$\%$\\4.2 $\pm$ 0.3$\%$\\27\\23 - 31} & \cell{c}{7.3 $\pm$ 1.0$\%$\\3.9 $\pm$ 0.3$\%$\\69\\56 - 78} & \cell{c}{11.1 $\pm$ 1.4$\%$\\9.8 $\pm$ 0.3$\%$\\7\\6 - 10} & \cell{c}{7.8 $\pm$ 1.5$\%$\\8.1 $\pm$ 0.8$\%$\\57\\57 - 57} & \cell{c}{13.7 $\pm$ 1.4$\%$\\1.3 $\pm$ 0.1$\%$\\57\\57 - 57} & \cell{c}{9.7 $\pm$ 1.1$\%$\\9.1 $\pm$ 0.3$\%$\\33\\26 - 41} & \cell{c}{6.5 $\pm$ 1.3$\%$\\5.4 $\pm$ 0.2$\%$\\36\\30 - 39} \\ 
   
   \bottomrule \end{tabular}

}
\caption{Accuracy and sparsity of all methods on UCI datasets. Here: test error denotes the 10-fold CV test error; train error denotes the 10-fold CV training error; model size corresponds to the 10-fold CV median model size; model range is the 10-fold minimum and maximum model-size. We have set free parameters to whatever values minimized the mean 10-fold CV error so as to reflect the most accurate model that was produced by each method.}
\label{Table::ExpResultsTable}
\end{table}

\clearpage
\bibliographystyle{plainnat}
\bibliography{MMforInterpretableClassification}
\end{document}